\documentclass[12pt]{article} 
\usepackage{epsfig}
\usepackage{amsfonts}
\usepackage{latexsym}
\usepackage{amsmath,amssymb}
\usepackage{verbatim}
\usepackage{setspace}
\usepackage[small]{caption}
\usepackage{graphicx}
\usepackage{feynmp}
\usepackage{color}
\definecolor{darkgreen}{rgb}{0,0.5,0}
\definecolor{darkblue}{rgb}{0,0,0.6}
\definecolor{purple}{rgb}{0.4,0.15,0.21}
\usepackage[colorlinks=true,citecolor=darkgreen,linkcolor=purple,urlcolor=darkblue]{hyperref}
\usepackage[textheight=9in, textwidth=6.5in, letterpaper]{geometry}

\numberwithin{equation}{section}

\def\tilde{\widetilde}
\newcommand{\bea}{\begin{eqnarray}}
\newcommand{\eea}{\end{eqnarray}}
\newcommand{\be}{\begin{equation}}
\newcommand{\ee}{\end{equation}}
\newcommand{\ba}{\begin{align}}
\newcommand{\ea}{\end{align}}

  \makeatletter
  \let\over=\@@over \let\overwithdelims=\@@overwithdelims
  \let\atop=\@@atop \let\atopwithdelims=\@@atopwithdelims
  \let\above=\@@above \let\abovewithdelims=\@@abovewithdelims

\renewcommand\section{\@startsection {section}{1}{\z@}%
                                   {-3.5ex \@plus -1ex \@minus -.2ex}%nn
                                   {2.3ex \@plus.2ex}%
                                   {\normalfont\large\bfseries}}

\renewcommand\subsection{\@startsection{subsection}{2}{\z@}%
                                     {-3.25ex\@plus -1ex \@minus -.2ex}%
                                     {1.5ex \@plus .2ex}%
                                     {\normalfont\bfseries}}

\def\Or[#1]{{\text{O}}\left({#1}\right)}
\def\dotl[#1,#2]{\left\langle #1, #2 \right\rangle}
\def\dotlb[#1,#2]{[ #1, #2 ]}
\def\dotp[#1,#2]{(#1) \cdot (#2)}
\def\aff[#1,#2]{\hat{#1}(#2)}
\def\n4sym{{\cal N}=4 SYM}
\def\>{\rangle}
\def\<{\langle}
\def\weight[#1,#2,#3]{\{(#1),#2,#3\}}
\def\ads[#1]{$\text{AdS}_{#1}$}

\begin{document}
\unitlength = 1mm
\ \\

\begin{center}

{ \LARGE {\textsc{Conformal Quivers and Melting Molecules}}}%The Wavefunction of Vassiliev's Universe}}}

\vspace{0.8cm}

Dionysios Anninos$^a$, Tarek Anous$^b$, Paul de Lange$^c$ and George Konstantinidis$^a$

\vspace{1cm}

\vspace{0.5cm}

$^a$ {\it Stanford Institute for Theoretical Physics, Stanford University}\\
%Stanford, CA 94305, USA}

$^b$ {\it Center for Theoretical Physics, MIT}\\

$^c$ {\it Institute for Theoretical Physics, University of Amsterdam}\\

\vspace{1.0cm}

\end{center}

\begin{abstract}

Quiver quantum mechanics describes the low energy dynamics of a system of wrapped D-branes. It captures several aspects of single and multicentered BPS black hole geometries in four-dimensional $\mathcal{N} = 2$ supergravity such as the presence of bound states and an exponential growth of microstates. The Coulomb branch of an Abelian three node quiver is obtained by integrating out the massive strings connecting the D-particles. It allows for a scaling regime corresponding to a deep AdS$_2$ throat on the gravity side. In this scaling regime, the Coulomb branch is shown to be an $SL(2,\mathbb{R})$ invariant multi-particle superconformal quantum mechanics. Finally, we integrate out the strings at finite temperature---rather than in their ground state---and show how the Coulomb branch `melts' into the Higgs branch at high enough temperatures. For scaling solutions the melting occurs for arbitrarily small temperatures, whereas bound states can be metastable and thus long lived. Throughout the paper, we discuss how far the analogy between the quiver model and the gravity picture, particularly within the AdS$_2$ throat, can be taken. 

\end{abstract}
% and we explicitly compute the generators
%\pagestyle{empty}

\pagebreak 
\pagestyle{plain}

\setcounter{tocdepth}{2}

\tableofcontents

\section{Introduction}

When studying the extremal limit of black hole solutions in gravity one immediately encounters a rather unusual geometric feature. The geometry develops a throat which becomes infinitely deep at extremality. This can be seen in the simple example of a $d$-dimensional Reissner-Nordstr\"{o}m black hole whose metric is:
\begin{equation}
ds^2 = - \frac{1}{r^2} \left(r - r_+ \right)\left(r - r_-\right) dt^2 + \frac{ r^2 dr^2}{\left(r - r_+ \right)\left(r - r_-\right)} + r^2 d\Omega^2_{d-2}~.
\end{equation}
The inner and outer horizon lie at $r = r_-$ and $r = r_+ > r_-$. In the extremal limit $r_+$ tends to  $r_-$ and the proper distance between some finite distance $r$ and the horizon diverges. One can isolate the geometry parametrically near the horizon within the infinitely deep throat and find a new solution to the Einstein-Maxwell system, namely the Bertotti-Robinson geometry AdS$_2 \times S^{d-2}$. The AdS$_2$ structure is not contingent upon supersymmetry but rather extremality, i.e. that the temperature of the horizon vanishes. Turning on the slightest temperature will destroy the precise AdS$_2$ structure and bring the horizon back to a finite distance. In the extremal near horizon region, the $\mathbb{R}\times SO(d-1)$ isometry group is enhanced to an $SL(2,\mathbb{R})\times SO(d-1)$. While the Hawking temperature of the black hole vanishes, the entropy, given by one-quarter of the size of the horizon in Planck units, is still macroscopically large. It is interesting that a full $SL(2,\mathbb{R})$ symmtetry (rather than just a dilatation symmetry) emerges in the near horizon region given that we are only loosing a single scale, the temperature. From a holographic point of view, the surprise stems from the fact that in quantum mechanics invariance under dilatations does not generally imply invariance under the full $SL(2,\mathbb{R})$ conformal transformations \cite{Michelson:1999zf}.

We  can study such extremal, and in addition supersymmetric, black holes in string theory. They are given by wrapping branes around compact dimensions such that they are pointlike in the non-compact directions. The electric and magnetic charges of these branes are (schematically) given by the number of times they wrap around various compact cycles. A simple and beautiful example \cite{Strominger:1996sh} is given by taking type IIA string theory compactified on a small $S^1 \times T^4$ and wrapping $N_4$ D4-branes around the $T^4$ and sprinkling $N_0$ D0-particles on top. In the large $N_0$ and $N_4$ limit, this gives an extremal D0-D4 black hole with an AdS$_2 \times S^3$ near horizon. Using $T$-duality on the $S^1$, we can map this system to a D1-D5 system in type IIB string theory. Since the circle becomes effectively non-compact in the IIB frame, our former black hole now becomes a black string. This black string has an AdS$_3 \times S^3$ near horizon with an $SL(2,\mathbb{R})_L\times SL(2,\mathbb{R})_R$ isometry group. We can express the AdS$_3$ as a Hopf fibration of the real line over an AdS$_2$ base space. The $SL(2,\mathbb{R})$ isometry of the AdS$_2$ base space is the left (or right) moving part of the isometries of the AdS$_3$. The real line becomes the $T$-dual direction and thus we see how the original AdS$_2$ naturally fits inside the AdS$_3$. From the point of view of the AdS/CFT correspondence, the AdS$_3$ is dual to a certain two-dimensional CFT and the large black hole degeneracy is counted by the supersymmetric ground states of the CFT which only excite one of the two $SL(2,\mathbb{R})$'s. The isometry of the near horizon supersymmetric AdS$_2$ reflects the remaining $SL(2,\mathbb{R})$.

There is another class of extremal black holes which arise in string theory which can be obtained by considering a Calabi-Yau compactification, with {\it generic} $SU(3)$ holonomy, of eleven dimensional M-theory down to five non-compact dimensions. Typically, such a Calabi-Yau will not have $U(1)$ isometries that we can easily $T$-dualize along. Yet, wrapping branes around the supersymmetric cycles of the Calabi-Yau yields extremal five-dimensional black holes \cite{Breckenridge:1996is,Katz:1999xq}, which again have an $SL(2,\mathbb{R})$ isometry in the near horizon. This AdS$_2$ does not naturally fit inside an AdS$_3$ with a two-dimensional CFT dual.\footnote{One might also consider this AdS$_2$ as a degenerate limit of the warped AdS$_3$/NHEK near horizon geometry of the rotating black hole, in the limit of vanishing angular momentum. The $SL(2,\mathbb{R})$ might then be a global subgroup of the full symmetries associated to the duals of such geometries \cite{Guica:2007gm,Anninos:2008fx,Anninos:2008qb,Guica:2008mu,Guica:2010ej,Detournay:2012pc}.} Remarkably, attempts to count the microstates of such supersymmetric black holes have faced significant difficulties \cite{Huang:2007sb}. In fact, whenever the precise counting of microstates has been successful it has involved a Cardy formula \cite{Strominger:1996sh,Vafa:1997gr,Maldacena:1997de}. Lacking the larger Virasoro structure, one may wonder where the `isolated' $SL(2,\mathbb{R})$ of these black holes originates and to what extent it is robust. 
  
Perhaps an additional motivation for understanding such an `isolated' AdS$_2$ geometry is the emergence of an $SL(2,\mathbb{R})$ symmetry in the worldline data of the static patch of de Sitter space \cite{Anninos:2011af,Anninos:2012qw}. It is interesting to note that the static patch of four-dimensional de Sitter space is conformally equivalent to AdS$_2 \times S^2$, whose `isolated' $SL(2,\mathbb{R})$ does not seem to reside within a larger structure containing a Virasoro algebra.
%It has been suggested that the underlying description of a single static patch of a de Sitter universe is itself a matrix-type theory \cite{}.

% (as oposed to matrix quantum field theories)
One particular way the $SL(2,\mathbb{R})$ isometeries of the black hole manifest themselves is in the worldine dynamics of D-particles propagating in the near horizon region. We might then ask whether there are microscopic models, such as matrix quantum mechanics models  with a large number of ground states, whose effective eigenvalue dynamics describe an $SL(2,\mathbb{R})$ invariant multiparticle theory.\footnote{The original $N\times N$ Hermitean matrix models \cite{Brezin:1977sv} in the double scaling limit has eigenvalue dynamics described by free fermions which naturally have an $SL(2,\mathbb{R})$ symmetry. Of course, such models contain only the eigenvalue degrees of freedom due to the $U(N)$ gauge invariance that allows for a diagonalization of the matrix, and hence do not have $\mathcal{O}(N^2)$ degrees of freedom. These models are dual to strings propagating in two dimensions (for some reviews see \cite{Klebanov:1991qa,Ginsparg:1993is,Polchinski:1994mb}).}
%emphasize quiver theories have a finite number of degrees of freedom and hence not field thoeries.

Some insight into these issues can be provided by studying certain quiver quantum mechanics models, which capture the low energy dynamics of strings connecting a collection of wrapped branes \cite{Douglas:1996sw,Denef:2002ru,Denef:2007vg,Bena:2012hf,Lee:2012sc,Manschot:2012rx,Manschot:2013sya} in a Calabi-Yau compactification of type IIA string theory to four-dimensions. Under certain conditions these quiver theories have an exponential number of (supersymmetric) ground states whose logarithm goes as the charge of the branes squared, which is the same scaling as the entropy of a supersymmetric black hole in $\mathcal{N}=2$ supergravity. It has been argued \cite{Denef:2007vg,Bena:2012hf} that the near horizon AdS$_2$ of these supersymmetric black holes is related to the exponential explosion in the number of ground states in the quiver quantum mechanics. The states in question are referred to as pure-Higgs states since they reside in the Higgs branch of the quantum mechanics, where all the branes sit on top of each other. Interestingly, going to the Coulomb branch after integrating out the massive strings stretched between the wrapped branes, leads a non-trivial potential and velocity dependent forces governing the wrapped brane position degrees of freedom in the non-compact space. Moreover, whenever the Higgs branch has an exponentially large number of ground states, the Coulomb branch exhibits a family of supersymmetric scaling solutions \cite{Denef:2002ru,Denef:2007vg,Bena:2007qc} continuously connected to its origin. The equations determining the positions of the wrapped branes in such supersymmetric zero energy scaling solutions are reproduced in four-dimensional $\mathcal{N} = 2$ supergravity \cite{Denef:2002ru}, believed to be the appropriate description of the system in the limit of a large number of wrapped branes.

In this note we would like to touch upon some of these issues. We do so by discussing two aspects of the Coulomb branch of such quiver theories, particularly those describing three wrapped branes containing scaling solutions.
\newline\newline
First we derive the Coulomb branch Lagrangian of a three node quiver model (see figure~\ref{quiver}), and establish the existence of a low energy scaling limit where the theory exhibits the full $SL(2,\mathbb{R})$ symmetry of conformal quantum mechanics \cite{jackiw,hagan,niederer,de Alfaro:1976je}. These scaling theories have velocity dependent forces, a non-trivial potential as well as a metric on configuration space. It is also worth noting that the full quiver quantum mechanics theory is itself {\it not} a conformal quantum mechanics (and most certainly not a two-dimensional conformal field theory). The emergence of a full $SL(2,\mathbb{R})$ symmetry rather than only a dilatation symmetry in the scaling limit is not guaranteed, and is reminiscent of the emergence of a full $SL(2,\mathbb{R})$ in the near horizon geometry of extremal black holes. 
\newline\newline
Second, we study the behavior of the Coulomb branch upon integrating out the strings in a thermal state, rather than in their ground state. At sufficiently high temperatures, the Coulomb branch melts into the Higgs branch. This is reminiscent of the gravitational analogue where increasing the temperature of a black hole increases its gravitational pull, or a particle falling back into the finite temperature de Sitter horizon.

%We would like to address the robustness of such scaling solutions upon deforming the theory in several ways. We consider adding a temperature, quantum corrections to the Coulomb branch from the superpotential, non-linear effects from integrating out auxiliary $D$-term fields. % and studying their presence in a mixed Higgs-Coulomb branch. 

\section{General Framework: Quiver quantum mechanics}

In this section we discuss the quiver quantum mechanics theory and its Coulomb branch. These theories constitute the low energy, non-relativistic and weakly coupled sector of a collection of branes along the supersymmetric cycles of a Calabi-Yau three fold. The wrapped branes look pointlike in the four-dimensional non-compact Minkowski universe.

\subsection{Full quiver theory}

\begin{figure}
\begin{center}
\includegraphics[width=3in]{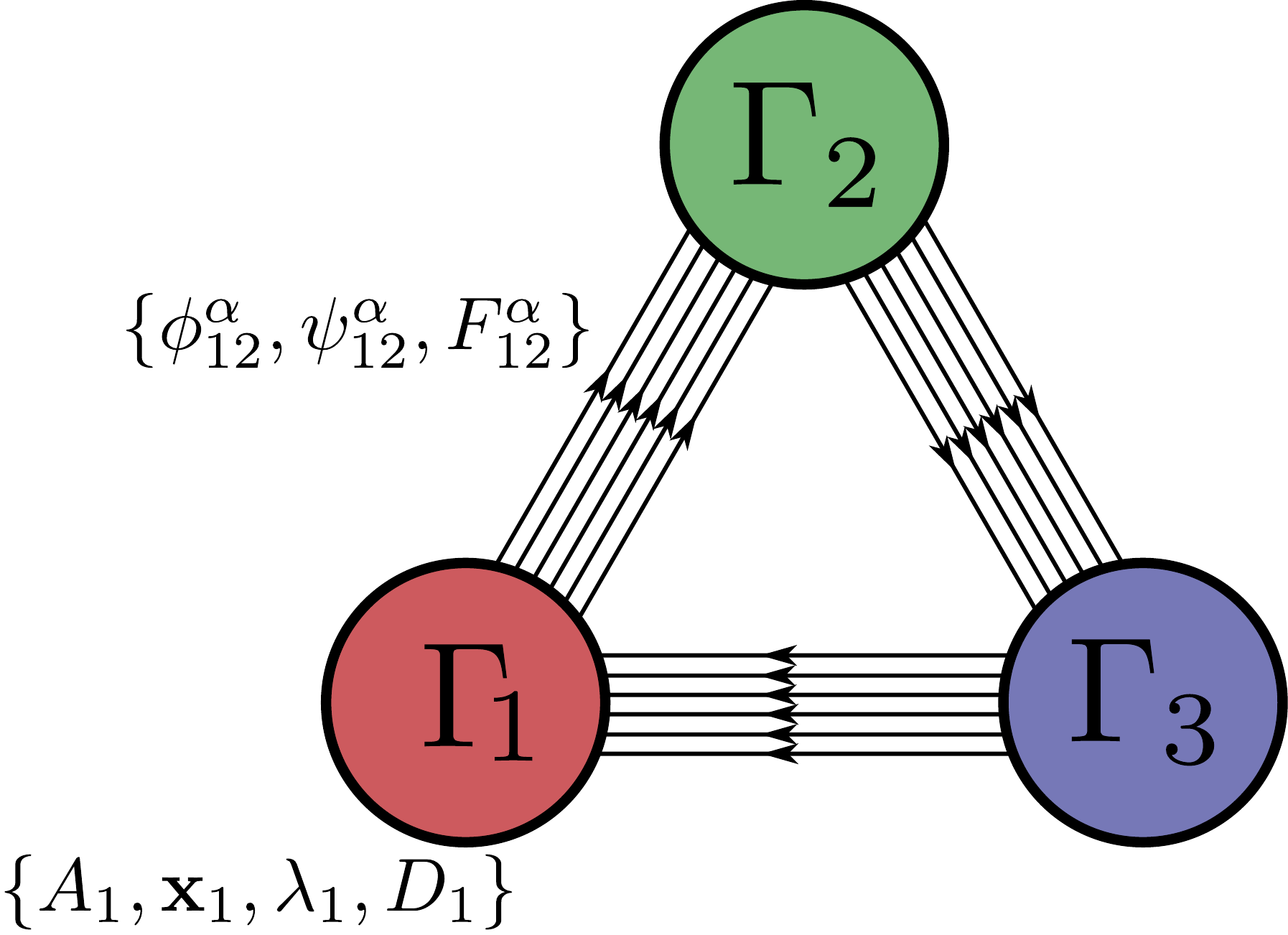} 
\end{center}
\caption{A 3-node quiver diagram which captures the field content of the Lagrangian $L=L_V+L_C+L_W$, each piece of which is given in (\ref{quiverqm}), (\ref{quiverqm2}), and (\ref{quiverqm3}). This quiver admits a closed loop if $\kappa^1$, $\kappa^2>0$ and $\kappa^3<0$.}\label{quiver}
\end{figure}

The $\mathcal{N} = 4$ supersymmetric quiver quantum mechanics comprises the following fields: chiral multiplets $\Phi^\alpha_{ij} = \{ \phi^\alpha_{ij}, \psi^\alpha_{ij}, F_{ij}^\alpha \}$ and vector multiplets $\bold{X}_{i} = \{ A_{i}, \bold{x}_{i}, \lambda_{i}, D_{i} \}$. The $\psi^\alpha_{ij}$ are the fermionic superpartners of the $\phi^\alpha_{ij}$, the $\lambda_i$ are the fermionic superpartners of the scalars $\bold{x}_i$, $A_i$ is a $U(1)$ connection, and $F^\alpha_{ij}$ and $D_i$ are auxiliary scalar fields. The $\Phi^\alpha_{ij}$ transform in the $(\bar{\bold{1}}_i,\bold{1}_j)$ of the $U(1)_i \times U(1)_j$. The index $\alpha = 1,2,\ldots,|\kappa_{ij}|$ denotes the specific arrow connecting node $i$ to node $j$ (see figure~\ref{quiver}). The chiral multiplets encode the low energy dynamics of strings stretched between the wrapped D-branes of mass $m_i$ sitting at three-vector positions $\bold{x}_i$ in the non-compact four-dimensions. The index $i=1,2,\ldots,N$ denotes the particular wrapped D-brane. The electric-magnetic charge vector, $\Gamma_i = (Q^{(i)}_I,P^{(i)}_I)$, of the wrapped branes depends on the particular cycles that they wrap, and the Zwanziger-Schwinger product of their charges are given by the $\kappa_{ij} =  (P^{(i)}_I Q^{(j)}_I - Q^{(i)}_I P^{(j)}_I)$. The $\kappa_{ij}$ count the number of intersection points in the internal manifold between wrapped branes $i$ and $j$. In what follows we measure everything in units of the string length $l_s$ which we have set to one. 

%We are working with the relative degrees of freedom $\bold{x}_{ij} \equiv (\bold{x}_i-\bold{x}_j)$, $D_{ij} = ( D_i -  D_j )$ since the center of mass degrees of freedom decouple and do not play a role in our discussions.

%~\tar\tar I don't know, but shouldn't we consider introducing this notation in the main text? Some people skip footnotes}

The Lagrangian $L = L_V + L_C+L_W$ for the three-node quiver quantum mechanics \cite{Denef:2002ru} describing the low energy non-relativistic dynamics of three wrapped branes which are pointlike in the $(3+1)$ non-compact dimensions  is given by:
%for the quiver in figure \ref{quiver} (with $\kappa^{1}, \kappa^2 > 0$ and $\kappa^3 < 0$)
%\tar\tar We don't include gauge fields in the kinetic terms of the chiral matter, now while this is kind of unimportant without a kinetic term on the gauge field, the gauge connections still enforce the conservation of a current, so should we include the gauge covariant derivative on $\phi$? \tar\tar
\begin{equation}\label{quiverqm}
L_V = \sum_{i=1}^3 \frac{\mu^i}{2}  \left(\dot{\bold{ q}}^i \cdot \dot{\bold{q}}^i + D^i D^i + 2 i \bar{\lambda}^i \dot{\lambda}^i \right)- \theta^i D^i~,
\end{equation}
and, %\\\nonumber
\begin{multline} \label{quiverqm2}
L_C=  \sum_{i=1}^3|\mathcal{D}_t{\phi}_\alpha^i |^2  - \left( \bold{q}^i \cdot \bold{q}^i + s_i D^i\right) |\phi_\alpha^i|^2 + |F_\alpha^i |^2+ i \bar{\psi}_\alpha^i \mathcal{D}_t{\psi}_\alpha^i  \\
- s_i \bar{\psi}_\alpha^i ({\boldsymbol \sigma} \cdot \bold{q}^i ) {\psi_\alpha^i} +  i \sqrt{2} \left( s_i \bar{\phi}_\alpha^i \lambda^i \epsilon \psi_\alpha^i - \text{h.c.} \right)~.
\end{multline}
%\\
%L_W=&   \sum_{i=1}^3\left( \frac{\partial W(\phi)}{\partial \phi^i_\alpha} F_\alpha^i + \text{h.c.} \right)+ \sum_{i,j=1}^3 %\left(\frac{\partial^2 W(\phi)}{\partial \phi^i_\alpha \partial \phi^j_\beta} \psi^i_\alpha \epsilon \psi^j_\beta + \text{h.c.} \right)\label{quiverqm3}~.
%\end{align}
{In~(\ref{quiverqm}-\ref{quiverqm2}) and what follows we will mostly work with the relative degrees of freedom (i.e. $\bold{x}_{ij} \equiv \bold{x}_i-\bold{x}_j$, $D_{ij} \equiv D_i -  D_j$, etc.) since the center of mass degrees of freedom decouple and do not play a role in our discussion. The notation we use is somewhat non-standard (for example $(\bold{q}^1,\bold{q}^2,\bold{q}^3) \equiv (\bold{x}_{12},\bold{x}_{23},\bold{x}_{13})$) and is given in appendix \ref{notation} along with all of our conventions. We note that the relative Lagrangian is only a function of two of the three vector multiplets, since $\bold{q}^3 = \bold{q}^1 + \bold{q}^2$, $D^3 = D^1 + D^2$ and $\lambda^3 = \lambda^1 +\lambda^2$. The $s_i$ encode the orientation of the quiver. For the majority of our discussion we choose $s_1 = s_2 = -s_3 = 1$, corresponding to a closed loop like the one in figure \ref{quiver}. The reduced masses $\mu^{i}$ (which we denote in superscript notation in~(\ref{quiverqm}-\ref{quiverqm2})) are related to the masses $m_i \sim 1/g_s$, where $g_s$ is the string coupling constant, of the wrapped branes sitting at the $\bold{x}_i$ by:
\begin{equation}
\mu^{1} = \frac{m_1 m_2}{m_1+m_2+m_3}~, \quad \mu^{2} = \frac{m_2 m_3}{m_1+m_2+m_3}~, \quad \mu^{3} = \frac{m_1 m_3}{m_1+m_2+m_3}~.
\end{equation}
The superpotential, which is allowed by gauge invariance only when the quiver has a closed loop, is given by:
\begin{align}
W(\phi) = \sum_{\alpha,\beta,\gamma} \omega_{\alpha\beta\gamma} \phi_\alpha^{1} \phi_\beta^{2} \phi_\gamma^{3}+\text{higher order terms}~,
\end{align}
%\begin{align}
%W(\phi) = \sum_{\alpha,\beta,\gamma} \omega_{\alpha\beta\gamma} \phi^\alpha_{12} \phi^\beta_{23} \phi^\gamma_{31}~,
%\end{align}
(where we take coefficients $\omega_{\alpha\beta\gamma}$ to be arbitrary) and contributes the following piece to the Lagrangian:
\begin{equation}\label{quiverqm3}
L_W =   \sum_{i=1}^3\left( \frac{\partial W(\phi)}{\partial \phi^i_\alpha} F_\alpha^i + \text{h.c.} \right)+ \sum_{i,j=1}^3 \left(\frac{\partial^2 W(\phi)}{\partial \phi^i_\alpha \partial \phi^j_\beta} \psi^i_\alpha \epsilon \psi^j_\beta + \text{h.c.} \right)~.
\end{equation} 
In this note we only consider cubic superpotentials and ignore the higher order terms. This is consistent so long as the $\phi^i_\alpha$ are small, which in turn can be assured by taking the $|\theta^i|$ sufficiently small \cite{Denef:2007vg}.

The theory contains a manifest $SO(3)$ global R-symmetry. In the absence of a superpotential, the Lagrangian is diagonal in the arrow (Greek) indices and thus the theory also exhibits a $U(|\kappa^1|)\times U(|\kappa^2|) \times U(|\kappa^3|)$ global symmetry under which the $\phi^i_\alpha$ transform as $U(|\kappa^i|)$ vectors. The superpotential explicitly breaks this symmetry down to the $U(1)_1 \times U(1)_2 \times U(1)_3$ gauge symmetry.

The theory can be obtained by dimensionally reducing an $\mathcal{N}=1$ gauge theory in four-dimensions to the $(0+1)$-dimensional worldline theory. It can also be viewed as the dimensional reduction of the $\mathcal{N}=2$ two-dimensional $\sigma$-models studied extensively, for example, in \cite{Witten:1993yc}. In order for the theory to have supersymmetric vacua we also demand that the Fayet-Iliopoulos constants sum to zero: $\theta_1 + \theta_2 + \theta_3 = 0$. 

%(\tar Changed the wording a bit) 
In units where $\hbar=1$ is dimensionless, and where we choose dimensions for which $[t] = 1$, the dimensions of energy are automatically set to $[E]=-1$. We also find the following dimensional assignments: $[\phi] = 1/2$, $[D]=-2$, $[\bold{x}]=-1$, $[\mu]=3$, $[\omega_{\alpha\beta\gamma}]=-3/2$, $[\psi]=0$, $[\lambda] = -3/2$, $[F] = -1/2$ and $[\theta] = 1$. The mass squared of the $\phi_\alpha^{i}$ fields upon integrating out the auxiliary $D$ fields is given by $M^2_{ij} = (|\bold{x}_{ij}|^2 + \theta_i/m_i - \theta_j/m_j)$. For physics whose energies obey $E^2/M^2 \ll 1$ we can integrate out the massive $\phi^i_\alpha$ fields and study the effective action on the Coulomb branch. Finally, notice also that the coupling $\omega_{\alpha\beta\gamma}$ has positive units of energy and is thus strong in the infrared limit since the natural dimensionless quantity is $\omega_{\alpha\beta\gamma} / E^{3/2}$. The contribution from $\phi_\alpha^{i}$ loops grows as $\kappa^{i}$ and thus the effective coupling constant at low energies is given by $g_{\rm eff} \sim g_s \kappa$. In the large $g_{\rm eff}$ limit, the wrapped branes backreact and the appropriate description of the system is given by four-dimensional $\mathcal{N} = 2$ supergravity \cite{Denef:2002ru}. 

\subsection{Some properties of the ground states}

An immediate question about the above model regards the structure of the ground states $\Psi_g \left[ \Phi^i_\alpha, \bold{Q}^i \right]$ of the theory, satisfying: $\hat{\bold{H}} \; \Psi_g \left[ \Phi^i_\alpha, \bold{Q}^i \right] = 0$. Though an explicit expression for the full $\Psi_g\left[ \Phi^i_\alpha, \bold{Q}^i\right] $ remains unknown, the degeneracy of ground states has been extensively studied \cite{Denef:2002ru,Denef:2007vg,Bena:2012hf,Lee:2012sc,Manschot:2012rx,Manschot:2013sya}. In particular, the degeneracy of ground states localized near $\bold{Q}^i = 0$, i.e. the ground states of the Higgs branch of the theory, were shown to grow exponentially in $\kappa^{i}$ when the theory contains a superpotential, the quiver admits closed loops (e.g. $\kappa^1, \kappa^2 > 0$ and $\kappa^3<0$) and the $\kappa^{i}$ obey the triangle inequality (i.e. $|\kappa^2| + |\kappa^3| \ge |\kappa^1|$ and cyclic permutations thereof). This growth is related to the exponential explosion in the Euler characteristic of the complete intersection manifold $\mathcal{M}$ \cite{Denef:2007vg} given by imposing the constraints from the $F$-term ($\delta_{F^i_\alpha} L|_{F^i_\alpha=0} = 0$) onto the $D$-term constraints ($\delta_{D^i} L|_{D^i=0} = 0$).\footnote{As an example, we can take $\theta^1,\theta^2 <0$. Then the complete intersection manifold is given by setting $\phi_\alpha^{3}=0$, imposing the $\kappa^{3}$ $F$-term constraints: $\omega_{\alpha\beta\gamma}\phi_\alpha^{1}\phi_\beta^{2} = 0$, inside a $\mathbb{C}\text{P}^{\kappa^{1}-1} \times \mathbb{C}\text{P}^{\kappa^{2}-1}$ space coming from the $D$-term constraints: $|\phi_\alpha^{1}|^2 = - \theta^1$ and $|\phi_\alpha^{2}|^2 = - (\theta^2-\theta^1)$. The space is a product of $\mathbb{C}\text{P}^k$'s since we have to identify the overall phase of the $\phi_\alpha^{i}$ due to the $U(1)$ gauge connection. When $|\kappa^{1}| + |\kappa^{2}| - 2 \ge |\kappa_{3}|$, which for large $\kappa$ amounts to the $\kappa^{i}$ satisfying the triangle inequality, the number of constraints become less or equal to the dimension of $\mathbb{C}\text{P}^{\kappa^{1}-1} \times \mathbb{C}\text{P}^{\kappa^{2}-1}$, allowing for more complicated topologies for $\mathcal{M}$.} Since $\kappa^{i}$ goes as the charge squared of the associated $U(1)$ gauge symmetry, the number of ground states scales in the same way as the Bekenstein-Hawking entropy of the associated black hole solutions in the large $g_{\rm eff}$ limit. Though a complete match between the ground states of a single Abelian quiver model and the entropy of a BPS black hole in $\mathcal{N}=2$ supergravity is not known,\footnote{Indeed, there are several quiver diagrams with the same net charges and one might suspect that all such quivers are required to obtain the correct entropy of the supersymmetric black hole (see for example \cite{Lee:2013yka}).} the vast number of microstates makes these systems potentially useful candidate toy models to study features of extremal or near extremal black holes. 

Other pieces of $\Psi_g$ localized near $\Phi^{i}_\alpha = 0$, i.e. the (quantum) Coulomb branch of the theory, have also been studied \cite{Denef:2002ru}. Unlike the Higgs branch quiver with a closed loop, superpotential and an exponential growth in its number of ground states, it was found that the number of Coulomb branch ground states grows only polynomially in the $\kappa^{i}$.  Interpreting the $\bold{q}^{i}$ as the relative positions of wrapped branes, these ground states can be viewed as describing various multiparticle configurations, as we will soon proceed to describe in further detail. To each ground state in the quantum Coulomb branch there exists a corresponding ground state in the Higgs branch, but the converse is not true. Another way to view this statement is that whenever a given $\Psi_g$ has non-trivial structure in the $\bold{Q}^i$ directions and peaks sharply about $\Phi^{i}_\alpha = 0$, it will also have a non-trivial structure in the $\Phi^{i}_\alpha$ directions and  peak sharply about $\bold{Q}^i = 0$ but not vice versa. 

% \left[ \Phi_\alpha^{i}, \bold{Q}^i\right]
%~\tar\tar~I had a question about this. In getting the coulomb branch lagrangian, we put the Higgs branch fields in their harmonic oscillator ground state and look at the remaining effective dynamics on the rest of the degrees of freedom, meaning we are assuming a product wavefunction. Could it be that this breaks down? Is this what we're trying to say? Because it seems like putting the Higgs fields in their HO ground state is \emph{not} giving them a peak for the total wavefunction. I suppose I'm a bit confused.~\tar\tar

\section{Coulomb branch and a scaling theory}

%\bold{x}_{ij} \equiv (\bold{x}_i - \bold{x}_j) We end up with the one-loop Coulomb branch.
%and the auxiliary $D$-fields (upon expanding the Lagrangian to quadratic order in $D^i$)

For large enough $|\bold{q}^i|$ we can integrate out the massive $\Phi^{i}_\alpha$'s (in their ground state) from the full quiver theory (\ref{quiverqm}). This can be done exactly given that the $\Phi^{i}_\alpha$ appear quadratically in (\ref{quiverqm}) whenever the superpotential vanishes. One finds the bosonic quantum effective Coulomb branch Lagrangian (up to quadratic order in $\dot{\bold{q}}^i$ and $D^i$):
\begin{equation}\label{supergoop}
L_{c.b.} =  \frac{1}{2} \sum_{i=1}^2  G_{ij} \;  \left( \dot{\bold{q}}^{i} \cdot \dot{\bold{q}}^{j}  + D^i \, D^i \right) -   \sum_{i=1}^3   s_i |\kappa^i| \bold{A}^d (\bold{q}^i)  \cdot \dot{\bold{q}}^i  - \left(\frac{s_i\,|\kappa^i |}{2|\bold{q}^i|}+ \theta^i \right) D^i~.%- V(\bold{q}^i)~. % + \text{fermions}
\end{equation}
%\tar\tar changed the sign in front of the $s_i|\kappa^i|D^i $ term to match equation D.14 in the Lagrangian. \tar\tar
The terms linear in $\dot{\bold{q}}^i$ and $D^i$ follow from a non-renormalization theorem \cite{Denef:2002ru}, whereas the quadratic piece in $\dot{\bold{q}}^i$ is derived in appendix \ref{cmetric}. 
%We have assumed that the $\kappa^i$ allow for a closed loop in the quiver diagram (e.g. $\kappa^1>0$, $\kappa^2>0$ and $\kappa^3 < 0$). 
Recall that the system is only a function of $\bold{q}^1$ and $\bold{q}^2$ since $\bold{q}^3 = \bold{q}^1 + \bold{q}^2$. The three-vector $\bold{A}^d$ is the vector potential for a magnetic monopole:
\begin{equation}\label{mon}
\bold{A}^d(\bold{x}) = \frac{ - y}{2r(z\pm r)} \hat{x} + \frac{x}{2r(z \pm r)} \hat{y} ~,
\end{equation}
and $G_{ij}$ is the two-by-two metric on configuration space:
\begin{equation}\label{metric}
\left[ G_{ij} \right] =  \left( \begin{array}{ccc}
\mu^{1} + \mu^{3} +  \frac{1}{4} \frac{|\kappa^{1}|}{|\bold{q}^1|^3} + \frac{1}{4} \frac{|\kappa^{3}|}{|\bold{q}^1 + \bold{q}^2 |^3} & \mu^{3} + \frac{1}{4} \frac{|\kappa^{3}|}{|\bold{q}^1 + \bold{q}^2 |^3} \\
\mu^{3}  + \frac{1}{4} \frac{|\kappa^{3}|}{|\bold{q}^1 + \bold{q}^2 |^3} & \mu^{2} + \mu^{3} + \frac{1}{4}\frac{|\kappa^{2}|}{|\bold{q}^2|^3} + \frac{1}{4} \frac{|\kappa^{3}|}{|\bold{q}^1 + \bold{q}^2 |^3}   \end{array} \right) ~.
\end{equation}
Upon integrating out the auxiliary $D^i$-fields, we obtain a multi-particle quantum mechanics with (bosonic) Lagrangian:
\begin{equation}\label{supergoop2}
L_{c.b.} =  \frac{1}{2} \sum_{i=1}^2  G_{ij} \;  \dot{\bold{q}}^{i} \cdot \dot{\bold{q}}^{j}  -   \sum_{i=1}^3   s_i |\kappa^i| \bold{A}^d (\bold{q}^i)  \cdot \dot{\bold{q}}^i  - V(\bold{q}^i)~. % + \text{fermions}
\end{equation}
%
%\begin{equation}\label{supergoop}
%L_{sg} = \frac{1}{2} \sum_{i} \left[  \dot{\bold{x}}^{i} \; g_{ij} \;  \dot{\bold{x}}^{j} + \bold{A}_i  \cdot \dot{\bold{x}}^i -  \frac{\left( \sum_{j} \frac{\kappa_{ij}}{2 r_{ij}} + \theta_i \right)^2}{\left(  m_i + \sum_{j} \frac{\kappa_{ij}}{r_{ij}^3} \right)} \right] + \text{fermions}
%\end{equation}
%\begin{equation}\label{supergoop}
%L_{sg} = \frac{1}{2} \sum_{i} \left[  \dot{\bold{x}}^2_i \left(  m_i + \sum_{j} \frac{\kappa_{ij}}{r_{ij}^3} \right) + \bold{A}_i  \cdot \dot{\bold{x}}_i -  \frac{\left( \sum_{j} \frac{\kappa_{ij}}{2 r_{ij}} + \theta_i \right)^2}{\left(  m_i + \sum_{j} \frac{\kappa_{ij}}{r_{ij}^3} \right)} \right] + \text{fermions}
%\end{equation}
%The above Lagrangian (\ref{supergoop2}) with the precise form of $G_{ij}$ is one of the technical results of our work. 
That the quantum effective Coulomb branch theory has a non-trivial potential $V(\bold{q}^i)$ should be contrasted with other supersymmetric cases such as interacting D0-branes or the D0-D4 system \cite{Douglas:1996yp} where the potential vanishes and the non-trivial structure of the Coulomb branch comes from the moduli space metric.
%The details for the derivation of the metric on configuration space $G_{ij}$ are rather involved and given in appendix \ref{cmetric}. 
The potential $V(\bold{q}^i)$ is also somewhat involved and is given in appendix \ref{potential}. 

%We note however that $V(\lambda q) = \lambda V(q)$
%$\bold{A}_i \equiv - 1/2 \sum_{j} \kappa_{ij} \left[ \bold{A}^d(\bold{x}_{ij}) + \bold{A}^d(\bold{x}_{ji})\right]$
% \mu^i  \delta_{ij} +  \frac{1}{4} \sum_{l} \frac{\kappa_{il}}{r_{il}^3}~.
%
%and $r \equiv |\bold{x}|$
%From the above Lagrangian we derive the Hamiltonian:
%\begin{equation}
%{H} =  \frac{1}{2} \sum_{i} \left( \left( \bold{p}_i - \bold{A}_i \right)^{\dag} g^{ij} \left( \bold{p}_j - \bold{A}_j \right) + \frac{\left( \sum_{j} \frac{\kappa_{ij}}{2 r_{ij}} + \theta_i \right)^2}{\left(  m_i + \sum_{j} \frac{\kappa_{ij}}{r_{ij}^3} \right)} \right) + \text{fermions}~.
%\end{equation}

\subsection{Supersymmetric configurations}

The supersymmetric configurations of the Coulomb branch consist of time independent solutions which solve the equations $V(\bold{q}^i) = 0$. For (\ref{supergoop}), this amounts to:
\begin{equation}\label{bound}
\frac{s_i \, |\kappa^{i}|}{|\bold{q}^{i}|} + \frac{s_3 \, |\kappa^{3}|}{|\bold{q}^{3}|} + 2 \theta^i = 0~, \quad  i = 1,2~.
%U^i\equiv
%U_i (\bold{q}^l) \equiv \sum_{j \neq i} \frac{\kappa^{j}}{2 |\bold{q}^{j}|} + \theta^i = 0~, \quad  i = 1,2~.
\end{equation}
In appendices \ref{sucos} and \ref{finiteD} we review that these supersymmetric configurations are robust against corrections of the Coulomb branch theory from the superpotential and from integrating out higher orders in the auxiliary $D$ fields.
%Depending on the $\kappa_{ij}$ these can come in two flavors. 

\subsubsection*{{\it Bound states}}

There are {\it bound state solutions} \cite{Denef:2002ru,Denef:2000nb} of (\ref{bound}) which are triatomic (or more generally $N$-atomic if dealing with $N$ wrapped branes) molecular like configurations. Of the original nine degrees of freedom, three can be removed by fixing the center of mass. Then the bound state condition (\ref{bound}) fixes another two-degrees of freedom. Thus, bound state solutions have a four-dimensional classical moduli space. Due to the velocity dependent terms in the Lagrangian, the flat directions in the moduli space are dynamically inaccessible at low energies -- the particles resemble electrons in a magnetic field. Several dynamical features of the three particles were studied in \cite{Nersessian:2007gc,Anninos:2012gk}.

%with an $four$-dimensional moduli space (after subtracting the center of mass degrees of freedom).  
%Had we been stud For instance if $N=2$ one finds a two-particle bound state with $r_{12} = -\kappa_{12}/(2\theta_1)$, so long as $\theta_1$ and $\kappa_{12}$ have opposite signs, and the two-dimensional moduli space is simply a two-sphere. In our case $N=3$ and we have a four-dimensional moduli space for the bound states.

\subsubsection*{{\it Scaling solutions}}

% and have cyclic signs with respect to the indices
There are also {\it scaling solutions} \cite{Denef:2007vg} of (\ref{bound}) which are continuously connected to the origin $|\bold{q^i}| = 0$. They occur whenever the $\kappa^{i}$ form a closed loop in the quiver diagram (e.g. $\kappa^1, \kappa^2 > 0$ and $\kappa^3 < 0$) and obey the same triangle inequality ($|\kappa^2| + |\kappa^3| \ge |\kappa^1|$ and cyclic permutations thereof) that the $|\bold{q}^{i}|$ are subjected to. These solutions can be expressed as a series: 
\begin{equation}
|\bold{q}^{i}| = |\kappa^{i}| \sum_{n=1}^\infty a_n \lambda^n~, \quad \lambda > 0~.
\end{equation}
 %\footnote{Remarkably, precisely when the $\kappa^{i}$ obey the triangle inequality and the Coulomb branch exhibits scaling solutions, the number of ground states of (\ref{quiverqm}) grows dramatically as $\sim e^{c \; \kappa_{ij}}$. Though the exponent scales like the charge square, which is the charge dependence of a single centered black hole's entropy in $\mathcal{N}=2$ supergravity, there is no known single quiver with given $\kappa_{ij}$'s that accounts for the full black hole entropy with the same charges.} 
The coefficient $a_1 = 1$, while the remaining $a_n$ can be obtained by systematically solving (\ref{bound}) in a small $\lambda$ expansion, and will hence depend on $\theta^i$.
%Consider again the case of three particles. 
%The nine position degrees of freedom of this system can be reduced to
%The six position degrees of freedom $\bold{q}by removing the center of mass degrees of freedom. 
The moduli space of the scaling solutions is given by the three rotations as well as the scaling direction parameterized by $\lambda$. Though the angular directions in the moduli space are dynamically trapped due to velocity dependent forces, the scaling direction is not and constitutes a flat direction even dynamically.
  %$|\bold{q}^{i}| = \lambda \, |\kappa^{i}|$
%Imposing further that one is in a scaling configuration reduces the (continuous) classical degrees of freedom to three. 

Requiring that the series expansion converges, i.e. $(a_{n+1}\lambda^{n+1})/(a_{n}\lambda^{n}) \ll 1$, leads to the condition: %we require that the following is satisfied (take the $\theta_i$'s, $\kappa^i$'s and $\mu^i$'s to have comparable magnitudes amongst each other):%\footnote{One might also imagine a limit where we can set the $\theta_i$ to zero while keeping the metric on configurations to be flat.}
%while having a scaling solution $|\bold{q}^{3}| = |\kappa^{3}| \sum_{n=1}^\infty \lambda^n a_n$ 
\begin{equation}\label{convergence}
%\frac{1}{\kappa} \left({\frac{\theta}{\mu}}\right)^{1/2} \ll 
\lambda \ll \frac{1}{\theta}~.
%\theta_3^2/m_3 % + \theta_1^2/m_1 \ll 1~, \quad ~.
\end{equation}
Because of this condition on the $\lambda$'s, one should be cautious when dealing with such scaling solutions. They occur in the near coincident limit of the branes where the bifundamentals that we have integrated out become light.  In order for the mass of the bifundamentals, $M^2_{ij} = \left( |\bold{x}_{ij}|^2 +  \theta_i/m_i - \theta_j/m_j \right)$, to remain large we require:
%(going momentarily back to the original notation): 
%$|\bold{x}_{ij}|^2 \gg (\theta_i/m_i - \theta_j/m_j)$. We expres
%Without loss of generality we can pick $\theta_1$ and $\theta_2$ to be greater than zero which immediately implies $\theta_3 < 0$. Thus, in order for the mass square of $\phi^{3}$ to be positive we require:
\begin{equation}\label{notach}
\left({\frac{\theta}{\mu}}\right)^{1/2}  \ll |\bold{q}^i|~. %\ll \frac{1}{\theta}~.
%\theta_3^2/m_3 % + \theta_1^2/m_1 \ll 1~, \quad ~.
\end{equation}
Taking $\mu^i = \nu \, \widehat{\mu}^i$ and $|\bold{q}^i| = |\widehat{\bold{q}}^i|/\nu^\alpha$ the inequalities (\ref{convergence}) and (\ref{notach}) can be satisfied in the limit $\nu \to \infty$, with $\widehat{\mu}^i$, $\widehat{\bold{q}}^i$, $\kappa^i$ and $\theta^i$ fixed and furthermore $\alpha \in (0,1/2)$. 
%For sufficiently large $\mu^i$ and $\kappa^{i}$ and small $\theta^i$ both (\ref{notach}) and (\ref{convergence}) can be satisfied. 
% $\kappa^i = \nu \, \tilde{\kappa}^i$, and $\theta^i = \tilde{\theta}^i/\nu$, $\tilde{\kappa}^i$, $ \tilde{\theta}^i$ and 

Notice that (\ref{convergence}) implies that the distances between particles in a scaling regime $\sim \lambda \kappa$ is much less than the typical inter-particle distance of a bound state $\sim \kappa/\theta$. 
%It is mostly the robustness of these scaling solutions that interests us in what follows. 

\subsection{Scaling Theory}\label{sec:scaltheor}

%We do this by taking $\bold{q}^i = {\nu} \; \widehat{\bold{q}}^i$ and $t = \widehat{t}/{\nu}$ and with $\nu \to 0$.
%

To isolate the physics of the Coulomb branch in the scaling regime we take an infrared limit of the Lagrangian (\ref{supergoop}), pushing the $\bold{q}^{i}$ near the origin and dilating the clock $t$. In particular, we would like the $\sim \kappa/|\bold{q}^i|^3$ part of the metric in configuration space to dominate over the $\sim \mu$ piece leading to:
\begin{equation}\label{scbounds}
|\bold{q}^i  | \ll \left( \frac{\kappa}{\mu} \right)^{1/3}~.
\end{equation}
Additionally we must satisfy the inequalities (\ref{convergence}) and (\ref{notach}). Again taking $\mu^i = \nu \, \widehat{\mu}^i$, $\bold{q}^i = \widehat{\bold{q}}^i/\nu^\alpha$ and in addition $t = \nu^{\alpha} \, \widehat{t}$ with fixed $\widehat{\mu}^i$, $\widehat{\bold{q}}^i$, $\kappa^i$ and $\theta^i$, we can also satisfy (\ref{scbounds}) in the limit $\nu \to \infty$, so long as we also ensure $\alpha \in (1/3,1/2)$.\footnote{Another limit one might imagine is given by: $|\bold{q}^i| \gg \left( {\kappa}/{\mu} \right)^{1/3}$, $|\bold{q}^i| \ll {\kappa}/{\theta}$. In this case the metric on configuration space remains flat while the potential scales like $\sim 1/|\bold{q}|^2$.} The rescaling of $t$ is required to maintain a finite action in the scaling limit. In type II string compactifications $\mu \sim 1/l_P \sim \sqrt{v}/g_s l_s$, where $l_P$ is the four-dimensional Planck length and $v$ is the volume of the Calabi-Yau in string units \cite{Denef:2002ru}, therefore the $\nu \to \infty$ limit corresponds to a parametrically small string coupling. Furthermore the scaling throat deepens as we increase the mass of the wrapped branes.
%$\kappa^i = \nu \, \tilde{\kappa}^i$,  $\theta^i = \tilde{\theta}^i/\nu$,
%  $\tilde{\kappa}^i/\tilde{\mu}^i \sim \mathcal{O}(1)$

The rescaling above amounts simply to setting the $\theta^i$ and $\mu^i$ to zero in (\ref{supergoop}) and replacing $\bold{q}^i$ and $t$ with $\widehat{\bold{q}}^i$ and $\widehat{t}$.
%the inequalities (\ref{notach}) and (\ref{convergence}) can be satisfied in the limit $\nu \to \infty$, with $\tilde{\kappa}^i$, $ \tilde{\theta}^i$ and $\tilde{\mu}^i$ fixed
%This is consistent so long as we impose the conditions:
%\begin{equation}\label{scbounds}
% \quad \quad |\bold{q}^i | \ll \frac{\kappa}{\theta}~, \quad \quad |\bold{q}^i | \gg \left({\frac{\theta}{\mu}}\right)^{1/2}~,
%\end{equation}
%\begin{equation}\label{ineq2}
%\theta \ll m^{1/3} \kappa^{2/3}~,
%\end{equation}
%ensuring that the $\mu^i$ and $\theta^i$ dependent pieces of (\ref{supergoop}) are indeed subleading.
%In fact (\ref{ineq2}) is already implied by (\ref{notach}). 
We call the remaining Lagrangian with vanishing $\theta^i$ and $\mu^i$ the {\it scaling theory}. Notice from equation~(\ref{scalpot}) that the potential $V(\bold{q}^i)$ in this limit becomes a homogeneous function of order one, i.e. $V(\nu \; \widehat{\bold{q}}^i) = \nu V(\widehat{\bold{q}}^i)$ and the linear in velocity term is retained. 

\subsubsection{Consistency of the small $D^i$ expansion}

One can infer from the supersymmetry variations \cite{Denef:2002ru} that supersymmetric configurations have vanishing $D^i$. Furthermore, as we have already noted, we have performed an expansion in small $D^i$ fields prior to integrating them out  (see appendix \ref{finiteD} for more details) in order to obtain the Coulomb branch. Thus, in order for a scaling theory to exist and be consistent with a small $D^i$ expansion, it must be the case that zero energy scaling configurations exist. 

Had we considered a two-particle theory, where no such scaling solutions can exist, taking a small $\bold{q}$ limit would be inconsistent with the small $D$ expansion. That is because the dimensionless small quantity in our perturbation series is actually $D/|\bold{q}|^2$ and the supersymmetric configuration $D=0$ occurs at $|\bold{q}| = -\kappa/2\theta$. Expanding the non-linear $D$ equation (\ref{twonodeD}) (see appendix \ref{finiteD}) in powers of $\epsilon \equiv D/|\bold{q}|^2$, while imposing $|\bold{q}| \gg ({\theta/\mu})^{1/2}$, one finds the following consistency condition:
\begin{equation}
\epsilon = \left( \frac{\theta}{\mu} \right) \frac{1}{|\bold{q}|^2} + \left( \frac{\kappa}{2\mu} \right) \frac{1}{|\bold{q}|^3} \left( 1 - \frac{1}{2} \epsilon + \mathcal{O}(\epsilon^2) \right)~.
\end{equation}
Indeed, the first term on the right hand side is negligible by construction, since we took $|\bold{q}| \gg ({\theta/\mu})^{1/2}$ to keep the strings massive. Smallness of the second term in the equation would require $|\bold{q}|^3 \gg \kappa/\mu$, in contradiction with the condition (\ref{scbounds}) required to isolate the scaling theory.

%In the three-node case, we expand the three-node non-linear $D^i$ equations (\ref{threenodeD}).
We now consider the three-node case for $s_i$ that admit a closed loop in the quiver. For small $\epsilon_2\equiv D^2/|\bold{q}^2|^2$ (taking all masses the same and $\theta_1=\theta_2=\theta$ and again imposing $|\bold{q}^2| \gg ({\theta/\mu})^{1/2}$) the equation of motion of the auxiliary $D^2$-field (in the form of (\ref{threenodeD})) is given by: 
\begin{equation}\label{delta}
\epsilon_2=\frac{3\theta}{2\mu|\bold{q}^2|^2}+\frac{|\kappa^2|}{2\mu|\bold{q}^2|^3}\left(\delta-\frac{1}{2}\epsilon_2+\mathcal{O}(\epsilon_2^2)\right)
%\epsilon \sim \left( \frac{\theta}{\mu} \right) \frac{1}{|\bold{q}|^2} + \left( \frac{\kappa}{\mu} \right) \frac{1}{|\bold{q}|^3} \left( \delta - \epsilon + \mathcal{O}(\epsilon^2) \right)~,
\end{equation}
where $\delta\equiv 1-\frac{|\bold{q}^2|}{2\,|\kappa^2|}\left(\frac{|\kappa^1|}{|\bold{q}^1|}+\frac{|\kappa^3|}{|\bold{q}^3|}\right)$ measures how close the configuration is to the scaling solution. For sufficiently small $\delta \sim \epsilon \ll 1$ we can consistently satisfy (\ref{delta}) in addition to imposing the scaling inequalities (\ref{scbounds}). 

In this sense the scaling theory is in fact a theory of the deep infrared configurations residing parametrically near the zero energy scaling solutions. This is consistent with the dilation of time required to obtain the scaling theory.

\section{Conformal Quivers: emergence of $SL(2,\mathbb{R})$}

In this section we uncover that the bosonic scaling theory action, i.e. (\ref{supergoop2}) with the $\theta^i$ and $\mu^i$ set to zero, has an $SL(2,\mathbb{R})$ symmetry. This is the symmetry group of conformal quantum mechanics \cite{de Alfaro:1976je}. The group $SL(2,\mathbb{R})$ is generated by a Hamiltonian $H$, a dilatation operator $D$ and a special conformal transformation $K$, with the following Lie algebra:
\begin{equation}\label{sl2ralg}
[H,D] = - 2 i H~, \quad \quad [H,K] = - i D~, \quad \quad [K,D] = 2i K~.
\end{equation}
As we have already mentioned, the full $SL(2,\mathbb{R})$ symmetry is not guaranteed by the existence of time translations and dilatations alone \cite{BrittoPacumio:1999ax}. This is suggested by the fact that the $H$ and $D$ operators form a closed subalgebra of the full $SL(2,\mathbb{R})$. The presence of a full $SL(2,\mathbb{R})$ is actually quite remarkable, particularly given the specific form of the scaling theory Lagrangian which has velocity dependent forces and a non-trivial potential. 

As discussed in the previous section, it is not true that, for any finite $\kappa^{i}$, $\mu^i$ and $\theta^i$, the Coulomb branch can be described precisely by the scaling theory action. There will always be small corrections that break its manifest scaling symmetry: ${{\bold{q}}}^i \to {\gamma} \; {{\bold{q}}}^i$, and ${t} \to {t}/{\gamma}$. This is comforting, given that the full quiver theory has a finite number of ground states---yet a conformal quantum mechanics has a diverging number of arbitrarily low-energy states, with a density of states that behaves as $dE/E$. The corrections serve as a cutoff for the infrared divergence in the number of states, such that the Coulomb branch can fit consistently inside the full quiver theory (related discussions can be found in \cite{Bena:2012hf,deBoer:2008zn}).

%In addition to time translation invariance, the scaling goop action, i.e. (\ref{supergoop}) with the $\theta_i$ and $m_i$ set to zero, exhibits a manifest dilatation symmetry: $\tilde{x}_i \to {\nu} \tilde{x}_i$, $\lambda_{ij} \to \sqrt{\nu} \lambda_{ij}$ and $\tilde{t} \to \tilde{t}/{\nu}$.

%We are interested in whether the theory is invariant under the full $SL(2,\mathbb{R})$ symmetry of conformal quantum mechanics \cite{}.
% (recall that the Hamiltonian and dilatation operators form a close subaglebra of $SL(2,\mathbb{R})$).
%This can be seen by recalling the 

\subsection{Conditions for an $SL(2,\mathbb{R})$ invariant action}
%\cite{de Alfaro:1976je}\cite{Michelson:1999zf}\cite{BrittoPacumio:1999ax}\cite{Papadopoulos:2000ka}

The conditions under which an action will be $SL(2,\mathbb{R})$ invariant (up to possible surface terms) have been studied extensively in \cite{Michelson:1999zf,BrittoPacumio:1999ax,Papadopoulos:2000ka}. %Using the results from section 2.2 of , we can write the conditions such that a theory has both a dilatation symmetry as well as a special conformal transformation. 
Showing that a general theory with bosonic Lagrangian describing $N$ degrees of freedom :
\begin{equation}
L = \frac{1}{2} \; \dot{q}^i \;  G_{ij}  \; \dot{q}^j  - A_i \; \dot{q}^i - V(q)~, \quad i = 1,2,\ldots,N
\end{equation}
has an $SL(2,\mathbb{R})$ symmetry is equivalent to finding a solution to the following equations \cite{Papadopoulos:2000ka}:
\begin{eqnarray}\label{conditions}
2 \; \nabla_{(i} Z_{j)} &=&  G_{ij}~, \label{homo1} \\
 - Z^i \partial_i V &=& V~, \label{homo2} \\
2 \; Z_i &=& \partial_i f~,   \label{closed} \\ %Z^i G_{ij}
Z^j F_{ji} &=& 0~, \quad F_{ij} \equiv \partial_{[i} A_{j]} \label{maxwell}~. 
\end{eqnarray}
Equations (\ref{homo1}) and (\ref{homo2}) ensure the existence of a dilatation symmetry. In particular, equation (\ref{homo1}) implies that the metric on configuration space allows for a conformal Killing vector field (also referred to as a homothetic vector field).  Equations (\ref{closed}) and (\ref{maxwell}) ensure the that the action remains invariant under special conformal transformations, where $f$ is an arbitrary function of the ${q}^i$. Indices are raised and lowered with the metric $G_{ij}$.

Interestingly, equation (\ref{closed}) imposes that the conformal Killing form $Z_i$ of the metric  be exact, which is generically not the case. Hence the existence of a dilatation symmetry does not necessarily imply the symmetry of the full conformal group.    

%Generally, one has no reason to expect that the conformal Killing vector of a particular metric must be a closed form, as required by (\ref{closed}), and so generically dilatations do not imply the full conformal group.  

Once a solution to (\ref{homo1}-\ref{maxwell}) is found, the three conserved quantities are then given by \cite{Papadopoulos:2000ka}:% (\tar corrected this):
\begin{equation}
\mathcal{Q}_n =  \frac{1}{2}t^{n+1} \, \dot{q}^i \, G_{ij} \, \dot{q}^j \, - (n+1) t^n Z^i \, G_{ij} \, \dot{q}^j  + t^{n+1} V(q) + F_n~,  \quad n =  -1,0,1
\end{equation}
where $F_{-1} = F_0 = 0$ and $F_1 = f$. The charge $\mathcal{Q}_{-1}$ is the Hamiltonian, whereas $\mathcal{Q}_{0}$ and $\mathcal{Q}_{1}$ are related to dilatations and special conformal transformations, respectively. These three charges generate the $SL(2,\mathbb{R})$ algebra (\ref{sl2ralg}) (up to factors of $i$) under the Poisson bracket. In what follows, we find $Z^i$ and $f$ for the scaling theory described in section \ref{sec:scaltheor}.  %We first consider the case with three particles constrained to live on a line, and finally we consider the full non-collinear three particle case.
%As a warm up exercise we consider the two particle case. XXXX not quite, because inconsistent with non-linear D term expansion of potential...

\subsection{Two particles}\label{d0d4}

%As a warm up, we consider the Lagrangian on the Coulomb branch describing the dynamics of the relative coordinate $\bold{x}$, upon integrating out the chiral matter and integrating out the auxiliary $D_{12}$ field expanded to second order. 
As a warm up, we can study a simple model consisting of a two node quiver with an equal number of arrows going to and from each node, with a total number $\kappa>0$ arrows altogether. This is the theory describing, for example, the low energy dynamics of a wrapped D4-D0 brane system (see \cite{Douglas:1996yp}). The bosonic Lagrangian for the relative position $\bold{q} = (q^x,q^y,q^z)$ on the Coulomb branch, in the scaling limit, is given by:
%(we are ignoring the linear in velocity term, as it does not affect the discussion at a qualitative level)
\begin{equation}
L = \frac{\kappa}{2} \; \frac{\dot{\bold{q}}^2}{|\bold{q}|^3}~.%\quad\quad \implies \quad\quad H = \frac{1}{2\kappa} q^3 p_q^2~.
\end{equation}
The above Lagrangian is also the non-relativistic limit of one describing a BPS particle in an AdS$_2 \times S^2$ background. The wordline theory has three degrees of freedom and a diagonal metric on configuration space: $g_{ij} = \kappa \; |\bold{q}|^{-3} \; \delta_{ij}$. In addition to the Hamiltonian $H$, the above theory has a dilatation operator $D$ and special conformal generator $K$ given by:
\begin{equation}
D = i \left( q^i \; \partial_i + {|\bold{q}|^{-9/2}} \, \partial_i \, |\bold{q}|^{9/2} \, q^i \; \right)~, \quad K = {2 \, \kappa} \, {|\bold{q}|^{-1}}~,
\end{equation}
such that the $SL(2,\mathbb{R})$ algebra is satisfied.
%: $[H,K] = - i D$, $[H,D] = - 2 i H$ and $[D,K] = -2 i K$ which is the $SL(2,\mathbb{R})$ algebra. 
It is thus a simple example of a conformal quantum mechanics. 
%This is precisely the metric on configuration space one finds when studying the D0-D4 system 

%It is given by:
%\begin{equation}
%L = \frac{1}{2}\left(\mu+\frac{\kappa}{4 |\bold{x}|^3}\right)\left(\dot{\bold{x}}^2  + 2 i \bar{\lambda} \dot{\lambda} \right) - \frac{1}{2}\frac{\left(\theta + \kappa/2|\bold{x}| \right)^2}{\left(\mu + \kappa/4|\bold{x}|^3 \right)} - \bold{A}^d(\bold{x}) \cdot \dot{\bold{x}} - \frac{\kappa \bold{x}}{2 | \bold{x}|^3} \cdot \bar{\lambda}\boldsymbol{\sigma} \lambda~.
%\end{equation}
%In the scaling limit where we can drop $\mu$ and $\theta$ the (bosonic) Lagrangian becomes:
%\begin{equation}\label{2node}
%L_{bos} = \frac{\dot{\bold{x}}^2}{8 |\bold{x}|^3} - \frac{\kappa |\bold{x}|}{2}- \bold{A}^d(\bold{x}) \cdot \dot{\bold{x}}~.
%\end{equation}
%We can easily confirm that equations (\ref{homo1})-(\ref{maxwell}) are satisfied with $\bold{Z} = -\bold{x}$ and $f_0 = 1 / 4|\bold{x}|$. Thus we conclude that (\ref{2node}) describes a conformal quantum mechanics.

\subsection{$SL(2,\mathbb{R})$ symmetry of the full three particle scaling theory}

For our particular problem, it is useful to note that the index structure of the relative coordinates, i.e. $i = 1,2,3$, is trivially tensored with the spatial index implicit in bold vector symbols (e.g. $\bold{q} = (q^x,q^y,q^z)$). Introducing  ${Q}^\alpha \equiv \left( \bold{q}^1, \bold{q}^2 \right)$, with $\alpha = 1,2,\ldots,6$, our Lagrangian (\ref{supergoop2}) (with $\mu^i$ and $\theta^i$ set to zero, and $s_i$ that admit a closed loop) takes the form: %$\kappa^1,\kappa^2 > 0$ and $\kappa^3 < 0$) takes the form:
\begin{equation}
L_{c.b.} = \frac{1}{2} \; \dot{Q}^\alpha \;  G_{\alpha \beta}  \; \dot{Q}^\beta  - A^{(Q)}_\alpha \; \dot{Q}^\alpha - V({Q}^\alpha)~,
\end{equation}
where $G_{\alpha\beta}$ and the six-dimensional vector potential can be extracted from (\ref{supergoop2}). An expression for the vector potential is simple to write down and is given by:
\begin{equation}
 A^{(Q)}_\alpha = \left( \; s_1\,|\kappa^1| \bold{A}^d (\bold{q}^1)+s_3\,|\kappa^3| \bold{A}^d(\bold{q}^1 + \bold{q}^2) \; , \;  s_2\, |\kappa^2| \bold{A}^d (\bold{q}^2)+s_3\, |\kappa^3| \bold{A}^d(\bold{q}^1 + \bold{q}^2) \;  \right)~.
\end{equation}
where $\bold{A}^d(\bold{x})$ is the vector potential for a magnetic monopole and is given in (\ref{mon}). It is straightforward to check, using Mathematica for example, that the conditions (\ref{homo1}-\ref{maxwell}) are indeed satisfied. We find $Z^\alpha = - Q^\alpha$ and $f = 2 \; Q^\alpha G_{\alpha \beta} Q^\beta$. The explicit generators of the $SL(2,\mathbb{R})$ are given by:
\begin{eqnarray}\label{generators}
H &=& -\frac{1}{2} \left( \frac{1}{\sqrt{G}} \partial_\alpha \sqrt{G} - i A^{(Q)}_\alpha \right) G^{\alpha\beta} \left( \partial_\beta - i A^{(Q)}_\beta \right) + V(Q^\alpha)~, \\
D &=&  i \left( Q^\alpha  \partial_\alpha  +  \frac{1}{\sqrt{G}} \partial_\alpha \sqrt{G}  Q^\alpha \right)~, \\
% i Q^\alpha \left( \partial_\alpha  - i A^{(Q)}_\alpha \right) + i \left( \frac{1}{\sqrt{G}} \partial_\alpha \sqrt{G} - i A^{(Q)}_\alpha \right) Q^\alpha~, \\
K &=& 2 \; Q^\alpha G_{\alpha \beta} Q^\beta~,
\end{eqnarray}
where we have used that $Q^\alpha A^{(Q)}_\alpha = 0$. The generator $K$ simplifies to:
\begin{equation}
K = \frac{|\kappa^1|}{2|\bold{q}^1|} +  \frac{|\kappa^2|}{2|\bold{q}^2|} +  \frac{|\kappa^3|}{2|\bold{q}^3|}~.
\end{equation}

We thus conclude that the bosonic scaling theory is an interacting multi-particle $SL(2,\mathbb{R})$ invariant conformal quantum mechanics with velocity dependent forces. The theory admits a further $SO(3)$ symmetry acting on the spatial three-vectors. 

\subsubsection{$N$-particle case}

Though we do not prove it here, it is natural to conjecture that the $N$-particle scaling theory will be an $N$-particle conformal quantum mechanics. Indeed, for any closed loop of $N_i$-particles we expect that in the corresponding throat there will be a conformal quantum mechanics with $D$ and $K$ as in (\ref{generators}) except $Q^\alpha$ runs over all the $3 N_i$ bosonic degrees of freedom. The one loop metric on configuration space in the Coulomb branch, i.e. the coefficient of $\dot{\bold{{q}}}^i \cdot \dot{\bold{{q}}}^j$, will be equivalent to that of $D^i \, D^j$ (which is far easier to compute) upon integrating out the massive strings for general quivers. Interestingly, when there are more than three nodes, there may be several distinct closed loops allowing for their own scaling throat and several distinct scaling throats may coexist, within its residing a decoupled conformal quantum mechanics theory.

\subsubsection{{Superconformal quantum mechanics}}

%Explain briefly why Coulomb branch is, ought to be a susyc qm.

It is worth remarking that the Coulomb branch is in fact a supersymmetric quantum mechanics with four supercharges $\mathcal{Q}_S^i$ and an $SO(3)$ global R-symmetry group. This follows from effective field theory. We are integrating out the heavy chiral $\Phi^i_\alpha$ multiplet in its supersymmetric ground state. Thus, the low energy effective theory of the vector multiplet will be endowed with the four supercharges of the parent quiver theory. Of course, there could be anomalies that arise in the process. For instance, the discrete time reversal symmetry of the full quiver theory is violated in the Coulomb branch by the linear in velocity terms. This is an anomaly which does not spoil the supersymmetry of the Coulomb branch, and occurs due to a zero mode in the functional determinant of the $\psi^i_\alpha$ fermions. As shown in \cite{smilga} for the two node case, one can compute the supercharges of the low energy effective theory in a systematic fashion using perturbation theory. 

We have shown above that the bosonic Lagrangian of this theory exhibits an $SL(2,\mathbb{R})$ symmetry. In order to establish that this extends to a superconformal quantum mechanics we must establish the existence of four supersymmetric special conformal generators $S^i$. These will be given by the commutator $S^i = i [\mathcal{Q}_S^i,K]$. As discussed in \cite{Michelson:1999zf}, one must also ensure that the commutator $[\mathcal{Q}_S^i,D] = -i \mathcal{Q}_S^i$ is satisfied in order to have a superconformal system. This is guaranteed due to the manifest scale invariance of the supersymmetric action: ${{\bold{q}}}^i \to {\gamma} \; {{\bold{q}}}^i$, $\lambda^{i} \to {\gamma}^{3/2} \, \lambda^{i}$ with ${t} \to {t}/{\gamma}$.\footnote{Though we have not presented the piece of the Lagrangian quadratic in $\lambda^i$, which will be of the form $\sim \kappa \; \bar{\lambda} \, \dot{\lambda} / |\bold{q}|^3$ in the scaling region, the linear piece is fixed by supersymmetry \cite{Denef:2002ru} to be (\ref{linear1}), which is enough to read off the scaling dimension of $\lambda^i$.} Hence the scaling theory is a superconformal multiparticle theory.
%$\lambda^{i} \to \sqrt{\gamma} \lambda^{i}$

\subsection*{\it{Gravity}}

As discussed in the introduction, the emergence of a full $SL(2,\mathbb{R})$ in the deep scaling regime is reminiscent of the emergence of a full $SL(2,\mathbb{R})$ in the deep AdS$_2$ throat in gravity, which could end at a horizon or cap off at the locations of entropyless D-particles. The symmetry group manifests itself in the dynamics of wrapped branes moving in a `geometry' resulting from integrating out the interconnecting strings and extends upon similar observations made for other multiparticle systems in \cite{Michelson:1999zf,BrittoPacumio:1999ax,Michelson:1999dx}. However, we must point out that the conformal quantum mechanics obtainde here comes directly from the low energy dynamics of strings interacting with wrapped branes, i.e. the quiver theory, rather than from a gravity calculation of the moduli space of a multi-black hole system. It would in fact be interesting to repeat such gravitational Ferrel-Eardley type calculations \cite{BrittoPacumio:1999ax,Michelson:1999dx,ferrel,gibbons,Maloney:1999dv} for the low energy velocity expansion of the corresponding one-half BPS scaling solutions \cite{Denef:2000nb} in $\mathcal{N}=2$ supergravity. This might give an operational meaning to the AdS$_2$/CQM correspondence \cite{Strominger:1998yg,Azeyanagi:2007bj,Sen:2011cn,Maldacena:1998uz,Chamon:2011xk} (particularly in the large $g_s \kappa$ limit and within the AdS$_2$ scaling region). Recall that in the classical supergravity limit, the gravitational backreaction of a brane becomes parametrically small given that its mass goes like $1/l_P$.\footnote{This approach is reminiscent of attempts to match the Coulomb branch of the BFSS matrix model \cite{Banks:1996vh,Becker:1997wh} with eleven-dimensional supergravity calculations. A basic difference is the presence of the ability to zoom into a deep AdS$_2$ throat in the geometry and that the microscopic quiver model is vector like rather than matrix like.} Furthermore, as emphasized in the introduction, our system is a quiver quantum mechanics that does not necessarily reside inside a larger two-dimensional conformal field theory, and thus the $SL(2,\mathbb{R})$ may be of the `isolated' type.

\subsection{{Wavefunctions in the scaling theory}}

Given the existence of an $SL(2,\mathbb{R})$ invariant scaling theory, one natural question is whether a quantum state will stay localized within the scaling regime or if its wavefunction will spread away from the scaling regime, or even fall back into the Higgs branch (where $\langle |\bold{q}^i|\rangle = 0$). One particular direction in which a wavefunction might easily spread is the scaling direction where the potential is identically zero and where there are (classically) no velocity dependent magnetic forces. 

Studying the quantum mechanics problem of the full three-particle scaling theory is difficult. Instead, we will look at the wavefunctions of the simpler wrapped D4-D0 model discussed in section \ref{d0d4}.
% consisting of a two node quiver with an equal number of arrows going to and from each node, with a total number of $\kappa$ arrows altogether. The bosonic Lagrangian for the relative position $\bold{q} = (q^x,q^y,q^z)$ on the Coulomb branch in the scaling limit is then given by:
%(we are ignoring the linear in velocity term, as it does not affect the discussion at a qualitative level)
%\begin{equation}
%L = \frac{\kappa}{2} \; \frac{\dot{\bold{q}}^2}{|\bold{q}|^3}~.%\quad\quad \implies \quad\quad H = \frac{1}{2\kappa} q^3 p_q^2~.
%\end{equation}
%The above Lagrangian is also the non-relativistic limit of a BPS particle in an AdS$_2 \times S^2$ background. The wordline theory has three degrees of freedom and a diagonal metric on configuration space: $g_{ij} = \kappa \; |\bold{q}|^{-3} \; \mathbb{I}_{3\times 3}$. This is precisely the metric on configuration space one finds when studying the D0-D4 system \cite{Douglas:1996yp}. 
The zero angular momentum piece of the Schr\"{o}dinger equation in spherical coordinates (i.e. $q^x = q \cos\theta\sin\phi$, $q^y = q \sin\theta\sin\phi$ and $q^z = q \cos\phi$, $q \equiv |\bold{q}|$) is:
\begin{equation}\label{conformals}
- \frac{1}{2 \sqrt{g}} \, \partial_i \, \sqrt{g} \, g^{ij} \, \partial_j \, \psi_E = - \frac{1}{2\kappa} \; q^{5/2} \; \frac{\partial}{\partial q} q^{1/2} \frac{\partial}{\partial q} \;  \psi_E = {E} \; \psi_E~,
\end{equation}
which can easily be solved: $\psi_E(q) \sim \sqrt{q} \exp \left( \pm 2 i \sqrt{2\kappa E/q} \right)$.\footnote{If we include the angular variables we would find $\psi_E^{lm} = Q^E_{l}(q)Y_l^m(\theta,\phi)$ with $Y_l^m$ spherical harmonics and $Q^E_l$ can be written in terms of Bessel functions. Here we only look at $l=0$ modes, $\psi_E(q)=Q^E_{0}(q)$.}
% In fact, numerical analysis suggests that for the operator $H+K$, only the $m=0$ modes are normalizable ~(\tar~ why do we say these are Bessel functions? It seems like they are simpler)
These energy eigenstates are non-normalizable at small $q$ with respect to the covariant inner product:
\begin{equation}\label{covnorm}
\langle \psi_1 | \psi_2 \rangle = \int d^3 q \; \sqrt{g} \; \psi_1^*(\bold{q}) \; \psi_2(\bold{q})~.
\end{equation}
Non-normalizability of energy eigenstates is common for scale invariant theories \cite{de Alfaro:1976je}, the non-normalizability means that energy eigenstates leak into the Higgs branch.

As in \cite{de Alfaro:1976je}, we can consider instead eigenstates $\psi_\lambda(q)$ of the $H+a K$ operator with eigenvalue $\lambda\equiv \sqrt{{a}} \left(n+\frac{1}{2}\right) \in \mathbb{R}$, where $a$ is a constant with appropriate dimensions. If we define a variable $x=q/(\kappa \sqrt{a})$, then the zero-angular momentum wavefunctions $\psi_\lambda(x)$ satisfy
\begin{equation}
\frac{1}{2} \left(-x^{5/2} \; \frac{\partial}{\partial x} x^{1/2} \frac{\partial}{\partial x} +\frac{4}{x}\right)\;  \psi_\lambda = \left(n+\frac{1}{2}\right)\; \psi_\lambda~,
\end{equation}
The normalizable wavefunctions are given by confluent hypergeometric functions:
\begin{equation}\label{confluent}
\psi_\lambda(x)=\mathcal{N}e^{-2/x}\,U\left(\frac{1-n}{2},\;\frac{3}{2},\;\frac{4}{x}\right)~,
\end{equation}
where $\mathcal{N}$ is a finite normalization factor that depends on $\kappa$ and $a$. 
%Calculating the norm according to~(\ref{covnorm}) gives
%\begin{equation}
%\frac{1}{\mathcal{N}^2}=4\pi\left(\frac{1}{a}\right)^{3/4}\int_0^\infty dx\; \frac{e^{-4/x}}{x^{5/2}}\,U\left(\frac{1-n}{2},\;\frac{3}{2},\;\frac{4}{x}\right)^2~,
%\end{equation}
%which is finite for all values of $n\in\mathbb{R}$~.
Whenever $n \in \mathbb{N}$, the expression for $U\left(\frac{1-n}{2},\;\frac{3}{2},\;\frac{4}{x}\right)$ simplifies significantly. We can also obtain asymptotic expressions near $x = 0$ and $x = \infty$:
\begin{eqnarray}\label{asymptotics}
U\left(\frac{1-n}{2},\;\frac{3}{2},\;\frac{4}{x}\right) &=& x^{-n/2}\left( 2^{n-1}\sqrt{  x} + \mathcal{O}\left(x^{3/2}\right) \right)~,  \quad \text{for } x \sim 0~,\\ \label{largex}
 &=& \frac{{\sqrt{\pi \, x}}}{2 \, {\Gamma}\left[(1-n)/2\right]}-\frac{2 \, \sqrt{\pi }}{{\Gamma}\left[-n/2\right]}+ \mathcal{O}\left(x^{-1/2}\right)~, \quad \text{for } x^{-1} \sim 0~.
\end{eqnarray}
The dependence on $\kappa$ and $a$ drops out of all expectation values of operators as a function of $x$ labeled $\mathcal{O}(x)$,
%This is simple to see; denoting $\psi_\lambda=\mathcal{N}\tilde{\psi}_\lambda$, then 
%\begin{equation}
%\langle\psi_\lambda|\mathcal{O}(x)|\psi_\lambda\rangle\equiv\langle\mathcal{O}(x)\rangle_\lambda=\dfrac{\int_0^\infty dx\,\dfrac{\tilde{\psi}_\lambda\mathcal{O}(x)\tilde{\psi}_\lambda}{x^{5/2}}}{\int_0^\infty dx\,\dfrac{\tilde{\psi}_\lambda^2}{x^{5/2}}}~,
%\end{equation}
hence we can instead normalize the wavefunctions $\tilde{\psi}_\lambda(x) \equiv \psi_\lambda(x) \rvert_{\kappa,a=1}$ and reinstate the dependence on $\kappa$ and $a$ in expectation values of operators by looking at the dependence of $\mathcal{O}(x)$ on the variable $x$ and how it transforms when we take $q\rightarrow x \kappa \sqrt{a}$. Examples of $\tilde{\psi}_\lambda(x)$ are displayed in figure \ref{wavefunctions}.
%We originally set out to analyze this simple Conformal quantum mechanics problem to see whether these wavefunctions have significant support near the Higgs branch or if they stay in the Coulomb branch. 
\begin{figure}
\begin{center}
{\includegraphics[width=3in]{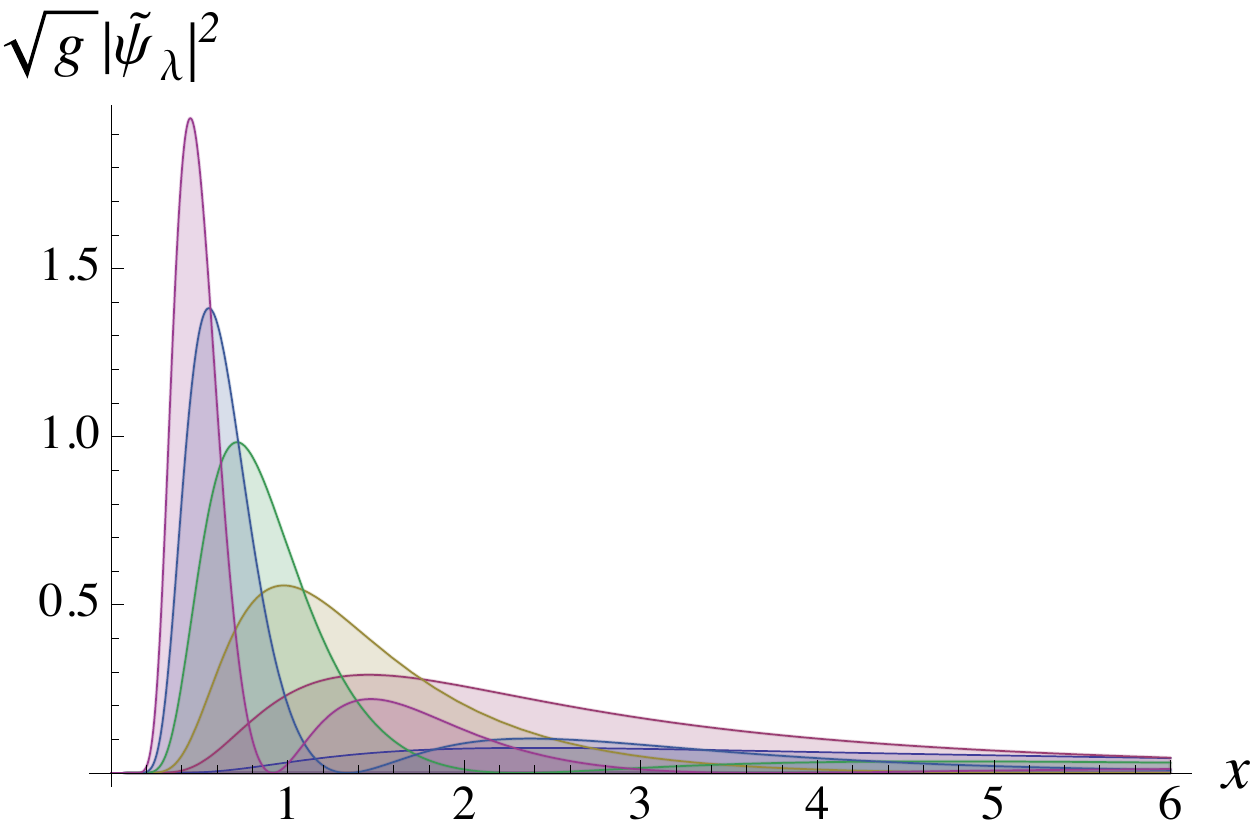} \quad\quad \includegraphics[width=3in]{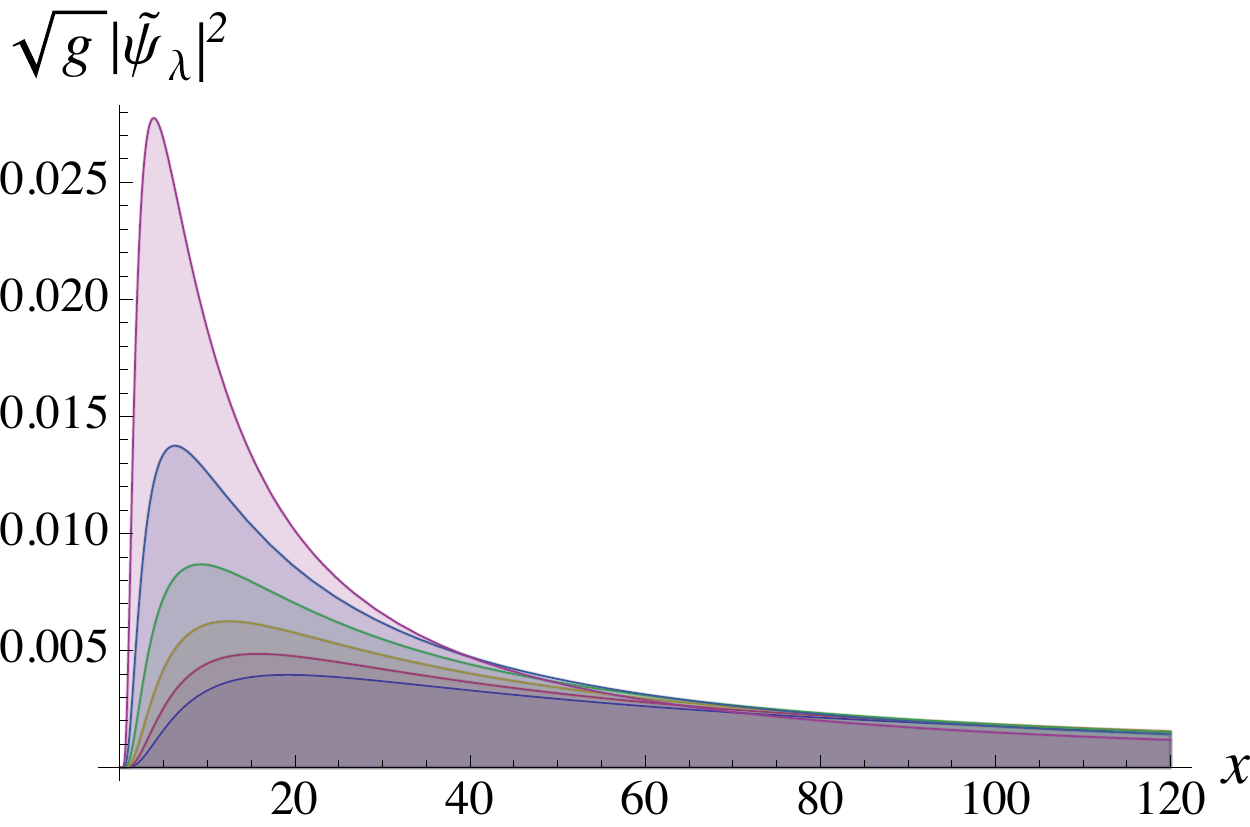} }
\end{center}
\caption{Examples of $H+a\,K$ eigenfunctions. Left: Plot of $\tilde{\psi}_\lambda(x)$ for $n \in (0.2,5.2)$ in unit increments. Right: Plot of $\tilde{\psi}_\lambda(x)$ for $n \in (-5.7,-0.7)$ in unit increments.}\label{wavefunctions}
\end{figure}

To understand whether these states leak back into the Higgs branch, we consider for example $\langle x \rangle_\lambda$. This is only finite if $n$ is an odd positive integer for which $\langle x \rangle_\lambda=8$ (or equivalently $\langle q \rangle_\lambda=8 \kappa \sqrt{a}$~), otherwise $\langle x \rangle_\lambda$ is infinite. This can be seen from the fact that for positive and odd $n$, $\Gamma[(1-n)/2]$ diverges and the $\mathcal{O}(\sqrt{x})$ term in the large $x$ expansion (\ref{largex}) dissappears. However, for any $\lambda(n)$ there exists a number $s<1$ such that $\langle x^s\rangle_\lambda$ is finite. Similarly, when $n$ is an odd positive integer, there exists a number $s>1$ such that $\langle x^s\rangle_\lambda$ is infinite. Thus a large class of $H+a K$ eigenstates do not leak into the Higgs branch for large $\kappa$. 

Finding the wavefunctions in the three-node Coulomb branch is clearly more complicated. There are now six-degrees of freedom and a non-trivial potential $V(\bold{q}^i)$. However, at the classical level, the scaling direction remains flat and motion along it is uninhibited, neither by the potential nor by any velocity dependent magnetic forces. So if the wavefunction has any chance of spreading, it will do so along the scaling direction. Our previous analysis of the two particle wavefunctions, which deals essentially with the scaling direction, suggests that at least some $H + a K$ eigenstates do not leak into the Higgs branch.\footnote{We also note that the quantization of the classical solution space of scaling configurations was considered in \cite{deBoer:2008zn}, where it was found that the expectation value of the total angular momentum operator was non-vanishing.}

\subsection*{ \it $SL(2,\mathbb{R})$ in the Higgs branch?}

We end this section with an important question. The $SL(2,\mathbb{R})$ symmetry we have been discussing so far resides in the Coulomb branch of the full quiver theory. In particular it resides in the scaling regime of the Coulomb branch which is  connected to the Higgs branch at its tip, i.e. where all the $\bold{q}^i$ become vanishingly small. How does the $SL(2,\mathbb{R})$ structure act on (if at all) the Higgs branch degrees of freedom? Perhaps a way to answer this question is via the Higgs-Coulomb map of \cite{Berkooz:1999iz} (further developed in the context of quiver theories in \cite{Bena:2012hf}). Indeed, an $SL(2,\mathbb{R})$ symmetry acting on the states residing in the Higgs branch would resonate closely with the appearance of an $SL(2,\mathbb{R})$ isometry of the AdS$_2$ in the near horizon region of the extremal black hole.

\section{Melting Molecules}

We have seen that the low energy excitations of the scaling regime in the Coulomb branch are described by a multiparticle $SL(2,\mathbb{R})$ invariant quantum mechanics. It was noted that this is reminiscent of the $SL(2,\mathbb{R})$ that appears in the near horizon geometry of an extremal black hole or the AdS$_2$ geometry outside a collection of extremal black holes that reside within a scaling throat. If we heat up such a collection of black holes (for example by making the centers slightly non-extremal) they will fall onto each other due to gravity's victory over electric repulsion. Naturally, then, we might ask what happens to our scaling theories and more generally the Coulomb branch configurations upon turning on a temperature?\footnote{Aspects of supersymmetric quantum mechanics at finite temperature are discussed in \cite{Fuchs:1984hc,das,Roy:1988xg}. An incomplete list of more recent studies including numerical simulations is \cite{Kabat:2000zv,Kabat:2001ve,Kawahara:2007fn,Catterall:2007fp,Anagnostopoulos:2007fw,Lin:2013jra,Iizuka:2013kha,Asplund:2011qj}.} This amounts to integrating out the massive strings, i.e. the chiral multiplet, in a thermal state as opposed to their vacuum state. In this section we assess the presence of bound states and scaling solutions as we vary the temperature. Roughly speaking, this is the weak coupling version of the calculations in \cite{Anninos:2011vn,Chowdhury:2011qu,Chowdhury:2013ema,Anninos:2013mfa} where the multicentered solutions were studied at finite temperature.\footnote{In fact, \cite{Anninos:2011vn,Anninos:2013mfa} explored how the effective potential of a D-particle/black hole bound state changed as the temperature of the black hole was increased. Hence, a closer analogy to \cite{Anninos:2011vn} would be to consider a four-node mixed Higgs-Coulomb branch as in figure \ref{quiver2}. One node is `pulled' away from the remaining three, such that one can integrate out all connecting heavy strings in some thermal state. The remaining three nodes, which we choose to contain a closed loop (and hence have exponentially many ground states), are in the Higgs branch which is also considered to be at some finite temperature. One could then study the Coulomb branch theory for the position degree of freedom of the far away wrapped brane as a function of the temperature.} We find that the bound (but non-scaling) configurations persist at finite temperature until a critical temperature after which they either classically roll toward the origin or become metastable. This amounts to the `melting' of the bound state. Scaling solutions on the other hand do not persist at finite temperature and instead the potential develops a minimum at the origin, even at small temperatures. 

As was alluded to, what we find is somewhat analogous to heating up a collection of extremally charged black holes, known as the Majumdar-Papapetrou geometries \cite{Majumdar:1947eu,papapetrou}, by making the masses of the centers slightly larger than their charges. Doing so will cause the black holes to collapse into a single center configuration upon the slightest deviation from extremality.

\subsection{Two nodes - melting bound states}

Consider the low energy dynamics of two wrapped branes, which is given by a two node quiver. There can be no gauge-invariant superpotential in this case since the quiver admits no closed loops. The relative Lagrangian is:
\begin{eqnarray}%+ |F^\alpha|^2
\nonumber L_V &=&\frac{\mu}{2} \left( \dot{\mathbf{q}}^2 + D^2 + 2 i \bar{\lambda} \dot{\lambda} \right) - \theta D~, \\
\nonumber L_C &=& |\mathcal{D}_t \phi_\alpha |^2 - \left( |\mathbf{q}|^2 + D \right) |\phi_\alpha|^2  + i \bar{\psi}_\alpha\mathcal{D}_t \psi_\alpha - \bar{\psi}_\alpha \mathbf{q} \cdot \boldsymbol{\sigma} \psi_\alpha - i \sqrt{2} \left( \bar{\phi}_\alpha \psi^\alpha \epsilon \lambda - \bar{\lambda} \epsilon \bar{\psi}_\alpha \phi_\alpha \right)~,
\end{eqnarray}
where $\mathcal{D}_t \phi_\alpha \equiv \left( \partial_t + i A \right) \phi_\alpha$. The matter content is given by a chiral multiplet $\Phi_\alpha = \{ \phi_\alpha,\psi_\alpha \}$ and a vector multiplet $\bold{Q} = \{ A,\mathbf{q},D,\lambda \}$ ($A$ is a one-dimenional $U(1)$ connection). The index $\alpha = 1,2,\ldots, \kappa$ (where $\kappa>0$) is summed over.

If we keep $|\mathbf{q}|$ constant and set to zero the fermionic superpartners $\lambda$, we can integrate out the $\phi_\alpha$ and $\psi_\alpha$ fields to get the effective bosonic action on the position degrees of freedom:
\begin{multline}\label{hotL}
L_{\rm eff} = \frac{\mu}{2} D^2 - \theta D - \kappa \log \det \left( -(\partial_t + i A)^2 + |\bold{q}|^2  + D \right) \\ + \kappa \log \det \left( -(\partial_t + i A)^2 + |\bold{q}|^2 \right)~.
\end{multline}
If we wish to study the system at finite temperature $T \equiv \beta^{-1}$, we must Wick rotate to Euclidean time $t \to i t_E$ and compactify $t_E \sim t_E + \beta$. Notice we cannot set the zero mode of $A$ to zero since one can have non-trivial holonomy around the thermal circle. The $\phi_\alpha$ fields have periodic boundary conditions and the $\psi_\alpha$ fields have anti-periodic boundary conditions around the thermal circle.
%where $A \equiv A_2 - A_1$.
%\begin{figure}
%\begin{center}
%{\includegraphics[width=2.5in]{class1.pdf}  \includegraphics[width=2.5in]{class3p2.pdf}  {\includegraphics[width=2.5in]{class3p5.pdf}  \quad\quad \includegraphics[width=2.4in]{class5.pdf}}}%{\includegraphics[width=2.5in]{T35thetam1.pdf} \quad\quad \includegraphics[width=2.5in]{T5thetam1.pdf}}}%{\includegraphics[width=2.5in]{T01thetam1.pdf} \quad\quad \includegraphics[width=2.5in]{T32thetam1.pdf}}
%\end{center}
%\caption{Thermal effective potentials for $\kappa=2\pi$, $\theta = -1$ and $\mu=1$ at $T=\{0.1,3.2,3.5,5 \}$ going from top left to bottom right. Notice the minimum is shifted toward $r=0$ at higher temperature.}
%\end{figure}
%\begin{figure}
%\begin{center}
%{\includegraphics[width=2.5in]{class3p5.pdf} \quad\quad \includegraphics[width=2.5in]{class5.pdf}}%{\includegraphics[width=2.5in]{T35thetam1.pdf} \quad\quad \includegraphics[width=2.5in]{T5thetam1.pdf}}
%\end{center}
%\caption{Thermal effective potentials for $\kappa=1$, $\theta = -1$ and $\mu=1$ at $T=3.5$ (left) and $T=5$ (right). Notice that the minimum has shifted all the way to $r=0$.}
%\end{figure}
The operator $\partial_{t_E}$ has as eigenvalues the Matsubara frequencies $\omega_n = 2\pi n T$ for the bosonic case and $\omega_n = 2\pi(n + 1/2) T$ for the fermionic case, where $n \in \mathbb{Z}$. Hence, it is our task to evaluate:
\begin{multline}
L^{(T)}_{\rm eff} = \frac{\mu}{2} D^2 - \theta D - \frac{\kappa}{\beta} \log \det \left( 4\pi^2(n + a)^2 T^2 + |\bold{q}|^2  + D \right) \\ + \frac{\kappa}{\beta} \log \det \left( 4\pi^2(n + 1/2 + a)^2 T^2 + |\bold{q}|^2 \right)~, 
\end{multline}
where we have defined $a \equiv A / 2\pi T$. Notice that $L^{(T)}_{\rm eff}$ depends on the connection $A$ which we eventually need to path-integrate over. %The effective thermal potential is given by:
%\begin{equation}
%V(T,a) =  -\frac{\mu}{2} D^2 + \theta D + \kappa \log \frac{\det \left( (n + a)^2 T^2 + r^2  + D %\right)}{\det \left( (n + 1/2 + a)^2 T^2 + r^2 \right)}~.
%\end{equation}
Upon evaluating the determinant (see appendix \ref{thermalDet} for details), one finds the effective thermal potential is given by:
\begin{equation}\label{thermalpot}
V(T,D,a) = -\frac{\mu}{2} D^2  + \theta D + \kappa \, T \log \left( \frac{{\cosh}\left(\frac{\sqrt{|\bold{q}|^2+D}}{T}\right)-{\cos}(2 \pi a  )}{ {\cosh}\left(\frac{|\bold{q}|}{T}\right)+{\cos}(2 \pi a ) } \right)~,
\end{equation}
%\begin{equation}
%V(T,a) = -\frac{\mu}{2} D^2  + \theta D + \kappa \frac{T}{2\pi} \log \left( \frac{-{\cos}(2 \pi a  )+{\cosh}\left(\frac{2 \pi  \sqrt{r^2+D}}{T}\right)}{{\cos}(2 \pi a ) + {\cosh}\left(\frac{2 \pi  r}{T}\right)} \right)~.
%\end{equation}
which has the correct low temperature limit $\lim_{T \to 0} V(T,D,a) =  -\mu D^2/2 + \theta D - \kappa |\bold{q}| + \kappa \sqrt{|\bold{q}|^2 + D}$, in accordance with the results of \cite{Denef:2002ru}. Our task is to now perform the path integral of $e^{-\beta V(T,a)}$ over $D$ and $a$. In doing so, we must be careful to ensure that $D > - |\bold{q}|^2$ such that integrating out the string is justifiable. Differentiating (\ref{thermalpot}) with respect to $a$ we find that $V(T,D,a)$ has a minimum at $\cos(2\pi a) =1$ for all $T$ and $D > -|\bold{q}|^2$. Solving the saddle point equations for $D$ must be done numerically. 

The potential $V$ has two scaling symmetries associated to it. Configurations related by:
\begin{equation}
T \to \gamma \, T~, \quad |\bold{q}| \to \gamma \, |\bold{q}|~, \quad D \to \gamma^2 \, D~, \quad \mu \to \gamma^{-2} \delta \, \mu ~, \quad \theta \to \delta \, \theta~, \quad \kappa \to \gamma \delta \, \kappa~,
\end{equation}
have respective potentials related by: $V(\mu,\theta,\kappa,|\bold{q}|,D,T) \to \delta \gamma^2 V(\mu, \theta, \kappa,|\bold{q}|,D,T)$. The qualitative features of the potential thus only depend on scale invariant quantities:
\begin{equation}
\tilde{\kappa}\equiv\kappa\left(\frac{\mu}{|\theta|^3}\right)^{1/2}~,\quad \tilde{T}\equiv T\left(\frac{\mu}{|\theta|}\right)^{1/2}~,\quad |\tilde{\bold{q}}|\equiv|\bold{q}|\left(\frac{\mu}{|\theta|}\right)^{1/2}~,
\end{equation}
and the scaling-invariant potential is given by $\tilde{V}\equiv V\frac{\mu}{|\theta|^2}$. Note that our finite temperature analysis breaks down for $T\sim 1$ in string units, where the effects of massive string modes start to kick in. However, we may still consider large $\tilde{T}$ as long as we restrict ourselves to parameter regions where our approximation remains valid, e.g. $T\ll1$ and $\mu/|\theta|\gg1$. We use the scaling symmetries to fix $\mu = |\theta| = 1$ and study the thermal phases of the theory as a function of $\tilde{T}$ and $\tilde{\kappa}$, noting that we can always extract physical quantities by properly rescaling $\mu$ and $\theta$.

\subsubsection{Thermal phases}

%\begin{figure}
%\begin{center}
%\includegraphics[width=3in]{stabbs.pdf} 
%%\includegraphics[width=3in]{minawayzero.pdf} 
%\end{center}
%\caption{Thermal effective potential of a two node quiver (for $\theta=-1$ and $\mu=1$). This is an example of phase type \ref{q0nonglobmin}.}\label{phases2}
%\end{figure}

Having derived the thermal effective potential we can now describe the different thermal configurations for the two-node quiver. The physically distinguishable phases have potentials $\tilde{V}$ with the following distinct properties:
%We have noted several different thermal configurations of the two-node system. The thermal potential can have:
\begin{enumerate}
\item A single stable minimum away from $|\tilde{\bold{q}}|=0$~.\label{singstab}
\item Two non-degenerate minima away from $|\tilde{\bold{q}}|=0$~.\label{twnondeg}
\item Two degenerate minima away from $|\tilde{\bold{q}}|=0$~.\label{twodeg}
\item Two minima with one at $|\tilde{\bold{q}}| = 0$~ where:
\begin{enumerate}
\item the minimum at $|\tilde{\bold{q}}| = 0$ is the global minimum,\label{q0globmin}
\item the minimum at $|\tilde{\bold{q}}| > 0$ is the global minimum.\label{q0nonglobmin}
\end{enumerate}
\item A single minimum at $|\tilde{\bold{q}}|=0$~.\label{hottest}
\end{enumerate}
Examples of each of these thermal configurations are displayed in figures \ref{phases} and \ref{phases3}. 

Depending on the location in the $(\tilde{\kappa},\tilde{T})$-plane, the system will be driven (either through thermal activation or quantum tunneling) to the most stable configuration. If the system is in a metastable configuration, such a process can take exponentially long. At high enough temperatures, the system will eventually fall back to $|\tilde{\bold{q}}|=0$ where the Higgs and the Coulomb branch meet, mimicking gravitational collapse or the melting of the molecule. 

\begin{figure}[t!]
\begin{center}
\includegraphics[width=2in]{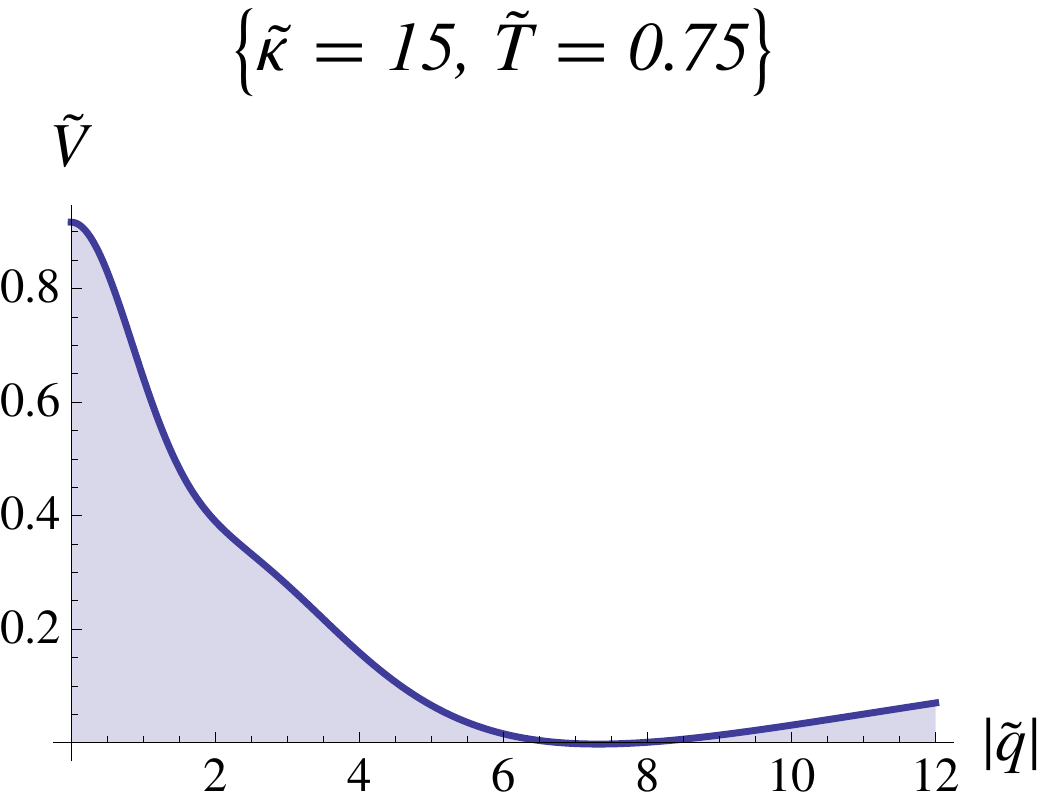} 
\includegraphics[width=2in]{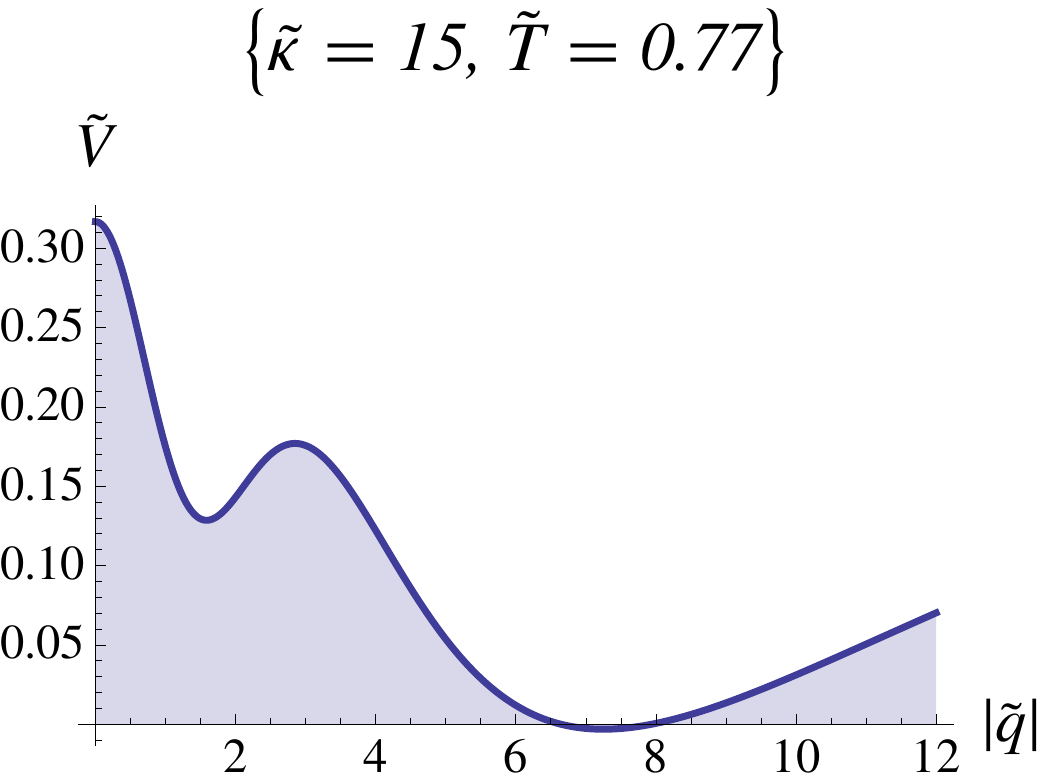}\vspace{0.7cm} 
\includegraphics[width=2in]{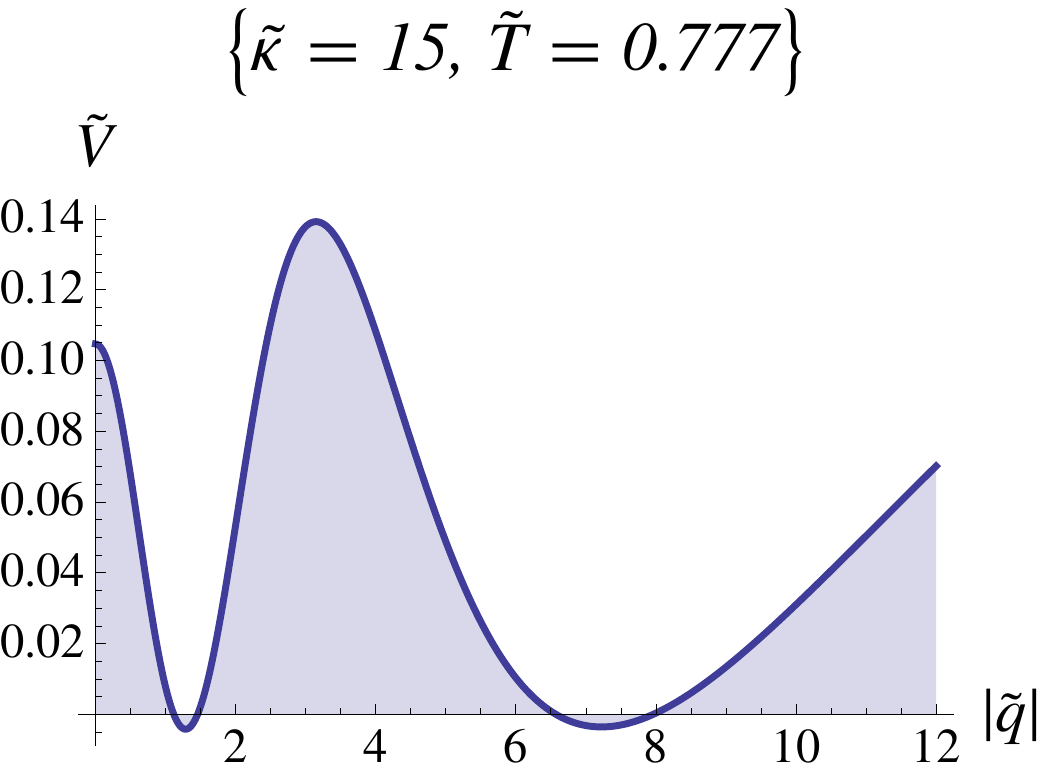} 
\includegraphics[width=2in]{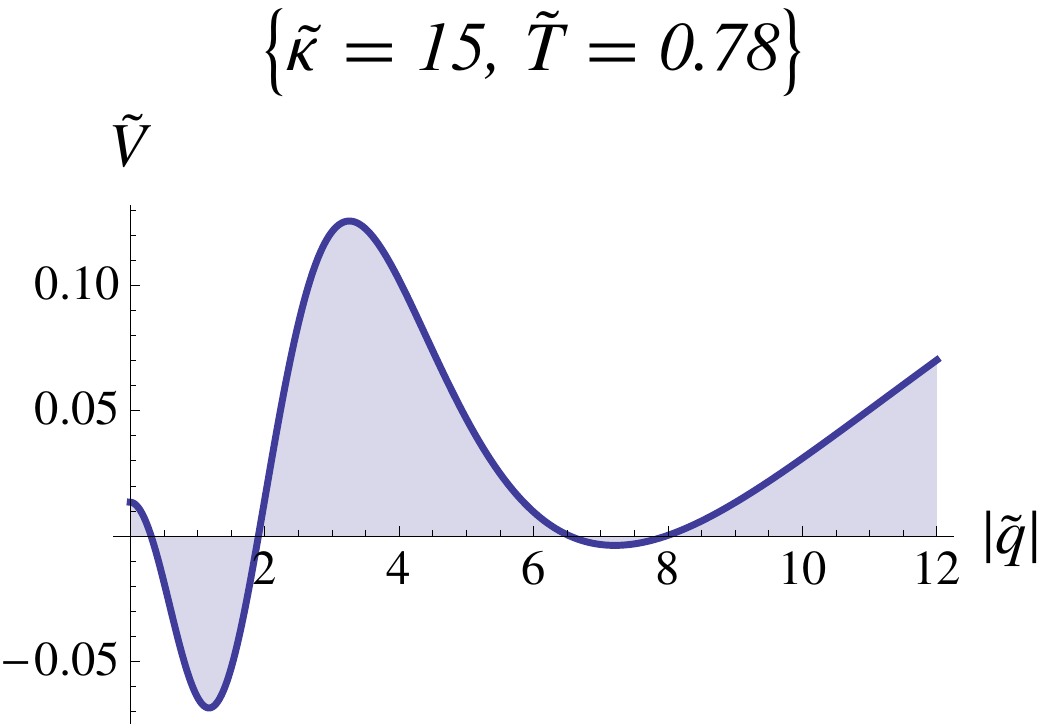}\vspace{0.7cm}
\includegraphics[width=2in]{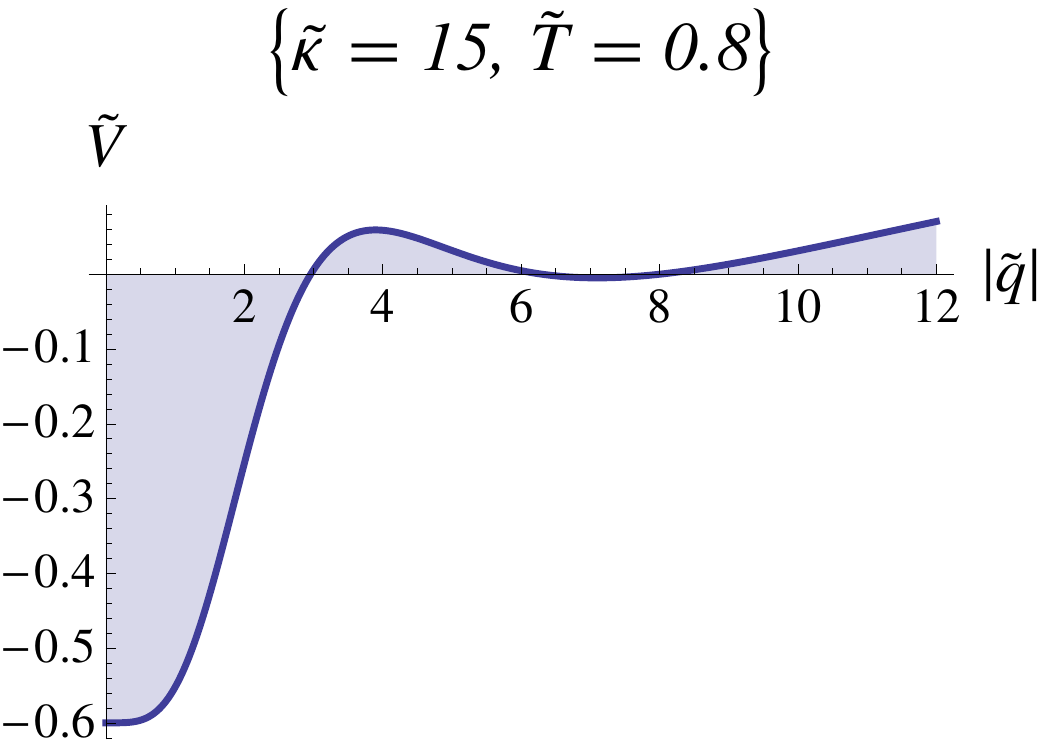} 
\includegraphics[width=2in]{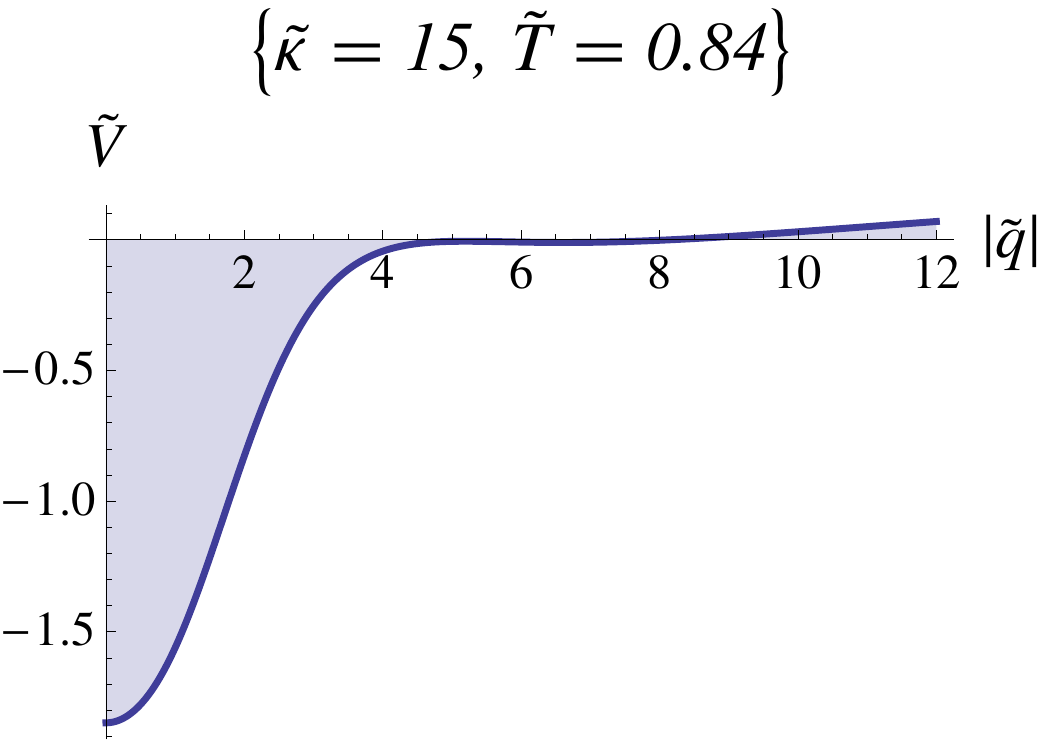} %\newline
\end{center}
\caption{Thermal effective potentials of a two node quiver ($\theta=-1$ and $\mu=1$). As the temperature is increased the system explores various thermal configurations of stable and metastable minima. From top left to bottom right the system is of type \ref{singstab} $\to$ \ref{twnondeg} $\to$ \ref{twodeg} $\to$ \ref{twnondeg} $\to$ \ref{q0globmin} $\to$ \ref{hottest}.}\label{phases}
\end{figure}

Note that for bound states in $\mathcal{N}=2$ supergravity, the effective potential of a small probe around a large hot black hole, as studied in \cite{Anninos:2011vn}, never exhibited two minima away from $|\tilde{\bold{q}}|=0$ (that is we never noticed potentials of types \ref{twnondeg}-\ref{q0nonglobmin}). Since supergravity is a good effective description at large $g_{\rm eff}\sim g_s\kappa$, it would be interesting to see if this behavior matches qualitatively in the Coulomb branch as $\tilde{\kappa}$ is increased. An example of this is shown in figure \ref{increasekappa}.

%The phase diagram is a function of three parameters $\kappa$, $\theta$ and $T$.
%We also find, for zero temperature bound states that are sufficiently far, the possible metastable bound states as well as the formation of two minima at finite temperature. An example is shown in figure \ref{bumpy}. In this case, unless the temperature is too high, the descent back into the Higgs branch will occur through a thermal or quantum fluctuation rather than a classical roll.

\begin{figure}[t!]
\begin{center}
\begin{tabular}{c|c}
\includegraphics[width=2in]{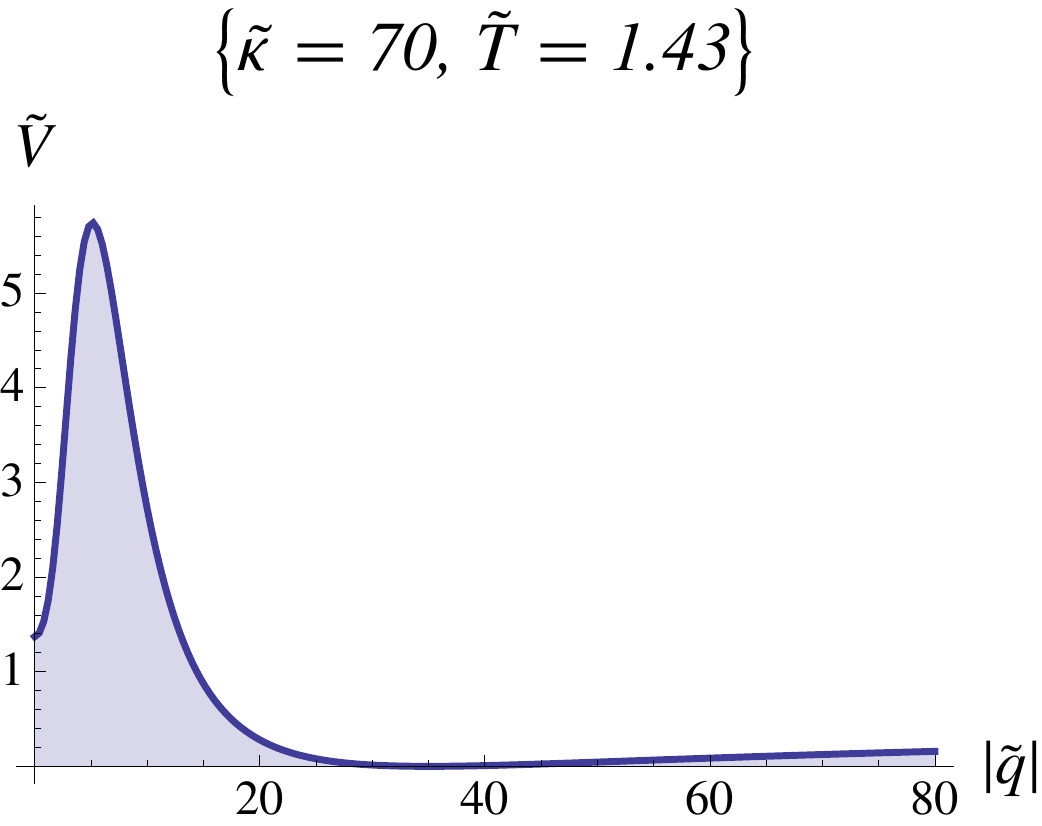} &
\includegraphics[width=2in]{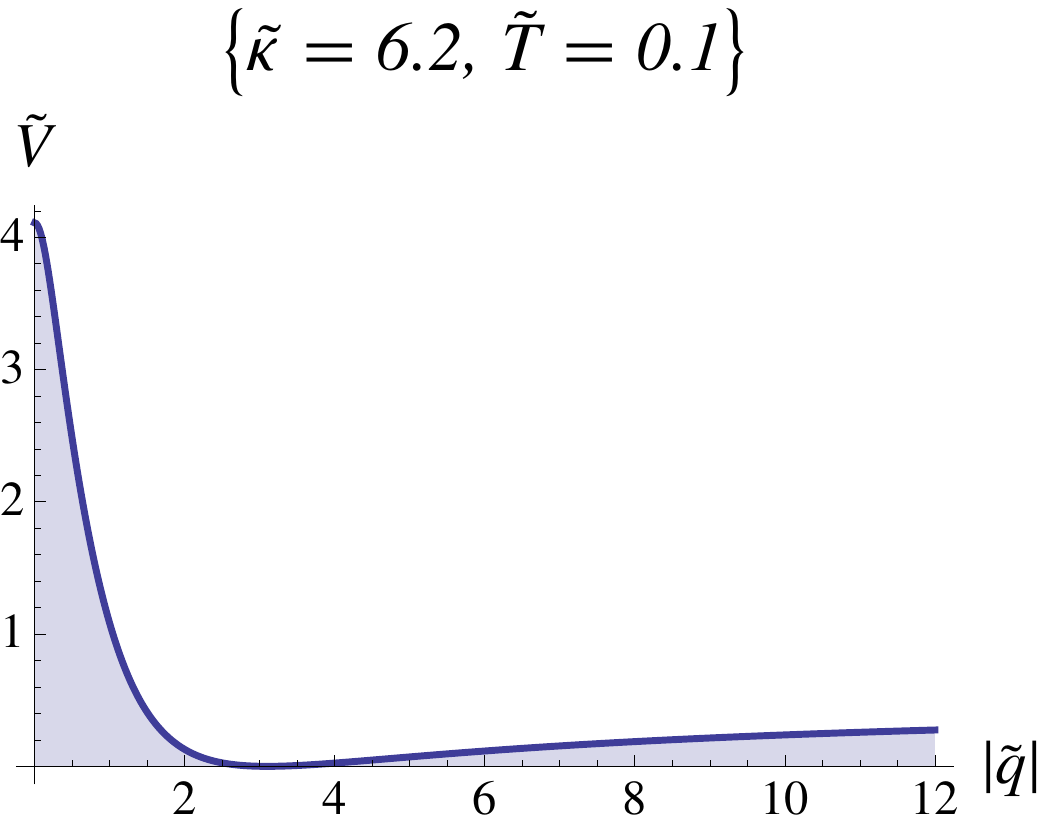} 
\includegraphics[width=2in]{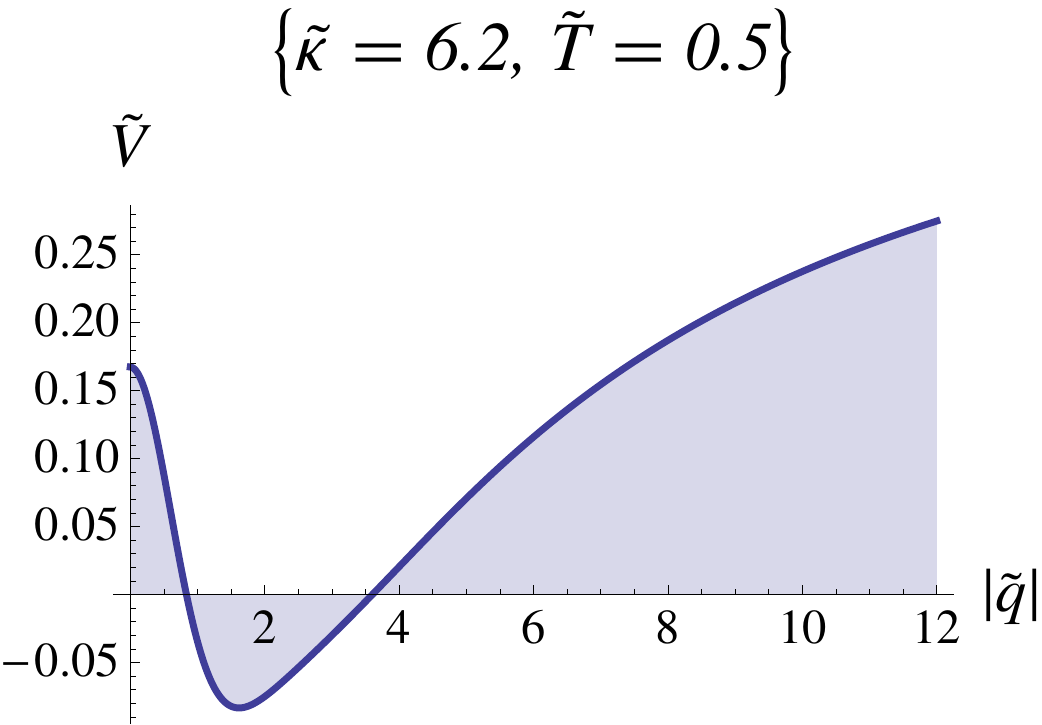} 
\end{tabular}
\end{center}
\caption{Thermal effective potentials of a two node quiver (for $\theta=-1$ and $\mu=1$). \emph{Left}: An example of phase type \ref{q0nonglobmin}. \emph{Right}: A case where the potential of the supersymmetric minimum decreases as the temperature is increased. A similar observation was made for supersymmetric bound states in \cite{Anninos:2011vn}.}\label{phases3}
\end{figure}

\subsubsection*{ \it High temperature behavior: an entropic effect}

%We can study the large temperature expansion of the potential more carefully. One finds upon expanding up to $\mathcal{O}(1/T)$:
%%Indeed, we already noticed how $V_0(r)=\lim\limits_{T \to 0} V(T)=-\mu D^2/2+\theta D-\kappa r+\kappa\sqrt{r^2+D}$ so that in the low-temperature limit the bound state is formed at the radius $\lim\limits_{T\to 0}r_{b.s.}(T)=-\kappa/(2\theta)$. 
%\be % V_{\infty}(r)=
%V(T)=-\frac{\mu}{2}D^2+\theta D+\frac{\kappa T}{2\pi}\log(\tan^2(a\pi))+\frac{\kappa\pi}{2T}\big( (D+r^2)\csc^2(a\pi)-r^2\sec^2(a\pi)\big)~.
%\ee 
%Upon integrating out the D-terms - which now can be done exactly - this yields: 
%\be
%V(T)=\frac{1}{2\mu}(\theta+\frac{\kappa\pi}{2T}\csc^2(a\pi))^2+\frac{\kappa T}{2\pi}\log(\tan^2 (a\pi))+\frac{\kappa\pi r^2}{2T}(\csc^2(a\pi)-\sec^2(a\pi))~.
%\ee 
%We wish to learn which $r$ minimizes this potential. To this end, note that $V=V(r,a)=F(a)+G(a)r^2$. To extremize this for both $a$ and $r$ we are to solve $\partial_rV(r,a)=\partial_aV(r,a)=0$. This gives the two equations $i)\hspace{4pt}2rG(a)=0$ and $ii)\hspace{4pt}F'(a)+r^2G'(a)=0.$ So indeed, one extremum is at $r=0$ where $F'(a)=0$. It's not hard to show that the latter equation is soluble, so indeed $r=0$ is a bona fide solution. 
%Figure \ref{melt} confirms this, and gives the region of intermediate temperature also, hinting at a phase transition.

A way to think about the high temperature melting transition is to integrate out the auxiliary $D$-fields first, which induces a quartic interaction $\left(|\phi_\alpha|^2 \right)^2/\mu$ for the bifundamentals. Since this interaction is relevant, it will play a minor effect at sufficiently large temperatures, namely for $T \gg \mu^{-1/3}$. Thus, at high temperatures we are dealing with a collection of $\kappa$ non-interacting complex bosonic and fermionic degrees of freedom with square masses $(|\bold{q}|^2 + \theta/m)$ and $|\bold{q}|^2$ respectively. In addition, due to the $U(1)$ connection we must only consider the gauge invariant states: the spectrum is constrained to those states annihilated by the $U(1)$ charge operator. A particularly convenient gauge is the temporal gauge, $A=0$.
%At sufficiently high temperatures this gauge constraint becomes less stringent since the number of allowed excited modes increases, and consequently so does the number of ways to construct gauge invariant states. Furthermore, t
At some fixed high temperature, the number of gauge invariant modes that aren't Boltzmann suppressed increases with decreasing mass and thus the dominant contribution to the free energy will come from for the lightest possible mass, i.e. $|\bold{q}|=0$. This can be viewed as an entropic effect \cite{Verlinde:2010hp}.

\subsection{Three nodes - unstable scaling solutions}

A similar analysis for three nodes gives rise to the following effective thermal potential:
\begin{equation}
V(T,D^i,a^i) = \sum_{i=1}^3 -\frac{\mu^i}{2} D^i D^i  + \theta^i D^i +  |\kappa^i| \, {T} \log \left( \frac{{\cosh}\left(\frac{ \sqrt{|\bold{q}^i|^2+s_i D^{i}}}{T}\right)-{\cos}(2 \pi a^i  )}{ {\cosh}\left(\frac{ |\bold{q}^{i}|}{T}\right)+{\cos}(2 \pi a^i  ) } \right)~.
\end{equation}
%where $r_{ij} \equiv |\bold{x}_{ij}|$. The sum goes over all $(i,j)$ pairs and $\mu_{ij} \equiv m_i m_j / (m_1+m_2+m_3)$. 
%Subsequently, we can expand in $D^{i} / |\bold{q}^i|^2$ and integrate out the $D^i$ fields.

Our main interest to understand the behavior of the scaling potential at finite temperature, and in particular the flat scaling direction. Thermal effects will kick in when $|\bold{q}^i| \lesssim T$. Since zero temperature scaling solutions occur for arbitrarily small $|\bold{q}^{i}|$, it is sufficient to perform a small temperature expansion. The dimensionless quantities capturing the thermal transition are $|\bold{q}^{i}| / T$.  As in the two-node case, one must integrate out the $U(1)$ connections $a^i$ and the $D^i$. For the case where the $\kappa^1 = \kappa^2 = - \kappa^3 >0$ we can identify a critical point along the scaling direction $|\bold{q}^i| = \lambda |\kappa^i|$: $a^1 = 1$, $a^2 = 0$, $D^1=0$ and $D^2 = 0$. On this saddle the potential becomes:
\begin{equation}\label{analytic}
V^*(T) = 2 \; T \; \sum_{i=1}^3 | \kappa^i | \, \log \left[ {{\tanh}\left(\frac{ {\lambda |\kappa^i|}}{2 T}\right)} \right]~,
\end{equation}
which is negative definite. Of course there could be more dominant saddles that make the potential even lower than the above. So more generally, this procedure must be done numerically, even in the small temperature expansion, as the equations governing the connections are intractable analytically. It amounts to numerically minimizing a function of four variables, namely $a^1$, $a^2$, $D^1$ and $D^2$ (recall that $a^3 = a^1 + a^2$ and  $D^3 = D^1 + D^2$). 

\begin{figure}[t!]
\begin{center}
\includegraphics[width=2in]{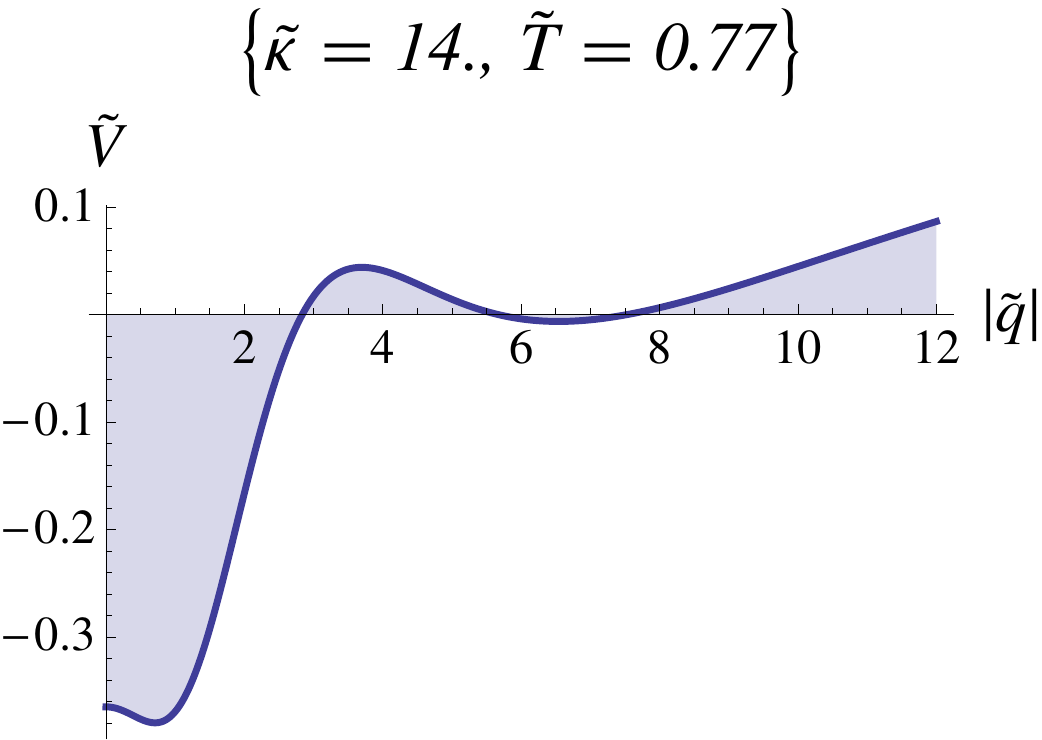}
\includegraphics[width=2in]{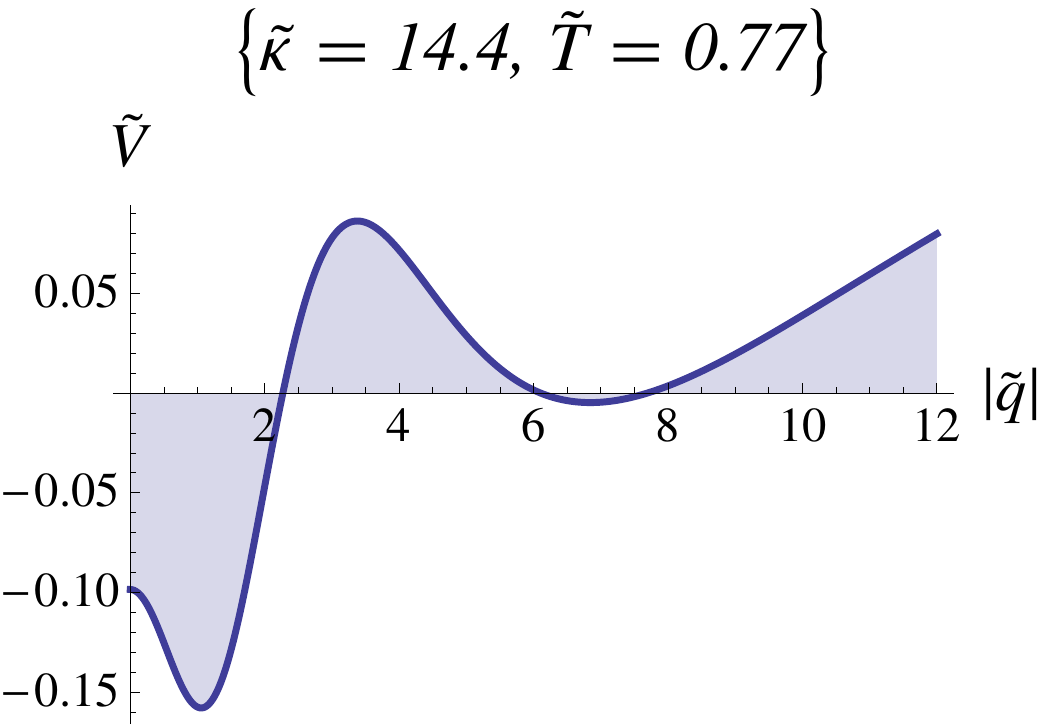} 
\includegraphics[width=2in]{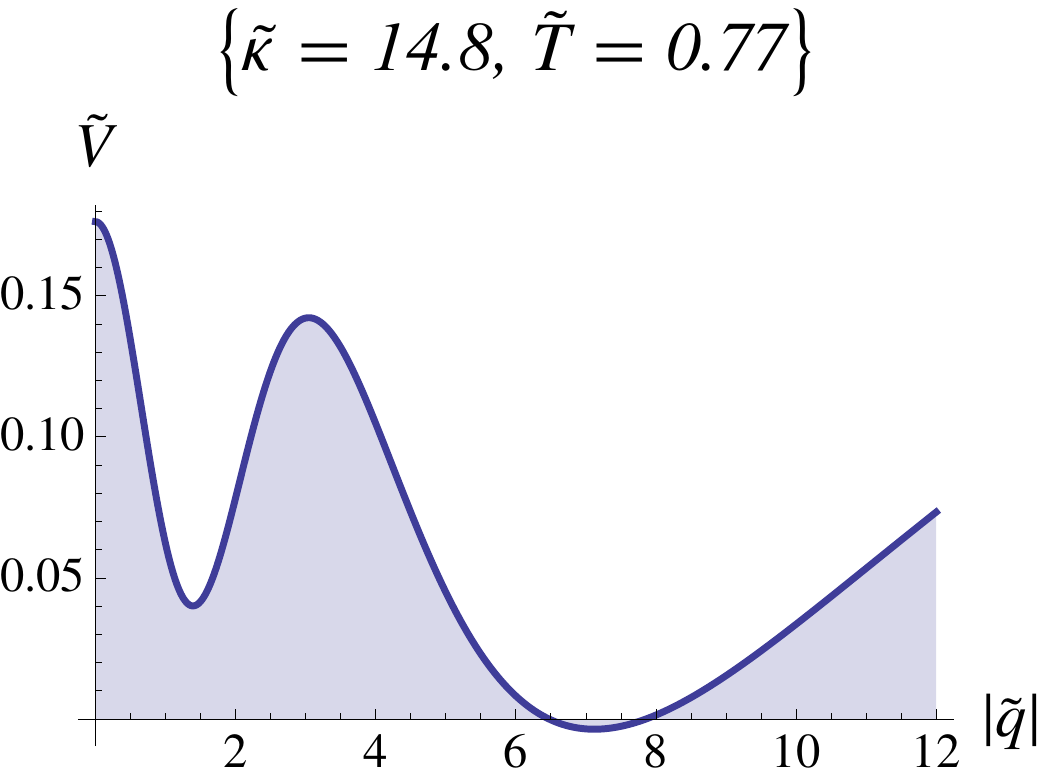} 
\includegraphics[width=2in]{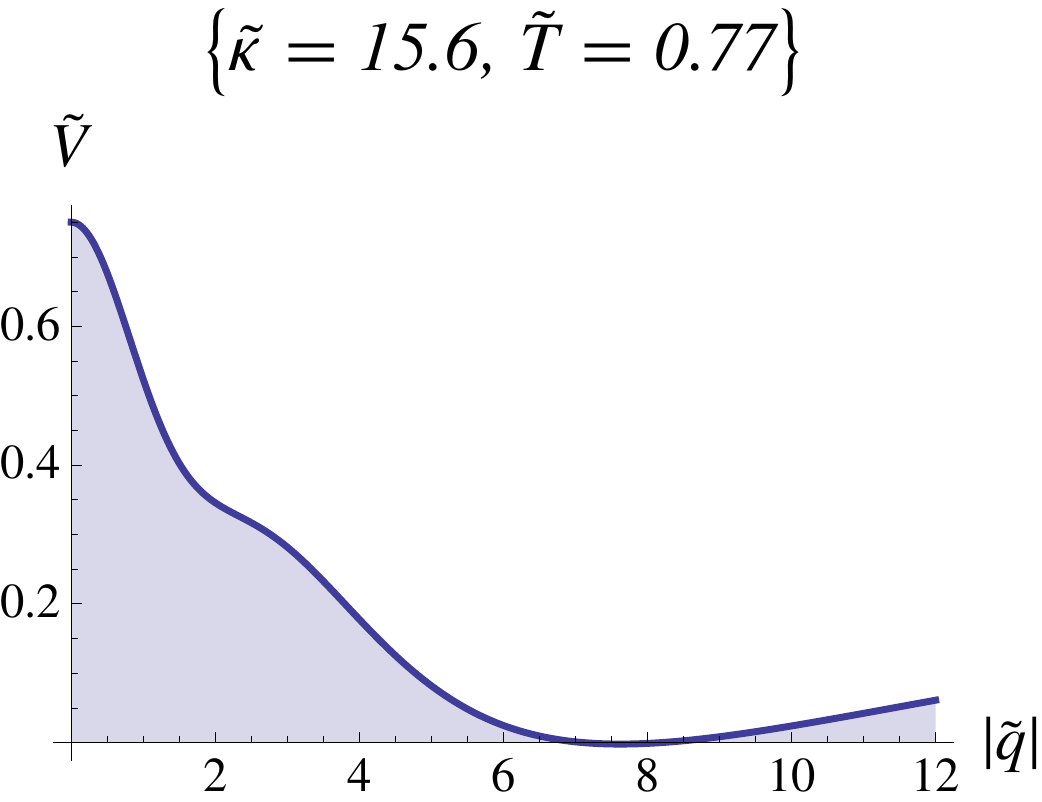} 
\end{center}
\caption{Thermal effective potentials of a two node quiver ($\theta=-1$ and $\mu=1$). As $\tilde{\kappa}$ is increased we note that the first minimum disappears.}\label{increasekappa}
\end{figure}

Since we are interested in the scaling branch we can set $\theta^i = \mu^i = 0$ in our analysis.\footnote{The reason we set $\theta^i = 0$ is for computational simplicity and numerical clarity. We could have started a Coulomb branch with non-zero $\theta^i$'s which has scaling solutions for $|\bold{q}^{i}| = \lambda \, \kappa^{i} + \mathcal{O}(\lambda^2)$ in the limit $\lambda \to 0$ and found similar results.} The resulting potential (with  $\theta^i = \mu^i = 0$) then exhibits the following scaling relation:
\begin{equation}
\kappa^i \to \delta \, \kappa^i~, \quad T \to \gamma \, T~, \quad |\bold{q}^i| \to \gamma \, |\bold{q}^i|~, \quad D^i \to \gamma^2 \, D^i~,
\end{equation}
such that $V(\kappa^i,|\bold{q}^i|,D^i,T) \to \gamma \, \delta \, V(\kappa^i, |\bold{q}^i|, D^i , T)$. From the scaling symmetries we observe that the numerical value of the temperature is of little meaning, all that matters for the thermal phase structure is whether or not it vanishes. This is to be expected since we are dealing with a scale invariant system.

In figure \ref{VoflambdaTfin} we display the potential at $T=0$ (left) and $T\neq 0$ (right)  in the scaling direction $|\bold{q}^{i}| = \lambda \; |\kappa^{i}|$ for $\kappa^{1} = \kappa^{2} = -\kappa^{3} = 1$. At $T=0$, we naturally find that it is vanishing for all $\lambda$. At $T \neq 0$ we see that for radial values larger than the temperature, the scaling direction is still effectively flat, but for radii of order the temperature the thermal potential quickly falls. The numerics are in good agreement with the analytic expression (\ref{analytic}). Hence, at finite temperature the system falls back into its Higgs branch. It would be interesting to explore this phenomenon in more generality for a larger number of nodes.
\begin{figure}[t!]
\begin{center}
{\includegraphics[width=3in]{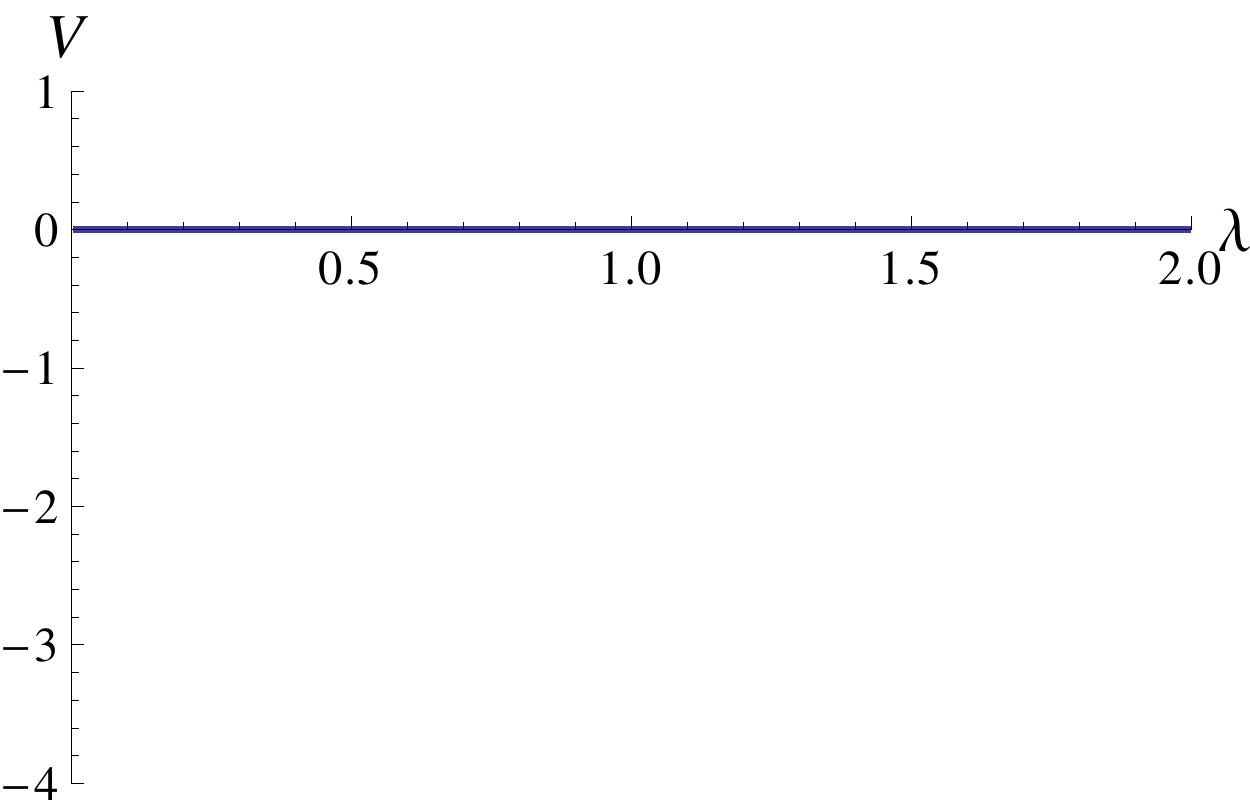} \quad \includegraphics[width=3in]{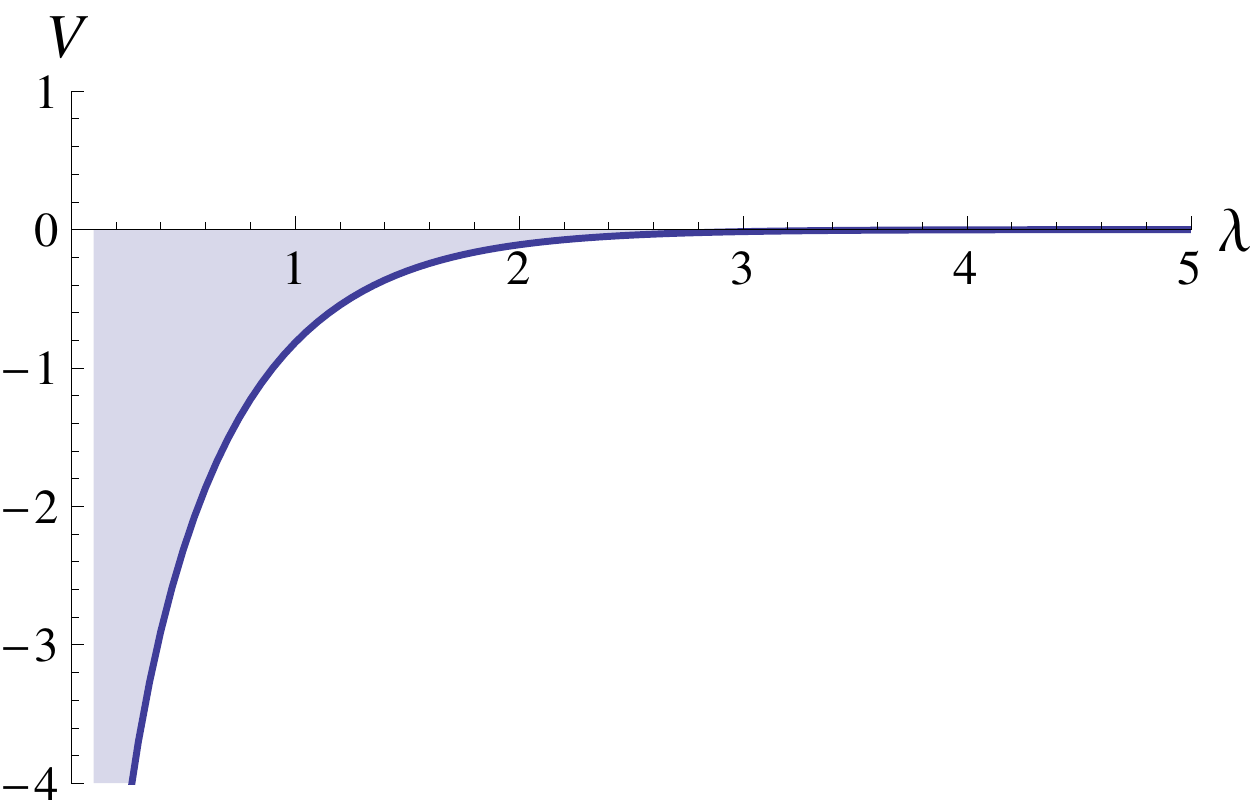}}
\caption{Thermal potential along the scaling direction $|\bold{q}^{i}| = \lambda \; \kappa^{i}$ for $\kappa^{1} = \kappa^{2}= -\kappa^{3} = 1$ at $T=0$ (left) and $T \neq 0$ (right).}\label{VoflambdaTfin}
\end{center}
\end{figure}
%Plot of the finite temperature ($T=0.5$) effective potential as a function of $\lambda$ for $\kappa_{ij}=1$ (left) and $\kappa_{ij}=5$ (right)
%\begin{figure}
%\begin{center}
%\includegraphics[width=3in]{V_of_lambda_T0_kappa1.pdf}
%\end{center}
%\caption{We set $x_{ij}=\lambda \kappa_{ij}$, $\kappa_{ij}=1$ and plot the effective potential at zero temperature as a function of $\lambda$. Naturally, it is a flat direction.}\label{VoflambdaT0}
%\end{figure}
%Increasing the magnitude of the $\kappa^{i}$ enhances the effect.
%In figure \ref{VoflambdaTfin} we the potential at finite temperature for two systems, one with $\kappa^{i} = 1$.

\section{Outlook} %: $SL(2,\mathbb{R})$, Holography and Wigner}
%ic Considerations and Wignerian

We have observed the emergence of a full $SL(2,\mathbb{R})$ symmetry from a quiver quantum mechanics model which itself is not a conformal quantum mechanics. The $SL(2,\mathbb{R})$ manifested itself in the effective theory of the position degrees of freedom of the D-particles, once the heavy strings were integrated out. Supersymmetry played an important role in the previous discussion given that the form of the quiver Lagrangian (\ref{quiverqm}) is heavily constrained by it. But as was mentioned in the introduction, the $SL(2,\mathbb{R})$ symmetry appears for geometries that need not be supersymmetric. In this final speculative section, we consider the idea that randomness (in the Wignerian sense) is behind the $SL(2,\mathbb{R})$ and observe that dilatations imply special conformal transformations for one dimensional systems whose dilatation symmetry is geometrized into an additional radial direction. Finally, we discuss some possible extensions.\footnote{We would like to acknowledge Frederik Denef, Diego Hofman and Sean Hartnoll for many interesting discussions leading to these ideas.} %We conclude with directions.

\subsection{Random Hamiltonians and emergent $SL(2,\mathbb{R})$?}\label{wignerian}
%Large number of nodes, 

It would be interesting to understand whether the emergence of such an $SL(2,\mathbb{R})$ from systems with large numbers of degrees of freedom, such as matrix quantum mechanics, could be more general. For instance, if there are a large number of almost degenerate vacua in the putative microscopic dual of AdS$_2$, one might imagine an approximate scale invariance emerging at large $N$ due to the formation of almost continuous bands of low energy eigenvalues. 

More generally, if the Hamiltonian is sufficiently complicated due to the multitude of internal cycles being wrapped by the branes, one might imagine drawing it from a random ensemble $\mathcal{H}_N$ of $N\times N$ Hermitean matrices. We can then formulate the following problem. Draw an $N\times N$ Hermitean matrix $H \in \mathcal{H}_N$ and assess (under some as of yet unspecified measure, perhaps the Frobenius norm is a possibility) how well the matrix equations (\ref{sl2ralg}) can be satisfied, given arbitrary Hermitean matrices $D$ and $K$. In particular, are the equations better satisfied as we increase $N$.\footnote{Of course, given the fact that $SL(2,\mathbb{R})$ has no finite dimensional unitary representations, the equations can only be satisfied exactly for $N = \infty$.} The emergence of an $SL(2,\mathbb{R})$ from a random ensemble of Hamiltonians, if true, would be similar in spirit to the emergence of the Wigner distribution of eigenvalue spacings from random matrices that is almost universal to quantum systems that become chaotic in the classical limit. %A simple 

Interestingly, an ensemble of random Hamiltonians with a scale invariant distribution of eigenvalues was found in \cite{amir}. We hope to return to this question in the future. 

\subsection{Holographic considerations}

If one assumes that there exists a gravitational dual to the theory, and in addition that a scaling symmetry $t \to \lambda^{a} t$ exists along with a radial transformation $z \to \lambda^2 z$ one obtains the two-dimensional metric (see for example \cite{Nakayama:2013is}):
\begin{equation}
ds^2 = -\frac{dt^2}{z^{a}}  + \frac{dz^2}{z^2}~,
\end{equation}
which is nothing more than AdS$_2$ (albeit in unusual coordinates). The isometry group of AdS$_2$ is $SL(2,\mathbb{R})$ and thus, holographically a dilatation symmetry seems to imply the existence a full $SL(2,\mathbb{R})$ symmetry. %(Recall that in quantum mechanics it is not the case that the existence of a dilatation symmetry $D$ implies the existence of a special conformal transformation $K$.) 
This is not true in higher dimensions unless one also assumes Lorentz invariance of the boundary metric. 

Another feature of conformal quantum mechanics that contrasts with higher dimensional conformal field theory is that there is no normalizable $SL(2,\mathbb{R})$ invariant ground state. On the other hand, $d$-dimensional conformal field theories have an $SO(d,2)$ invariant ground state wavefunctional. We view this as a hint that whatever the holographic description of AdS$_2$ is, it is not necessarily a conformal quantum mechanics off the bat. Indeed, the quiver quantum mechanics whose ground state degeneracies count a large fraction of the microstates of a black hole with an AdS$_2$ near horizon are {\it not} conformal quantum mechanics in and of themselves. Instead, the $SL(2,\mathbb{R})$ symmetry of AdS$_2$ might only become exact in some kind of large $N$ limit,\footnote{This $SL(2,\mathbb{R})$ might persist in a perturbative treatment of a supergravity solution but not at finite $N$, unless of course it resides in a much larger Virasoro structure.} as opposed to the $SO(d,2)$ symmetries of AdS$_{d+1}$ which should persist at finite $N$ (assuming the $\beta$-function of the dual CFT vanishes at finite $N$ as is the case for $\mathcal{N}=4$ SYM).

\subsection{Possible extensions}

As a final note, we would like to mention some open questions and extensions to our discussion. We have left question of superpotential corrections to the Coulomb branch of the three-node quiver untouched. Though the superpotential will not affect the ground state energy or existence of scaling solutions, it will correct the higher powers in the velocity expansion and it would be interesting to understand whether the $SL(2,\mathbb{R})$ symmetry is preserved by such corrections. The quartic interaction coming from integrating out the $D$-terms, $\left( |\phi_\alpha|^2 \right)^2 / \mu$, is dominated by an expansion in cactus diagrams in the large $\kappa$ limit. There is no such cactus diagram expansion for the interaction coming from integrating out the $F$-term, the quartic bosonic interaction being: $\sum_{\alpha, \beta, \gamma, \tilde{\beta}, \tilde{\gamma}} \omega_{\alpha\beta\gamma} \omega^*_{\alpha\tilde{\beta}\tilde{\gamma}} \phi^i_\beta \bar{\phi}^i_{\tilde{\beta}}  \phi^j_\gamma \bar{\phi}^j_{\tilde{\gamma}}$. Thus we are confronted with how (if at all) does one organize a large $\kappa$ expansion in this case. By dimensional analysis, and inspection of the two-loop Feynman diagrams (see figure \ref{dxdx2loop}}), we expect the velocity squared piece of the Coulomb branch Lagrangian in the scaling regime to become (up to $\mathcal{O}(1)$ coefficients):
\begin{equation}
\delta L_{c.b.} \sim |\kappa| \; {\bold{\dot{q}}} \cdot {\bold{\dot{q}}} \left( \frac{1}{|\bold{q}|^3} + \frac{(|\kappa| \, |\omega|)^2}{|\bold{q}|^6} + \mathcal{O}\left(\frac{|\kappa|^5 |\omega|^4}{|\bold{q}|^9}\right) \right)~,
\end{equation}
where $|\omega|^2 \sim |\kappa|^{-3} \sum_{\alpha, \beta,\gamma} |\omega_{\alpha\beta\gamma}|^2$. In order for the superpotential contribution to be subleading we would further require:
\begin{equation}
|\bold{q}| \gg (|\kappa| \omega_{\alpha\beta\gamma})^{2/3}~,
\end{equation}
in addition to the condition (\ref{scbounds}) that forces the system into the scaling regime. Thus, we clearly see that the effect of the superpotential (a relevant deformation) becomes important in the deep infrared region of the scaling regime, where detailed stringy physics begins to manifest itself and potentially destroys the $SL(2,\mathbb{R})$.% as it should given that it is a relevant deformation.

\begin{figure}[!ht]
\begin{center}
\setlength{\tabcolsep}{10mm}
\begin{tabular}{cccc}

\begin{fmffile}{dxdx2loop1}
\begin{fmfgraph*}(20,20) \fmfpen{thick}
\fmfleft{i1}
\fmflabel{$\delta x$}{i1}
\fmfright{o1}
\fmflabel{$\delta x$}{o1}
\fmf{photon}{i1,v1}
\fmf{plain,right,tension=.4}{v1,v2}
\fmf{plain,right,tension=.4,label=$\phi$}{v2,v1}
\fmf{plain,right,tension=.4}{v2,v3}
\fmf{plain,right,tension=.4}{v3,v2}
\fmf{photon}{v3,o1}
\end{fmfgraph*}
\end{fmffile}
& 
\begin{fmffile}{dxdx2loop2}
\begin{fmfgraph*}(20,20) \fmfpen{thick}
\fmfleft{i1}
\fmflabel{$\delta x$}{i1}
\fmfright{o1}
\fmflabel{$\delta x$}{o1}
\fmftop{t1}
\fmfbottom{b1}
\fmf{photon,tension=2}{i1,v1}
\fmf{plain,label=$\phi$}{v1,v2}
\fmf{plain}{v2,v3}
\fmf{plain}{v3,v1}
\fmf{phantom}{v2,v4}
\fmf{phantom}{v4,v1}
\fmf{plain,right,tension=1}{v3,t1,v3}
\fmf{phantom,right,tension=1}{v4,b1,v4}
\fmf{photon,tension=2}{v2,o1}
\end{fmfgraph*}
\end{fmffile}
&
\begin{fmffile}{dxdx2loop3}
\begin{fmfgraph*}(20,20) \fmfpen{thick}
\fmfleft{i1}
\fmflabel{$\delta x$}{i1}
\fmfright{o1}
\fmflabel{$\delta x$}{o1}
\fmftop{t1}
\fmfbottom{b1}
\fmf{photon,tension=4}{i1,v1}
\fmf{plain}{v1,v3}
\fmf{phantom}{v1,v5}
\fmf{plain,label=$\phi$}{v1,v2}
\fmf{phantom}{v3,t1}
\fmf{phantom}{v4,t1}
\fmf{phantom}{v5,b1}
\fmf{phantom}{v6,b1}
\fmf{plain}{v2,v4}
\fmf{phantom}{v2,v6}
\fmf{dots,right,tension=.1}{v3,v4}
\fmf{dots,right,tension=.1,label=$\psi$}{v4,v3}
\fmf{phantom,right,tension=.1}{v5,v6,v5}
\fmf{photon,tension=4}{v2,o1}
\end{fmfgraph*}
\end{fmffile}
&
\begin{fmffile}{dxdx2loop7}
\begin{fmfgraph*}(20,20) \fmfpen{thick}
\fmfleft{i1}
\fmflabel{$\delta x$}{i1}
\fmfright{o1}
\fmflabel{$\delta x$}{o1}
\fmftop{t1}
\fmfbottom{b1}
\fmf{photon,tension=4}{i1,v1}
\fmf{plain,right}{v1,v2,v1}
\fmf{phantom,right}{v1,v3,v1}
\fmf{plain,right}{v2,t1}
\fmf{plain,right,label=$\phi$}{t1,v2}
\fmf{phantom,right}{v3,b1}
\fmf{phantom,right}{b1,v3}
\fmf{photon,tension=4}{v1,o1}
\end{fmfgraph*}
\end{fmffile}
\\
\begin{fmffile}{dxdx2loop4}
\begin{fmfgraph*}(20,20) \fmfpen{thick}
\fmfleft{i1}
\fmflabel{$\delta x$}{i1}
\fmfright{o1}
\fmflabel{$\delta x$}{o1}
\fmftop{t1}
\fmfbottom{b1}
\fmf{photon,tension=3}{i1,v1}
\fmf{phantom,tension=4}{v3,t1}
\fmf{phantom,tension=4}{v4,b1}
\fmf{plain}{v1,v3}
\fmf{plain,label=$\phi$}{v4,v1}
\fmf{dots}{v2,v3}
\fmf{dots,label=$\psi$}{v4,v2}
\fmf{dots}{v3,v4}
\fmf{photon,tension=3}{v2,o1}
\end{fmfgraph*}
\end{fmffile}
& 
\begin{fmffile}{dxdx2loop5}
\begin{fmfgraph*}(20,20) \fmfpen{thick}
\fmfleft{i1}
\fmflabel{$\delta x$}{i1}
\fmfright{o1}
\fmflabel{$\delta x$}{o1}
\fmftop{t1}
\fmfbottom{b1}
\fmf{photon,tension=3}{i1,v1}
\fmf{phantom,tension=4}{v3,t1}
\fmf{phantom,tension=4}{v4,b1}
\fmf{dots}{v1,v3}
\fmf{dots}{v4,v1}
\fmf{dots}{v2,v3}
\fmf{dots,label=$\psi$}{v4,v2}
\fmf{plain,label=$\phi$}{v3,v4}
\fmf{photon,tension=3}{v2,o1}\end{fmfgraph*}
\end{fmffile}
&
\begin{fmffile}{dxdx2loop6}
\begin{fmfgraph*}(20,20) \fmfpen{thick}
\fmfleft{i1}
\fmflabel{$\delta x$}{i1}
\fmfright{o1}
\fmflabel{$\delta x$}{o1}
\fmftop{t1}
\fmfbottom{b1}
\fmf{photon,tension=4}{i1,v1}
\fmf{dots}{v1,v3}
\fmf{phantom}{v1,v5}
\fmf{dots,label=$\psi$}{v1,v2}
\fmf{phantom}{v3,t1}
\fmf{phantom}{v4,t1}
\fmf{phantom}{v5,b1}
\fmf{phantom}{v6,b1}
\fmf{dots}{v2,v4}
\fmf{phantom}{v2,v6}
\fmf{dots,right,tension=.1}{v3,v4}
\fmf{plain,right,tension=.1,label=$\phi$}{v4,v3}
\fmf{phantom,right,tension=.1}{v5,v6,v5}
\fmf{photon,tension=4}{v2,o1}
\end{fmfgraph*}
\end{fmffile}
&
\begin{fmffile}{dxdx2loop8}
\begin{fmfgraph*}(20,20) \fmfpen{thick}
\fmfleft{i1}
\fmflabel{$\delta x$}{i1}
\fmfright{o1}
\fmflabel{$\delta x$}{o1}
\fmftop{t1}
\fmfbottom{b1}
\fmf{photon,tension=4}{i1,v1}
\fmf{plain}{v1,v2}
\fmf{plain}{v1,v3}
\fmf{phantom}{v1,v4}
\fmf{phantom}{v1,v5}
\fmf{phantom,tension=4}{v2,t1}
\fmf{phantom,tension=4}{v3,t1}
\fmf{phantom,tension=4}{v2,i1}
\fmf{phantom,tension=4}{v3,o1}
\fmf{phantom,tension=4}{v4,b1}
\fmf{phantom,tension=4}{v5,b1}
\fmf{phantom,tension=4}{v4,i1}
\fmf{phantom,tension=4}{v5,o1}
\fmf{dots}{v2,v3}
\fmf{phantom}{v4,v5}
\fmf{plain}{v2,t1}
\fmf{plain,label=$\phi$}{t1,v3}
\fmf{phantom}{v4,b1,v5}
\fmf{photon,tension=4}{v1,o1}
\end{fmfgraph*}
\end{fmffile}
\end{tabular}
\end{center}
\caption{2-loop Feynman diagrams contributing to the $\delta x \delta x$ term of the effective Lagrangian. Solid lines represent $\phi$, while dotted lines correspond to $\psi$ propagators.}
\label{dxdx2loop}
\end{figure}
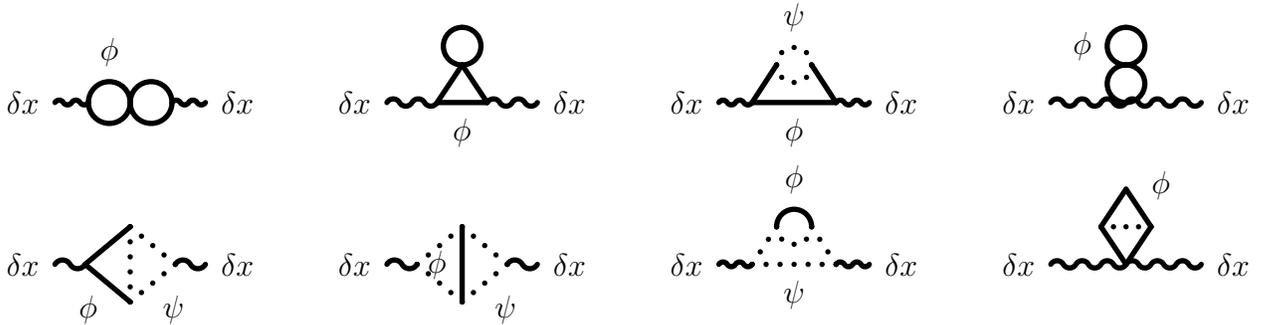

We have also left untouched the issue of larger numbers of nodes. In such a case one could consider mixed Higgs-Coulomb branches. For example, we could consider a four-node quiver where three nodes are in their Higgs branch and the remaining one residing far away in the Coulomb branch (see also \cite{Iizuka:2001cw,Iizuka:2008hg}), as shown in figure \ref{quiver2}. Given the exponentially large number of ground states in the Higgs branch of a three node closed loop quiver, heating such a system up might provide a useful toy model for a D-particle falling into a black hole. 

\begin{figure}
\begin{center}
\includegraphics[width=5in]{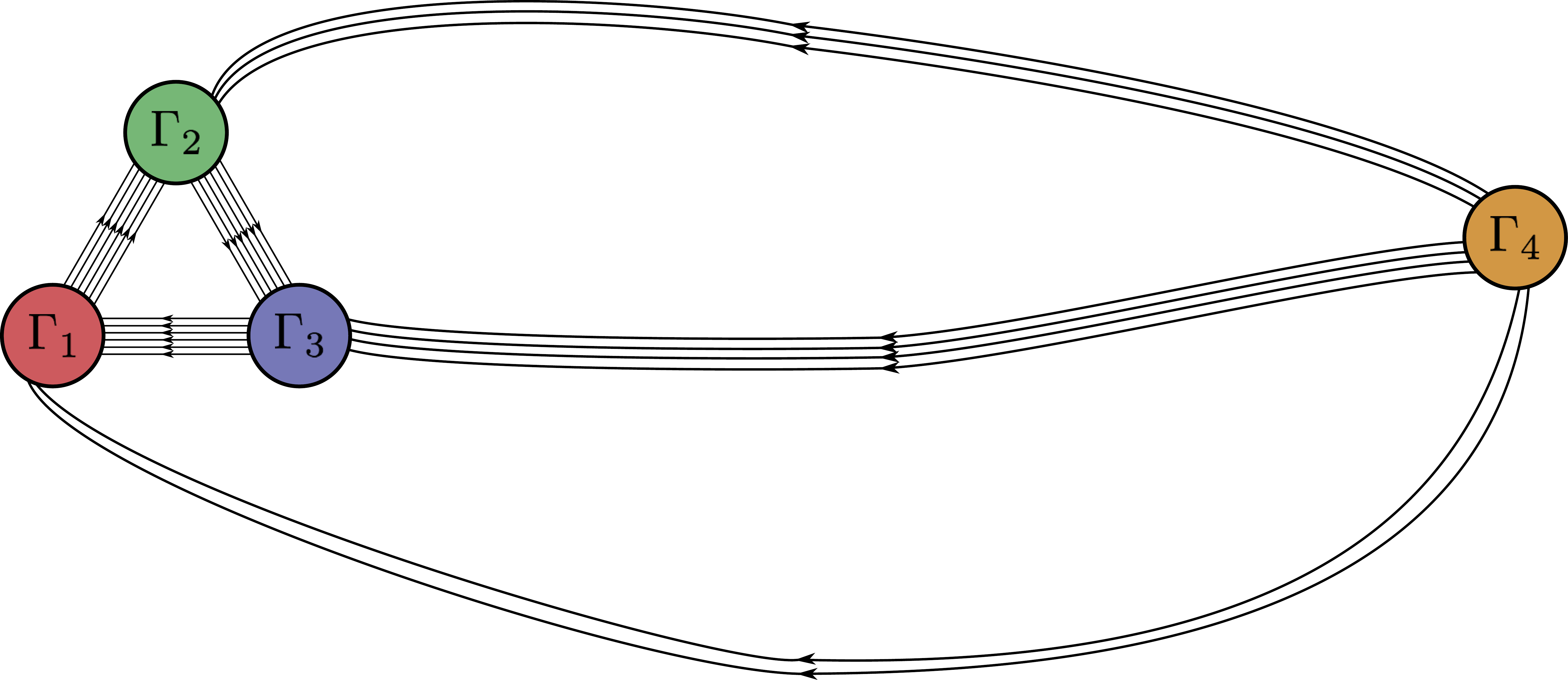} 
\end{center}
\caption{A schematic representation of a system in a mixed Higgs-Coulomb branch. The long arrows represent very massive strings. Note that there is a closed loop connecting $\Gamma_1$, $\Gamma_2$ and $\Gamma_3$. }\label{quiver2}
\end{figure}

%A hint toward constraining this Wignerian approach to AdS$_2$ is that scale invariance must imply $SL(2,\mathbb{R})$ invariance in the large $N$ limit for our ensemble of random Hamiltonians.

\section*{Acknowledgements}

It is a great pleasure to thank Alejandra Castro, Jan de Boer, Sheer El Showk, Dan Freedman, Alessandra Gnecchi, Umut Gursoy, Sean Hartnoll, Diego Hofman, C.~Martes, Edgar Shaghoulian, Andy Strominger, Bert Vercnocke and Erik Verlinde for useful discussions. We are particularly grateful for uncountable insightful discussions with Frederik Denef throughout the course of this work. The authors are also grateful to the {\it Cosmology and Complexity} conference in Hydra for their hospitality while this work was in progress. D.A. would like to acknowledge the hospitality of the Aspen Center for Physics, the Crete Center for Theoretical physics and the Open Questions in an Open Universe Workshop in Istanbul where part of this work was performed. T.A. would also like to thank the Centro de Ciencias de Benasque for their warm hospitality while this work was in progress. This work has been partially funded by DOE grants DE-FG02-91ER40654 and DE-FG02-05ER41360 and by a grant of the John Templeton Foundation. The opinions expressed in this publication are those of the authors and do not necessarily reflect the views of the John Templeton Foundation.
%\section{Mixed Coulomb-Higgs branch}

%\newpage 

\appendix

\section{Notation}\label{notation}

%and $s_1 = 1$, $s_2 = 1$ and $s_3 = -1$
For convenience, we have used a compact notation whereby latin superscripts denote relative degrees of freedom or arrow directions in the quiver, for example: $(\bold{q}^1,\bold{q}^2,\bold{q}^3) \equiv (\bold{x}_{12},\bold{x}_{23},\bold{x}_{13})$ or $(\phi^1_\alpha,\phi^2_\beta,\phi^3_\gamma  )\equiv (\phi_{12}^\alpha,\phi_{23}^\beta,\phi_{13}^\gamma)$. The supermultiplets in this notation are: $\Phi^i_{\alpha} = (\phi^i_\alpha,\psi^i_\alpha,F^i_\alpha)$ and the relative vector multiplet is denotes as $\bold{Q}^i = (A^i,\bold{q}^i,\lambda^i,D^i)$. The only exception to the rule is given by $\theta^1 \equiv \theta_1$, $\theta^2 \equiv -\theta_3$ and $\theta^3 = 0$. 
\newline\newline
Furthermore, we have: $\bold{q}^3 =  \bold{q}^1 + \bold{q}^2$, $D^3 = D^1 + D^2$, $\lambda^3 = \lambda^1 + \lambda^2$. So with respect to the vector multiplet fields, the {\it relative} Lagrangian is only a function of $\bold{q}^1$, $\bold{q}^2$, $D^1$, $D^2$, $\lambda^1$ and $\lambda^2$. We also have: $|\phi_\alpha^i |^2 \equiv \sum_{\alpha} |\phi_\alpha^i |^2$.
\newline\newline
The orientation of the quiver is encoded by the $s^i$. For three nodes we deal with the case of a quiver with a closed loop, and without loss of generality the particular choice $s_1 = 1$, $s_2 = 1$ and $s_3 = -1$, corresponding to $\kappa^1, \kappa^2 > 0$ and $\kappa^3 < 0$. An example of a quiver without closed loops is $s_1 = -1$, $s_2 = 1$, $s_3 = -1$.
\newline\newline
The spinors $(\psi^i_\alpha)_a$ and $(\lambda^i)_a$ with $a=1,2$ transform in the $\bold{2}$ of the $SO(3)$ and the anti-symmetric symbol $\epsilon^{12} = +1$ is such that ${\lambda}^i \epsilon \psi^i_\alpha \equiv (\lambda^i)_a \epsilon^{ab} (\psi^i_\alpha)_b$. The $\boldsymbol{\sigma} = (\sigma^x,\sigma^y,\sigma^z)$ are the Pauli matrices: 
\begin{equation}
\sigma^x = \left( \begin{array}{cc}
0 & 1 \\
1 & 0 \end{array} \right)~, \quad \sigma^y =  \left( \begin{array}{cc}
0 & i \\
-i & 0 \end{array} \right)~, \quad \sigma^z =  \left( \begin{array}{cc}
1 & 0 \\
0 & -1 \end{array} \right)~.
\end{equation}
For example $\bar{\psi}^i_\alpha {\sigma^x} \psi^i_\alpha \equiv (\bar{\psi}^i_\alpha)^a {({\sigma^x})_{a}}^{b} (\psi^i_\alpha)_b$. Finally the $U(1)$ covariant derivative $\mathcal{D}_t \phi^i_\alpha \equiv \left(\partial_t + i A^i \right) \phi^i_\alpha$.

We will occasionally revert back to the original $\bold{x}_{i}$ notation, particularly in the appendices.

\section{Superpotential corrections of the Coulomb branch}\label{sucos}

It is interesting to compute the effects on the Coulomb branch dynamics due to quantum corrections from the superpotential. This amounts to a two-loop calculation. It is important to note that due to a non-renormalization theorem in \cite{Denef:2002ru}, the linear piece of the $\mathcal{N} = 4$ supersymmetric quantum mechanics Lagrangian is constrained to be of the form:
\begin{equation}\label{linear1}
L^{(1)} = \sum_i  \left( - U_i(\bold{x}) D_i + \bold{A}_i(\bold{x}) \cdot \dot{\bold{x}}_i  \right) + \sum_{i,j} \left( C_{ij}(\bold{x}) \bar{\lambda}_i \lambda_j + \bold{C}_{ij}(\bold{x}) \cdot \bar{\lambda}_i \boldsymbol{\sigma} \lambda_j \right)~,
\end{equation}
with:
\begin{equation}\label{linear2}
\bold{C}_{ij} = \nabla_i U_j = \nabla_j U_i = \frac{1}{2}\left( \nabla_i \times \bold{A}_j + \nabla_j \times \bold{A}_i \right)~, \quad\quad C_{ij} = 0~. 
\end{equation}
This form receives {\it{no}} corrections due to quantum effects originating from the superpotential. This means, in particular, that the supersymmetric configurations of the Coulomb branch which solve $U_i = 0$ for all $i$, both bound and scaling, are unmodified in the presence of a superpotential. 

\begin{figure}[!ht]
\begin{center}
\setlength{\tabcolsep}{10mm}
\begin{tabular}{c}
\begin{fmffile}{DD2loop}
\begin{fmfgraph*}(30,30) \fmfpen{thick}
\fmfleft{i1}
\fmfright{o1}
\fmf{dashes}{i1,v1}
\fmf{plain,left,tension=.3}{v1,v2}
\fmf{plain,left,tension=.3,label=$\phi_{31}$}{v2,v1}
\fmf{dashes}{v3,o1}
\fmf{plain,left,tension=.3}{v3,v2}
\fmf{plain,left,tension=.3,label=$\phi_{23}$}{v2,v3}
\fmflabel{$D_1$}{i1}
\fmflabel{$D_{2}$}{o1}
\end{fmfgraph*}						
\end{fmffile}
\end{tabular}
\end{center}
%\setlength{\tabcolsep}{10mm}
%\begin{tabular}{cc}
%\begin{fmffile}{D1selfenergy}
%\begin{fmfgraph*}(30,30) \fmfpen{thick}
%\fmfleft{i1}
%\fmflabel{$D_1$}{i1}
%\fmfright{o1}
%\fmflabel{$D_1$}{o1}
%\fmf{dashes}{i1,v1}
%\fmf{plain,right,tension=.4}{v1,v2}
%\fmf{plain,right,tension=.4,label=$\phi_{12}$}{v2,v1}
%\fmf{dashes}{v2,o1}
%\end{fmfgraph*}
%\end{fmffile}
%\end{tabular}
\caption{Example of Feynman diagram contributing to $D_i D_j$ from the superpotential.}\label{wfeynman}
\end{figure}
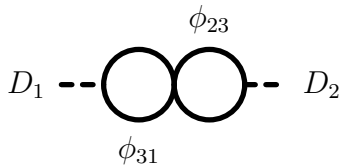

On the other hand, the quadratic piece can receive quantum corrections in the presence of a superpotential. An example of a Feynman diagram contributing to the $D_{i} D_{j}$ term is given in figure \ref{wfeynman}.

%below is NOT TRUE
%On the other hand, the quadratic piece can receive quantum corrections, however $\mathcal{N}=4$ supersymmetry \cite{smilga,Diaconescu:1997ut} constrains the coefficient of $\dot{\bold{x}}_i \cdot \dot{\bold{x}}_j$ to be equal to that of $D_i D_j$ and of the Kahler form $\nabla^2 \mathcal{K}(\bold{x}_i)$. Higher powers of the velocity expansion will receive corrections originating from the superpotential.
% non-renormalization theorem

\section{Finite $D$-terms}\label{finiteD}

When going to the Coulomb branch one integrates out the massive scalars, expands in small $D/r^2$, solves for the $D$ equations of motion and feeds the solution back into the action. We would like to show here that the supersymmetric configurations are in fact preserved for {\it finite} $D$. 

In the case of two node quivers, upon integrating out the chiral multiplet degrees of freedom $(\phi_\alpha,F_\alpha,\psi_\alpha)$ one finds the following non-linear equations for $D$ (we take $\theta_1 + \theta_2 = 0$ and $\kappa>0$):
\begin{equation}\label{2nD}
\mu D - \theta = \frac{\kappa}{2\sqrt{|\bold{q}|^2+D}}~, \quad \theta \equiv \theta_1 - \theta_2~.
\end{equation}
Given that the above equation is a cubic equation in $D$, one can find analytic solutions, which we are given in appendix \ref{twonodeD}. It is straightforward to see that at $|\bold{q}| = -\kappa/(2\theta)$, $D = 0$ is a solution to the non-linear equations. Plugging $D = 0$ back into the Lagrangian (which for zero velocity is the potential itself) shows that at $|\bold{q}| = -\kappa/(2\theta)$ the system has a zero energy. Away from $|\bold{q}| = -\kappa/(2\theta)$ the Lagrangian receives finite $D$ corrections. It is not hard to convince one's self, however, that the potential will never acquire another minimum due to finite $D$ effects. We argue this in appendix \ref{twonodeD}.

For three nodes the non-linear equations of $D^i$ become (we take $\theta_1 + \theta_2+\theta_3 = 0$ and pick the $\kappa$'s to form a closed loop in the quiver diagram):
\begin{equation}\label{threenodeD}
\mu^{1} D^1 +\mu^3D^3-\theta^1 =  \frac{|\kappa^1|}{2\sqrt{|\bold{q}^{1}|^2+D^1}}-\frac{|\kappa^3|}{2\sqrt{|\bold{q}^{3}|^2-D^{3}}}~,
%\mu^{1} D^1 +\mu^3 D^3-\theta^1 =  \frac{\kappa^1}{2\sqrt{|\bold{q}^{1}|^2+D^1}}+\frac{\kappa^3}{2\sqrt{|\bold{q}^{3}|^2+D^{3}}}~,
%m_{1} D_{1} -\theta_1 =  \frac{\kappa_{12}}{2\sqrt{|\bold{x}_{12}|^2+D_{12}}}+\frac{\kappa_{13}}{2\sqrt{|\bold{x}_{13}|^2+D_{13}}}~,
%m_{1} D_{1} -\theta_1 =  \frac{\kappa_{12}}{2\sqrt{r_{12}^2+D_{12}}}+\frac{\kappa_{23}}{2\sqrt{r_{23}^2+D_{23}}}-\frac{\kappa_{31}}{2\sqrt{r_{13}^2-D_{31}}}~,
\end{equation}
%\begin{eqnarray}
%m_{1} D_{12} -\theta_1 =  \frac{\kappa_{12}}{2\sqrt{r_{12}^2+D_{12}}}+\frac{\kappa_{23}}{2\sqrt{r_{23}^2+D_{23}}}-\frac{\kappa_{31}}{2\sqrt{r_{13}^2-D_{31}}}~, \\
%m_{2} D_{23} -\theta_2 =  \frac{\kappa_{23}}{2\sqrt{r_{23}^2+D_{23}}}~, \\
%m_{3} D_{31} -\theta_3 =  \frac{\kappa_{31}}{2\sqrt{r_{13}^2-D_{31}}}~.
%\end{eqnarray}
and cyclic permutations thereof. These equations can no longer be solved analytically. If we set the $\theta$'s to zero we can see that at $|\bold{q}^{i}| = \lambda \, |\kappa^{i}|$, $D^{i} = 0$ solves the linear equations and so the scaling solutions persist at finite $D$ for zero $\theta$. For non-zero $\theta$'s, setting $|\bold{q}^{i}| = \lambda \, |\kappa^{i}| + \mathcal{O}(\lambda^2)$ and expanding in small $\lambda$, we find that again $D^{i} = 0$ is a consistent solution order by order in the $\lambda$ perturbation theory. The existence of new bound states or scaling solutions which are non-supersymmetric due to finite $D$ effects becomes far more intricate in this case. A preliminary numerical scan seems to suggest there are none and we hope to report further on this in the future.

\subsection{Non-linear $D$-term solutions for two-node quivers}\label{twonodeD}

%Recall from section \ref{finiteD} that
For two-nodes, the $D$-term equation is given by (\ref{2nD}).
%\begin{equation}\label{2nD}
%\mu D - \theta = \frac{\kappa}{2\sqrt{|\bold{q}|^2+D}}~.
%%\mu D - \theta = \frac{\kappa}{2\sqrt{D+r^2}}~.
%\end{equation}
This equation can be solved analytically for $D$. Supersymmetric bound states are found when $\kappa \, \theta < 0$ at $|\bold{q}| = - \kappa/(2\theta)$. In what follows, we choose $\kappa>0$ and allow $\theta$ to have any sign. Equation (\ref{2nD}) is equivalent to solving
\begin{equation}
\left(D-\frac{\theta}{\mu}\right)^2\left(D+|\bold{q}|^2\right)-\frac{\kappa^2}{4\mu^2}=0~,
\end{equation}
which can be turned into a depressed cubic of the form $x^3+px+s=0$ using the subsitution $D\rightarrow x-\tfrac{1}{3}\left(|\bold{q}|^2-\tfrac{2 \theta}{\mu}\right)$ and identifying
\begin{equation}
p\equiv-\frac{1}{3}\left(|\bold{q}|^2+\frac{\theta}{\mu}\right)^2~,\quad\text{and}\quad s\equiv-\frac{\kappa^2}{4\mu^2}+\frac{2}{27}\left(|\bold{q}|^2+\frac{\theta}{\mu}\right)^3~.
\end{equation}
Since $p<0$, the three roots of this equation may be written as \cite{wikicubic}:
\begin{equation}
D_k=-\frac{1}{3}\left(|\bold{q}|^2-\frac{2 \theta}{\mu}\right)+2\sqrt{-\frac{p}{3}}\cos\left(\frac{1}{3}\arccos\left(\frac{3s}{2p}\sqrt{-\frac{3}{p}}\right)-k\frac{2\pi}{3}\right)~,
\end{equation}
for $k=0,1,2$. For all roots to be real, the arguments of the $\arccos$ must be between $[-1,1]$, which implies $4p^3+27s^2\leq 0$ or $|\bold{q}|^2\geq -\tfrac{\theta}{\mu}+\tfrac{3}{2}\left(\tfrac{\kappa^2}{2\mu^2}\right)^{1/3}$. The $k=0$ branch is real for all values of $|\bold{q}|$.

\begin{figure}
\begin{center}
\includegraphics[width=2.5in]{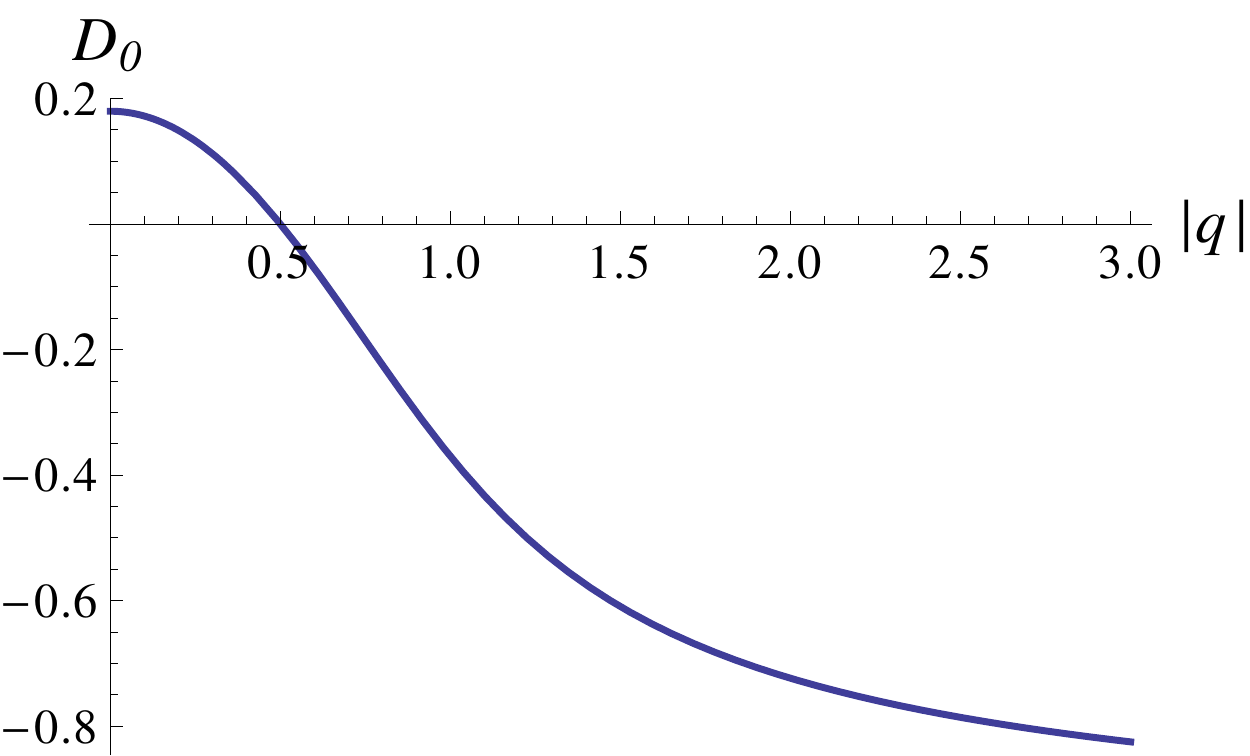}
\end{center}
\caption{Plot of $D_0$ for $\mu = -\theta = \kappa = 1$. Notice that the solution is real for all values of $|\bold{q}|$.}\label{dsol1}
\begin{center}
{\includegraphics[width=2.5in]{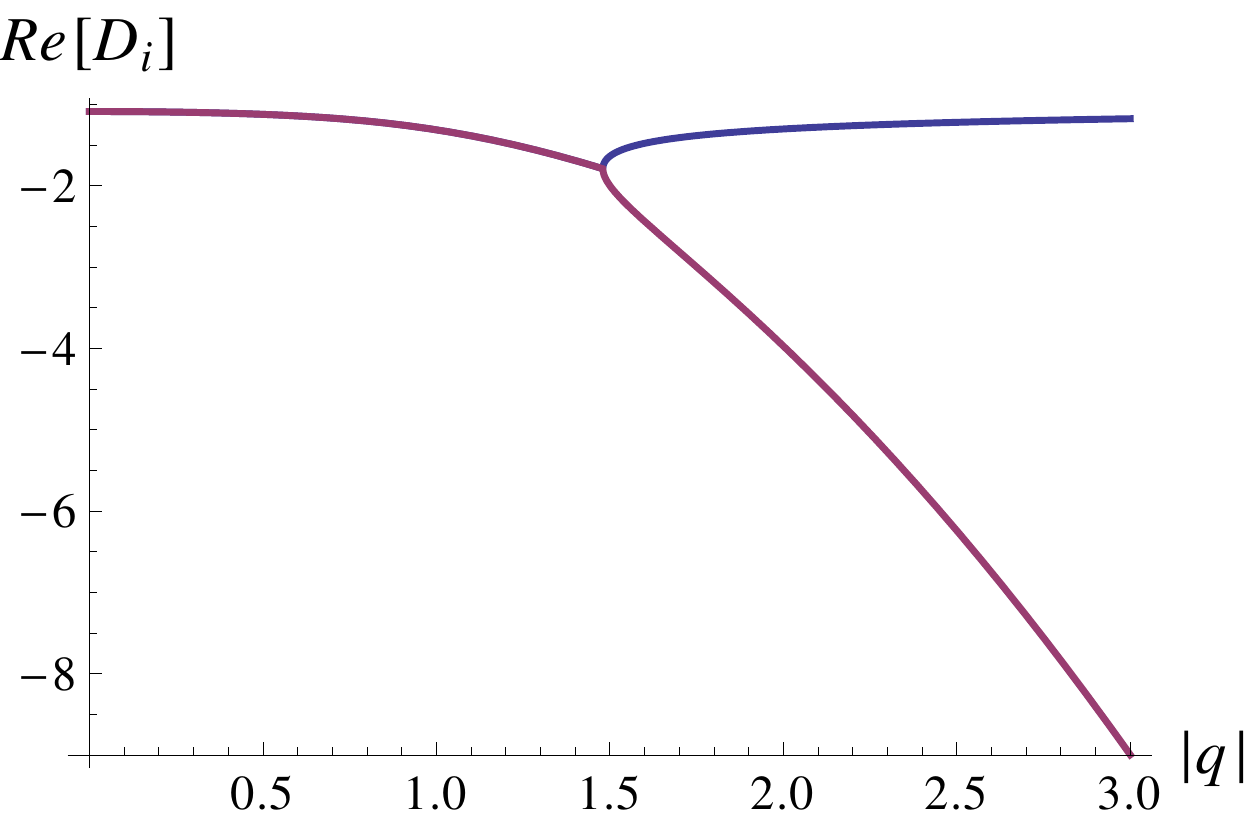} \quad \includegraphics[width=2.5in]{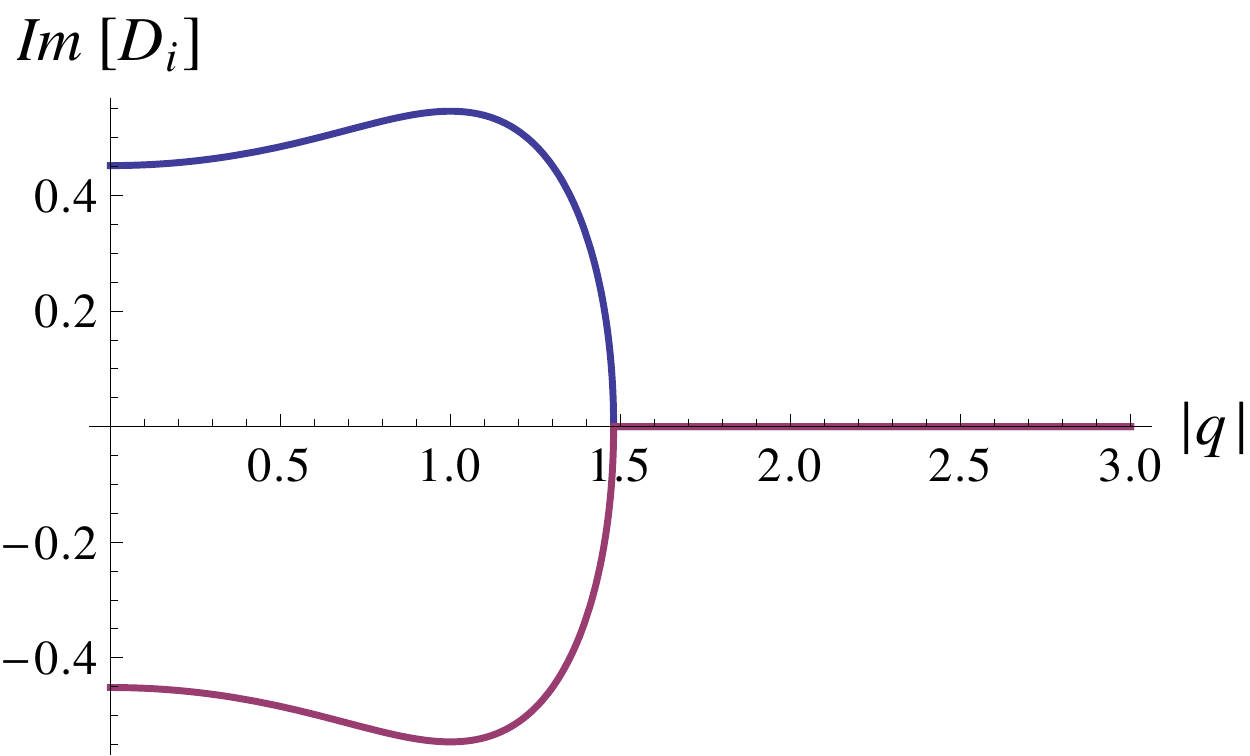}}
\caption{Left: Plot of $\Re[D_1]$ (blue) and $\Re[D_2]$ (violet). Right: Plot of $\Im[D_1]$ (blue) and $\Im[D_2]$ (violet). Both plots are for $\mu = -\theta = \kappa = 1$. Notice that when complex, the solutions form a conjugate pair.}\label{dsol2}
\end{center}
\end{figure}

\begin{figure}
\begin{center}
{\includegraphics[width=3in]{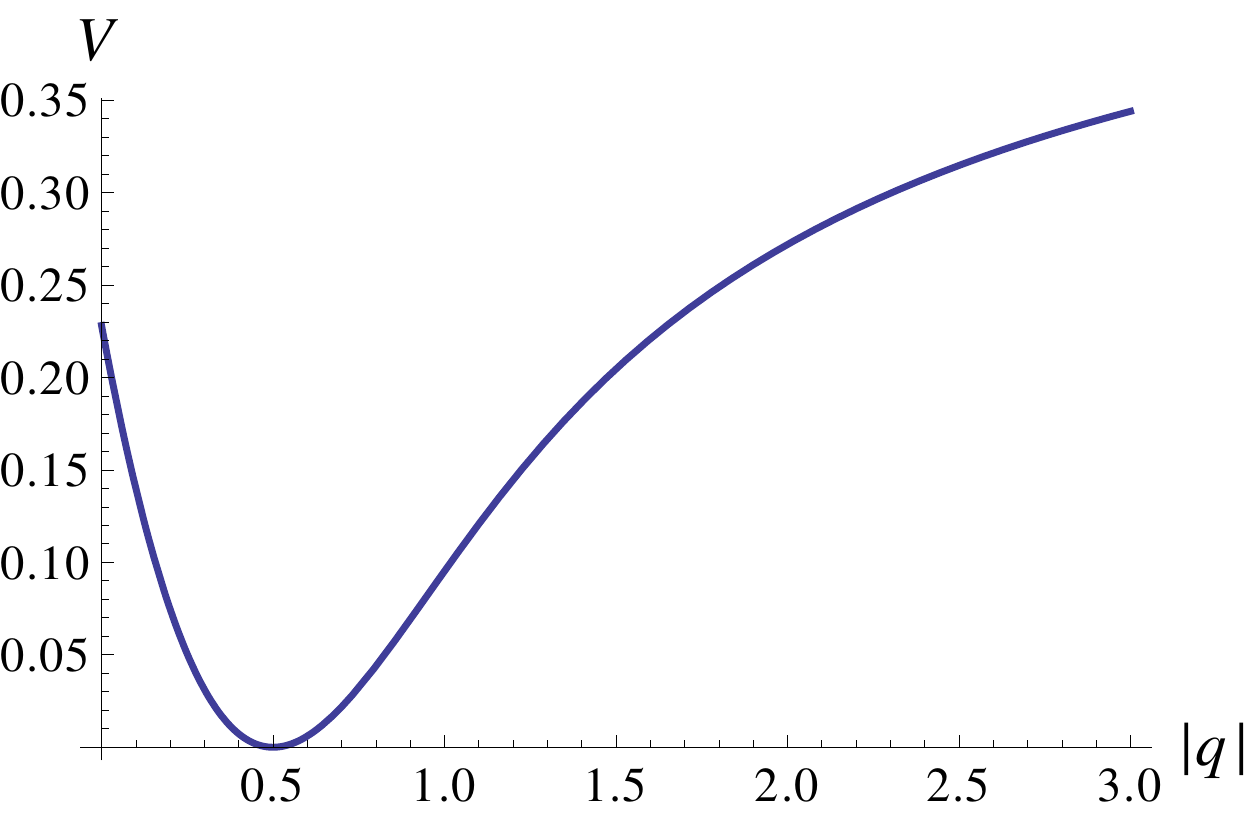} \quad \includegraphics[width=3in]{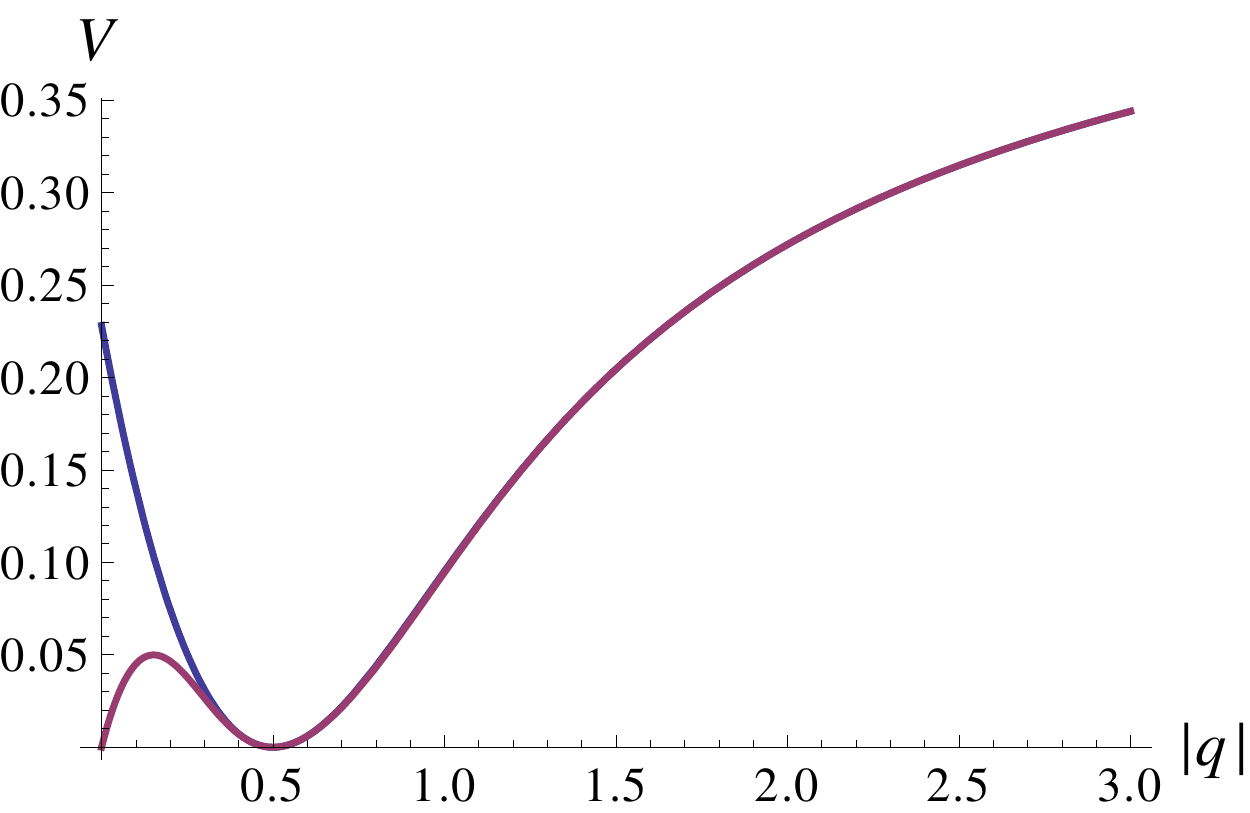}}
\end{center}
\caption{Left: Plot of $V$ evaluated on $D_0$ for $\mu = -\theta = \kappa = 1$. Right: Plot of $V$ evaluated on the perturbative $D$ solution in violet, compared with the full non-perturabative $D_0$ in blue.}\label{dsmall}
\end{figure}

In figures \ref{dsol1} and \ref{dsol2} we show some plots of the solutions.
In figure \ref{dsmall} we display the effective potential for the $D_0$ solution and its comparison to the effective potential obtained by expanding $D$ to second order in $D/|\bold{q}|^2$ and then evaluating the potential on-shell. We notice that the position
of the minimum is unaffected. Evaluating the potential on the $D_1$ and $D_2$ results in an unphysical complex potential for small values of $|\bold{q}|$. Scanning the parameters for all possible combinations of signs of $(\kappa,\theta)$ results in no other bound states.

\section{Three node Coulomb branch}

In this appendix we present some details leading to the Coulomb branch Lagrangian of a three-node quiver. As was shown in \cite{Denef:2002ru}, the $\mathcal{N}=4$ supersymmetric quantum mechanics of a $U(1)$ vector multiplet has a Lagrangian whose linear piece in the velocity and $D$ fields is completely fixed by supersymmetry. Explicit expressions are given in (\ref{linear1}) and (\ref{linear2}). 

In section \ref{cmetric} below, we show that the coefficient of the $\dot{\bold{x}}_i \cdot \dot{\bold{x}}_j$ term of the Lagrangian is the same as that of the $D_i D_j$ term, upon integrating out the chiral matter. In section \ref{potential} we integrate out the auxiliary $D$ fields in the Coulomb theory and obtain an expression for the effective potential on the position degrees of freedom.

\subsection{Second order Lagrangian for three-node quiver}\label{cmetric}

Recall the Lagrangian of a $n$-node quiver theory, after setting the fermionic $\lambda$ fields to zero, is given by:\footnote{For the purposes of this subsection we use the original notation, i.e. $\bold{x}_i$, $\lambda_i$, $\phi_{ij}$, $\psi_{ij}$, $\bold{x}_{ij} \equiv (\bold{x}_i - \bold{x}_j)$. and so on. We also supress the Greek indices on the chiral multiplet fields. Also, we do not decouple the center of mass degrees of freedom and switch off the superpotential.}
\begin{multline}
L = \sum_i \frac{m_i}{2}\left(|\dot{\bold{x}}_i |^2+D_i^2\right)-\theta_i D_i
+\sum_{j\rightarrow i}\left(|\dot\phi_{ij}|^2+F_{ij}^2+i\bar \psi_{ij}\dot\psi_{ij}\right) \\ 
-\sum_{j\rightarrow i}\left[\left(|\bold{x}_{ij}|^2+D_{ij}\right)|\phi_{ij}|^2+\bar \psi_{ij}\; \boldsymbol{\sigma} \cdot \bold{x}_{ij} \;\psi_{ij}\right]~.
%+\sum_{j\rightarrow i} \left( \frac{\partial W(\phi)}{\partial \phi_{ij}^\alpha}F_{ij}^\alpha +h.c.\right)
%+\left(\frac{\partial^2 W(\phi)}{\partial \phi_{ij}^\alpha \partial \phi_{ij}^\beta}\psi^\alpha \epsilon\psi^\beta+h.c.\right).
\label{pnodelagrangian}
\end{multline}
%\begin{eqnarray}
%L&=&\sum_p \frac{m_p}{2}\left(\dot x_p^2+D_p^2\right)-\theta_p D_p
%+\sum_{q\rightarrow p}\left(|\dot\phi_{pq}|^2+F_{pq}^2+i\bar \psi_{pq}\dot\psi_{pq}\right) \nonumber\\
%&-&\sum_{q\rightarrow p}[\left(x_{pq}^2+D_{pq}\right)|\phi_{pq}|^2+\bar \psi_{pq}\sigma^ix_{pq}^i\psi_{pq}]\nonumber\\
%&+&\sum_{q\rightarrow p} \left( \frac{\partial W(\phi)}{\partial \phi_{pq}^\alpha}F_{pq}^\alpha +h.c.\right)
%+\left(\frac{\partial^2 W(\phi)}{\partial \phi_{pq}^\alpha \partial \phi_{pq}^\beta}\psi^\alpha \epsilon\psi^\beta+h.c.\right).
%\label{pnodelagrangian}
%\end{eqnarray}
For the case $n=3$ there are three pairs of $(i,j)$ to be considered: $(1,2)$, $(2,3)$ and $(3,1)$. We do not consider contributions from the superpotential in this appendix. We take $\kappa_{12}>0$, $\kappa_{23}>0$ and $\kappa_{31}>0$.%The superpotential can consistently be assumed to have the form $W=w_{\alpha \beta \gamma}\phi_{12}^\alpha \phi_{23}^\beta \phi_{31}^\gamma$.
\newline\newline
The $\phi_{ij}$ propagator is given by:
\begin{equation}
D_{\phi_{ij}}(\omega)=\frac{i}{\omega^2-|\bold{x}_{ij}|^2}
\end{equation}
and the $\psi_{ij}$ propagator is given by:
\begin{equation}
D_{\psi_{ij}}(\omega)=\frac{-i}{\omega^2-|\bold{x}_{ij}|^2}
\begin{pmatrix}
\omega-z_{ij} & -{x}_{ij}+i y_{ij}\\
-i y_{ij}-x_{ij} & \omega+z_{ij}
\end{pmatrix},
\end{equation}
where $\bold{x}_{ij}=(x_{ij},y_{ij},z_{ij})$ (recall that each lower index denotes its corresponding node). We evaluate our momentum integrals on the imaginary axis, which is guaranteed to give the same result had we evaluated the integral along a real contour with the usual Feynman pole prescription, to lowest order in the the external momenta $l$. Higher order terms in the external momenta will differ by factors of $i$.  
%We will also use $r_{pq}=|\vec x_{p}-\vec x_{q}|$.
% and $r_{pq}^i$ denotes the $i$th component of $\vec r_{pq}$

\subsubsection*{$D_i \; D_j$ term}

\begin{figure}[!ht]
\begin{center}
\setlength{\tabcolsep}{10mm}
\begin{tabular}{cc}
\begin{fmffile}{D1selfenergy}
\begin{fmfgraph*}(30,30) \fmfpen{thick}
\fmfleft{i1}
\fmflabel{$D_1$}{i1}
\fmfright{o1}
\fmflabel{$D_1$}{o1}
\fmf{dashes}{i1,v1}
\fmf{plain,right,tension=.4}{v1,v2}
\fmf{plain,right,tension=.4,label=$\phi_{12}$}{v2,v1}
\fmf{dashes}{v2,o1}
\end{fmfgraph*}
\end{fmffile}
& 
\begin{fmffile}{D1selfenergy2}
\begin{fmfgraph*}(30,30) \fmfpen{thick}
\fmfleft{i1}
\fmflabel{$D_1$}{i1}
\fmfright{o1}
\fmflabel{$D_1$}{o1}
\fmf{dashes}{i1,v1}
\fmf{plain,right,tension=.4}{v1,v2}
\fmf{plain,right,tension=.4,label=$\phi_{31}$}{v2,v1}
\fmf{dashes}{v2,o1}
\end{fmfgraph*}
\end{fmffile}\\
$(i)$&$(ii)$
\end{tabular}
\end{center}
\caption{1-loop Feynman diagrams contributing to the $D_1^2$ term of the effective Lagrangian.}
\label{D1^2}
\end{figure}
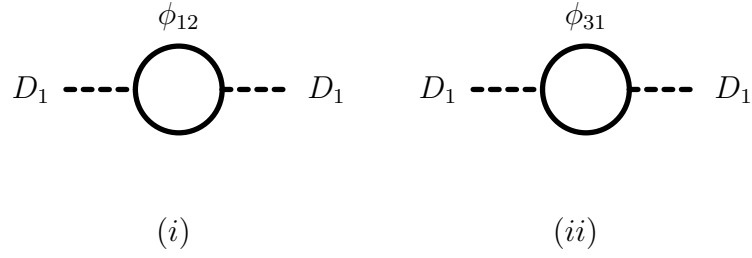

First let us consider the diagonal terms and more specifically the contribution to $D_1 D_1$. The two diagrams that contribute are shown in figure \ref{D1^2}.
The first diagram, series expanded for small external momentum $l$, is given by
\begin{equation}
\frac{1}{2}(-i)^2\kappa_{12}\int \frac{d\omega}{2\pi}D_{\phi_{12}}(\omega)D_{\phi_{12}}(\omega+l)=\frac{\kappa_{12}}{8 |\bold{x}_{12}|^3}-\frac{l^2\kappa_{12} }{32 |\bold{x}_{12}|^5}~,
\end{equation}
while the second is given by
\begin{equation}
\frac{1}{2}(-i)^2\kappa_{31}\int \frac{d\omega}{2\pi}D_{\phi_{31}}(\omega)D_{\phi_{31}}(\omega+l)=\frac{\kappa_{31}}{8 |\bold{x}_{31}|^3}-\frac{l^2\kappa_{31} }{32 |\bold{x}_{31}|^5}~.
\end{equation}
Therefore, ignoring the $l^2$ terms, the total contribution to the effective Lagrangian is
\begin{equation}
L_{eff,D_1^2}=\frac{D_{1}^2}{8}\left(\frac{\kappa_{31}}{|\bold{x}_{31}|^3}+\frac{\kappa_{12}}{|\bold{x}_{12}|^3}\right)~.
\end{equation}
The $D_2^2$ and $D_3^2$ terms can be obtained by cyclic permutation.

Now let us consider the off diagonal terms and more specifically $D_1D_2$. The only diagram that contributes to 1-loop order is:
\begin{center}
\setlength{\tabcolsep}{10mm}
\begin{tabular}{cc}
\begin{fmffile}{D1D2selfenergy}
\begin{fmfgraph*}(30,30) \fmfpen{thick}
\fmfleft{i1}
\fmflabel{$D_1$}{i1}
\fmfright{o1}
\fmflabel{$D_2$}{o1}
\fmf{dashes}{i1,v1}
\fmf{plain,right,tension=.4}{v1,v2}
\fmf{plain,right,tension=.4,label=$\phi_{12}$}{v2,v1}
\fmf{dashes}{v2,o1}
\end{fmfgraph*}
\end{fmffile}
\end{tabular}
\end{center} 
This diagram differs only by a sign and a factor of 2 compared to $(i)$ of figure \ref{D1^2}. Therefore, the $D_1D_2$ term of the effective Lagrangian is
\begin{equation}
L_{eff,D_1D_2}=-D_1D_2\left(\frac{\kappa_{12}}{4|\bold{x}_{12}|^3}\right)~.
\end{equation}
Other off diagonal terms can be obtained by cyclic permutation.

\subsubsection*{$\delta \bold{x}_i \cdot \delta \bold{x}_j$ term}

In order to obtain the correction to the quadratic velocity terms in the Lagrangian, we consider fluctuations the bosonic degrees of freedom $\bold{x}_i+\delta\bold{x}_i$.  Terms of this form can be divided in terms that are diagonal in the quiver such as $\delta \bold{x}_1 \cdot \delta \bold{x}_1$ and others that are off diagonal such as $\delta \bold{x}_1 \cdot \delta \bold{x}_2$. For each of these there is a further subdivision to terms that are diagonal in the vector component such as $\delta {x}_1 \delta {x}_1$ and $\delta {x}_1 \delta {x}_2$ and terms that are off diagonal in the vector component such as $\delta {x}_1 \delta {y}_1$ and $\delta {x}_1 \delta {y}_2$.

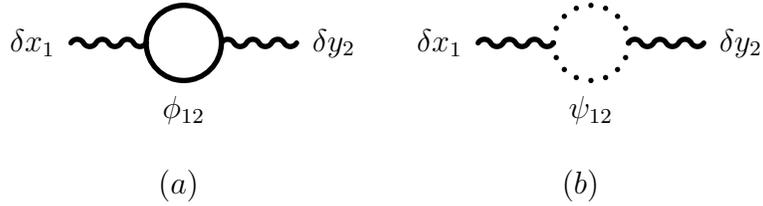
\begin{figure}[!ht]
\begin{center}
\setlength{\tabcolsep}{10mm}
\begin{tabular}{cc}
\begin{fmffile}{dx11dx22phi}
\begin{fmfgraph*}(30,30) \fmfpen{thick}
\fmfleft{i1}
\fmflabel{$\delta x_1$}{i1}
\fmfright{o1}
\fmflabel{$\delta y_2$}{o1}
\fmf{photon}{i1,v1}
\fmf{plain,right,tension=.5,label=$\phi_{12}$}{v1,v2}
\fmf{plain,right,tension=.5}{v2,v1}
\fmf{photon}{v2,o1}
\end{fmfgraph*}
\end{fmffile}
&
\begin{fmffile}{dx11dx22psi}
\begin{fmfgraph*}(30,30) \fmfpen{thick}
\fmfleft{i1}
\fmflabel{$\delta x_1$}{i1}
\fmfright{o1}
\fmflabel{$\delta y_2$}{o1}
\fmf{photon}{i1,v1}
\fmf{dots,right,tension=.5,label=$\psi_{12}$}{v1,v2}
\fmf{dots,right,tension=.5}{v2,v1}
\fmf{photon}{v2,o1}
\end{fmfgraph*}
\end{fmffile}\\
$(a)$&$(b)$
\end{tabular}
\end{center}
\caption{1-loop Feynman diagrams contributing to the $\delta  x_1 \delta y_2$ term of the effective Lagrangian.}
\label{dxpdxqij}
\end{figure}

We begin by considering terms that are off diagonal in the quiver index, and off diagonal in the vector index. For concreteness we will study $\delta x_1 \delta y_2$. The diagrams that contribute are shown in figure \ref{dxpdxqij}.
The first diagram, series expanded for small external momentum $l$, is given by:
\begin{equation}
4 {x}_{12} {y}_{12} (-i)i\kappa_{12}\int\frac{d\omega}{2\pi}D_{\phi_{12}}(\omega)D_{\phi_{12}}(\omega+l)=
-\kappa_{12}\left(
\frac{{x}_{12} {y}_{12}}{|\bold{x}_{12}|^3}
-\frac{l^2 {x}_{12} {y}_{12}}{4|\bold{x}_{12}|^5}\right)
+\mathcal{O}(l^3)~.
\label{eqdxpdxqij}
\end{equation}
The second diagram, series expanded for small external momentum $l$, is given by:
\begin{equation}
\kappa_{12}(-i)i\int\frac{d\omega}{2\pi}\mbox{tr}(\sigma_1D_{\psi_{12}}(\omega)\sigma_2D_{\psi_{12}}(\omega+l))=-\kappa_{12}\left(
-\frac{{x}_{12} y_{12}}{|\bold{x}_{12}|^3}
+\frac{l z_{12}}{2|\bold{x}_{12}|^3}
+\frac{l^2 x_{12} y_{12}}{4|\bold{x}_{12}|^5} \right)
+\mathcal{O}(l^3)~.
\end{equation} 
Summing the two, the only term that remains uncancelled is the term linear in $l$.

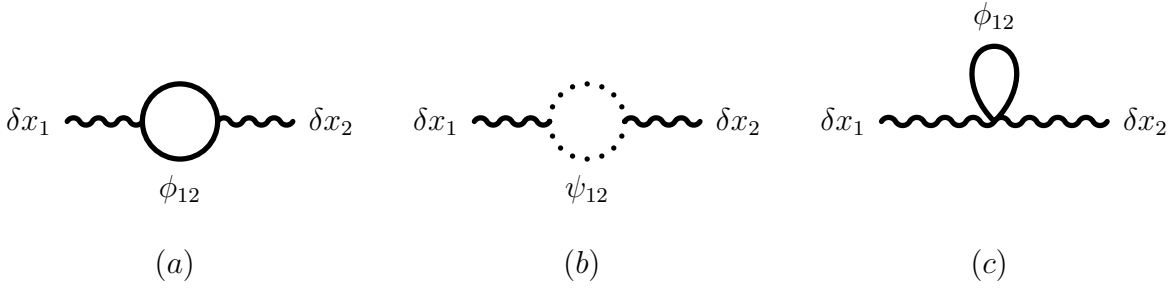
\begin{figure}[!ht]
\begin{center}
\setlength{\tabcolsep}{10mm}
\begin{tabular}{ccc}
\begin{fmffile}{dx11dx21phi}
\begin{fmfgraph*}(30,30) \fmfpen{thick}
\fmfleft{i1}
\fmflabel{$\delta x_1$}{i1}
\fmfright{o1}
\fmflabel{$\delta x_2$}{o1}
\fmf{photon}{i1,v1}
\fmf{plain,right,tension=.5,label=$\phi_{12}$}{v1,v2}
\fmf{plain,right,tension=.5}{v2,v1}
\fmf{photon}{v2,o1}
\end{fmfgraph*}
\end{fmffile}
&
\begin{fmffile}{dx11dx21psi}
\begin{fmfgraph*}(30,30) \fmfpen{thick}
\fmfleft{i1}
\fmflabel{$\delta x_1$}{i1}
\fmfright{o1}
\fmflabel{$\delta x_2$}{o1}
\fmf{photon}{i1,v1}
\fmf{dots,right,tension=.5,label=$\psi_{12}$}{v1,v2}
\fmf{dots,right,tension=.5}{v2,v1}
\fmf{photon}{v2,o1}
\end{fmfgraph*}
\end{fmffile}
&
\begin{fmffile}{dx11dx21loop}
\begin{fmfgraph*}(30,30) \fmfpen{thick}
\fmfleft{i1}
\fmflabel{$\delta x_1$}{i1}
\fmfright{o1}
\fmflabel{$\delta x_2$}{o1}
\fmf{photon}{i1,v1}
\fmf{plain,right,tension=1,label=$\phi_{12}$}{v1,v1}
%\fmf{plain,right,tension=.5}{v2,v1}
\fmf{photon}{v1,o1}
\end{fmfgraph*}
\end{fmffile}
\\
$(a)$&$(b)$&$(c)$
\end{tabular}
\end{center}
\caption{1-loop Feynman diagrams contributing to the $\delta x_1 \delta x_2$ term of the effective Lagrangian.}
\label{dxpdxqii}
\end{figure}

Next we consider term off diagonal in the quiver index, but diagonal in the vector index, for example $\delta \bold {x}_1^1\delta \bold{ x}_2^1$. The contributing diagrams are shown in figure \ref{dxpdxqii}.
The first diagram is given by (\ref{eqdxpdxqij}) with $y_{12}$ replaced by $x_{12}$. The second diagram gives:
\begin{equation}
\kappa_{12}(-i)i\int\frac{d\omega}{2\pi}\mbox{tr}(\sigma_1D_{\psi_{12}}(\omega)\sigma_1D_{\psi_{12}}(\omega+l))=
-\kappa_{12}\left(|y_{12}|^2+|z_{12}|^2\right)
\left(\frac{1}{|\bold{x}_{12}|^3}
-\frac{l^2}{4|\bold{x}_{12}|^5}\right)
+\mathcal{O}(l^3)~.
\end{equation}
%Remember that $|\bold{x}_{ij}^a|^2$ is simply the square of the $i$ component of $\bold{x}_i-\bold{x}_j$. 
The third diagram is given by:
\begin{equation}
2i\kappa_{12}\int\frac{d\omega}{2\pi}D_{\phi_{12}}(\omega)=\frac{\kappa_{12}}{|\bold{x}_{12}|}~.
\end{equation}
%We can now write the effective Lagrangian for off diagonal quiver index terms
%\begin{equation}
%\frac{1}{\kappa_{12}}L_{eff,\delta\vec x_1\delta\vec x_2}=
%\frac{l^2 \delta \vec x_1\cdot\delta \vec x_2}{4|\vec r_{12}|^3}
%-\frac{lr_{12}^3\delta \vec x_1^1\delta \vec x_2^2}{2|\vec r_{12}|^3}
%-\frac{lr_{12}^1\delta \vec x_1^2\delta \vec x_2^3}{2|\vec r_{12}|^3}
%-\frac{lr_{12}^2\delta \vec x_1^3\delta \vec x_2^1}{2|\vec r_{12}|^3}
%\end{equation}
%It is now straightforward to write the other two off diagonal terms.

Let us now consider terms that are diagonal in the quiver index. The Feynman diagrams contributing to $\delta \bold{x}_1^a \; \delta \bold{x}_1^b$ are shown in figure \ref{dxpdxpij}.
\begin{figure}[!ht]
\begin{center}
\setlength{\tabcolsep}{10mm}
\begin{tabular}{ccc}
\begin{fmffile}{dx1idx1jphi}
\begin{fmfgraph*}(30,30) \fmfpen{thick}
\fmfleft{i1}
\fmflabel{$\delta x_1^a$}{i1}
\fmfright{o1}
\fmflabel{$\delta x_1^b$}{o1}
\fmf{photon}{i1,v1}
\fmf{plain,right,tension=.5,label=$\phi_{12}$}{v1,v2}
\fmf{plain,right,tension=.5}{v2,v1}
\fmf{photon}{v2,o1}
\end{fmfgraph*}
\end{fmffile}
&
\begin{fmffile}{dx1idx1jpsi}
\begin{fmfgraph*}(30,30) \fmfpen{thick}
\fmfleft{i1}
\fmflabel{$\delta x_1^a$}{i1}
\fmfright{o1}
\fmflabel{$\delta x_1^b$}{o1}
\fmf{photon}{i1,v1}
\fmf{dots,right,tension=.5,label=$\psi_{12}$}{v1,v2}
\fmf{dots,right,tension=.5}{v2,v1}
\fmf{photon}{v2,o1}
\end{fmfgraph*}
\end{fmffile}
&
\begin{fmffile}{dx1idx1iloop}
\begin{fmfgraph*}(30,30) \fmfpen{thick}
\fmfleft{i1}
\fmflabel{$\delta x_1^a$}{i1}
\fmfright{o1}
\fmflabel{$\delta x_1^b$}{o1}
\fmf{photon}{i1,v1}
\fmf{plain,right,tension=1,label=$\phi_{12}$}{v1,v1}
%\fmf{plain,right,tension=.5}{v2,v1}
\fmf{photon}{v1,o1}
\end{fmfgraph*}
\end{fmffile}
\\
$(a)$&$(b)$&$(c)$
\\ & & \\
\begin{fmffile}{dx1idx1jphi31}
\begin{fmfgraph*}(30,30) \fmfpen{thick}
\fmfleft{i1}
\fmflabel{$\delta x_1^a$}{i1}
\fmfright{o1}
\fmflabel{$\delta x_1^b$}{o1}
\fmf{photon}{i1,v1}
\fmf{plain,right,tension=.5,label=$\phi_{31}$}{v1,v2}
\fmf{plain,right,tension=.5}{v2,v1}
\fmf{photon}{v2,o1}
\end{fmfgraph*}
\end{fmffile}
&
\begin{fmffile}{dx1idx1jpsi31}
\begin{fmfgraph*}(30,30) \fmfpen{thick}
\fmfleft{i1}
\fmflabel{$\delta x_1^a$}{i1}
\fmfright{o1}
\fmflabel{$\delta x_1^b$}{o1}
\fmf{photon}{i1,v1}
\fmf{dots,right,tension=.5,label=$\psi_{31}$}{v1,v2}
\fmf{dots,right,tension=.5}{v2,v1}
\fmf{photon}{v2,o1}
\end{fmfgraph*}
\end{fmffile}
&
\begin{fmffile}{dx1idx1iloop31}
\begin{fmfgraph*}(30,30) \fmfpen{thick}
\fmfleft{i1}
\fmflabel{$\delta x_1^a$}{i1}
\fmfright{o1}
\fmflabel{$\delta x_1^b$}{o1}
\fmf{photon}{i1,v1}
\fmf{plain,right,tension=1,label=$\phi_{31}$}{v1,v1}
%\fmf{plain,right,tension=.5}{v2,v1}
\fmf{photon}{v1,o1}
\end{fmfgraph*}
\end{fmffile}
\\
$(d)$&$(e)$&$(f)$
\end{tabular}
\end{center}
\caption{1-loop Feynman diagrams contributing to the $\delta \bold{x}_1^a \; \delta \bold{x}_1^b$ term of the effective Lagrangian.}
\label{dxpdxpij}
\end{figure}
%The effective Lagrangian for the $\delta \vec x_1\delta \vec x_1$ terms is given by:
One finds after similar calculations the following contribution to the effective Lagrangian:
\begin{multline}
%L_{eff,\delta\vec x_1\delta\vec x_1}=
\kappa_{12}\left(\frac{l z_{12} \delta {x}_1 \delta y_1}{4|\bold{x}_{12}|^3} 
+\frac{l x_{12} \delta y_1 \delta z_1}{4|\bold{x}_{12}|^3}
+\frac{l y_{12} \delta z_1 \delta x_1}{4|\bold{x}_{12}|^3}\right) \\
+\kappa_{31}\left(\frac{l z_{31} \delta  x_1\delta y_1}{4|\bold{x}_{31}|^3}
+\frac{l x_{31} \delta y_1 \delta z_1}{4|\bold{x}_{31}|^3}
+\frac{l y_{31} \delta z_1 \delta x_1}{4|\bold{x}_{31}|^3}\right) 
-\frac{\kappa_{12}l^2 \delta \bold{x}_1\cdot \delta \bold{x}_1}{8|\bold{x}_{12}|^3}
-\frac{\kappa_{31}l^2 \delta \bold{x}_1\cdot \delta \bold{x}_1}{8|\bold{x}_{31}|^3} ~.
\end{multline}

%\tar\tar It seems as if we don't mention that the linear in $l$ piece contributes to the magnetic monopole like vector field and that the quadratic in l piece gives the correction to the metric, but are the signs even right? It seems like, if the minus sign in front of the quadratic in $l$ piece is correct (comming from an $e^{ilt}$ type time dependence), then there should be an $i$ in front of the linear piece? Am I confused?\tar\tar

\subsubsection*{Quadratic Lagrangian}

Having computed all relevant Feynman diagrams that amount to integrating out the $\phi_{ij}$ and $\psi_{ij}$ fields we can now write the resulting quadratic piece of the effective Lagrangian: %It is given in Eq. \ref{3nodeeffective}.
\begin{equation}
L^{(2)}_{eff} =
\sum_i \frac{m_i}{2}\left( |\mathbf{\dot x}_i|^2+D_i^2 \right)
+\frac{1}{8}\sum_{i\rightarrow j}\frac{|\kappa_{ij}|}{|\bold{x}_{ij}|^3} \; \left( |\mathbf{\dot x}_i- \mathbf{\dot x}_j|^2 +  (D_i-D_j)^2 \right)~.
\label{3nodeeffective}
%\left(|\mathbf{\dot x}_i|^2-2\mathbf{\dot x}_i \cdot \mathbf{\dot x}_j+ |\mathbf{\dot x}_j|^2\right)
\end{equation}
Notice that the coefficient of $\dot{\bold{x}}_i \cdot \dot{\bold{x}}_j$ indeed matches that of $D_i \; D_j \;$. This can be simply generalized to the $N$-particle case, where the Lagrangian takes the same form as (\ref{3nodeeffective}). As was noted before, the linear piece was fixed by supersymmetry and given by (\ref{linear1}) and (\ref{linear2}) \cite{Denef:2002ru}. 

%\frac{\kappa_{ij}}{| \bold{x}_{ij}|^3}
%\frac{1}{8}\sum_{j\rightarrow i
%(D_i^2-2 D_i D_j+D_j^2)
%\begin{eqnarray}
%L_{eff}&=&
%\sum_p \frac{m_p}{2}\left( \mathbf{\dot x}_p^2+D_p^2 \right)-\theta_p D_p
%+\frac{1}{8}\sum_{q\rightarrow p}(\mathbf{\dot x}_p^2-2\mathbf{\dot x}_p \mathbf{\dot x}_q+\mathbf{\dot x}_q^2)\frac{\kappa_{pq}}{|\vec r_{pq}|^3}\nonumber\\
%&+&\frac{1}{2}\sum_{q\rightarrow p}(D_q-D_p)\frac{\kappa_{pq}}{|\vec r_{pq}|}+\frac{1}{8}\sum_{q\rightarrow p}(D_p^2-2 D_p D_q+D_q^2)\frac{\kappa_{pq}}{|\vec r_{pq}|^3}.
%\label{3nodeeffective}
%\end{eqnarray}

\subsection{Coulomb branch potential}\label{potential}

In this appendix we give the necessary formulae for the three-node Coulomb branch potential. Upon integrating out the chiral fields $\Phi^\alpha$ in (\ref{quiverqm}), while keeping the $\bold{q}^i$ fixed and time independent, and expanding up to quadratic order in the $D^{i}$ fields, the $D$-dependent piece of the Lagrangian is:
\begin{equation}
L_D =  \sum_{i=1}^3 \left( \frac{\mu^i}{2} D^i \; D^i - \theta^i D^i - s_i \, \frac{|\kappa^{i}|}{2 |\bold{q}^{i}|}D^{i} + \frac{|\kappa^{i}|}{8 |\bold{q}^{i}|^3} D^{i} \; D^i \right)~.
\end{equation}
Noting that $D^3 = D^1 + D^2$, we must integrate out $D^1$ and $D^2$ from $L_D$. The resulting on-shell Lagrangian is simply the potential $-V(\bold{q}^i)$. The equations of motion for $D^i$, using $s_1=s_2=-s_3=1$, are given by:
\begin{eqnarray}
\left( \mu^1 + \mu^3+ \frac{|\kappa^1|}{4 |\bold{q}^1|^3 }+\frac{|\kappa^3|}{4 |\bold{q}^3|^3 }\right)D^1 + \left( \mu^3 + \frac{|\kappa^3|}{4 |\bold{q}^3|^3 }\right)D^2 &=&  \theta^1 +\frac{s_1|\kappa^1|}{2 |\bold{q}^1|} + \frac{s_3|\kappa^3|}{2 |\bold{q}^3|}  \\
\left( \mu^2 + \mu^3 + \frac{|\kappa^2|}{4 |\bold{q}^2|^3 }+ \frac{|\kappa^3|}{4 |\bold{q}^3|^3 }\right)D^2 + \left( \mu^3 + \frac{|\kappa^3|}{4 |\bold{q}^3|^3 }\right)D^1 &=&  \theta^2 +\frac{s_2|\kappa^2|}{2 |\bold{q}^2|} + \frac{s_3|\kappa^3|}{2 |\bold{q}^3|} 
\end{eqnarray}
Writing the above equations as: $a_{ij} D^i = c_i$, the solution is simply given by:
\begin{equation}
D^1 = \frac{a_{12} c_2 - a_{22}c_1}{a_{12}^2 - a_{11}a_{22}}~, \quad \quad
D^2 = \frac{a_{12} c_1 - a_{11}c_2}{a_{12}^2 - a_{11}a_{22}}~. 
\end{equation}
Notice that the supersymmetric solution $D^i = 0$ is indeed given by $c_1 = c_2 = 0$ which is nothing more than equation (\ref{bound}).

\subsubsection*{Scaling Limit}

In the scaling regime, where $\mu^i=0,\,\theta^i=0$, and $\kappa^1>0$, $\kappa^2>0$ and $\kappa^3<0$ (or equivalently $s_1=s_2=-s_3=1$), the potential can be written as:
\begin{multline}\label{scalpot}
V(\bold{q}^i)=\frac{1}{\alpha+\beta+\gamma}\biggl(\frac{\alpha+\beta}{2\gamma}|\bold{q}^3|^4+\frac{\alpha+\gamma}{2\beta}|\bold{q}^2|^4+\frac{\beta+\gamma}{2\alpha}|\bold{q}^1|^4\\-|\bold{q}^1|^2|\bold{q}^2|^2-|\bold{q}^1|^2|\bold{q}^3|^2-|\bold{q}^2|^2|\bold{q}^3|^2\biggr)~,
\end{multline}
where
\begin{equation}
\alpha=\frac{|\bold{q}^1|^3}{\kappa^1}~,\quad\beta=\frac{|\bold{q}^2|^3}{\kappa^2}~,\quad\text{and }\gamma=-\frac{|\bold{q}^3|^3}{\kappa^3}~.
\end{equation}

\section{Thermal determinant}\label{thermalDet}

The thermal effective potential for the two-node quiver at finite temperature can be derived from (\ref{hotL}):
\begin{equation}\nonumber
L_{\rm eff} = \frac{\mu}{2} D^2 - \theta D - \kappa \ln \det \left( -(\partial_t + i A)^2 + |\bold{q}|^2  + D \right) + \kappa \ln \det \left( -(\partial_t + i A)^2 + |\bold{q}|^2 \right)~.
\end{equation}
Wick rotating $t\mapsto i t_E$ and periodically identifying $t_E\sim t_E+\beta$, introduces the thermal ensemble. The operator $\partial_{t_E}$ has eigenvalues on the Matsubara frequencies: $\omega_n=2\pi nT$ for bosons and $\omega_n=2\pi (n+\frac{1}{2})T$ for fermions, with $n\in \mathbb{Z}$. With $a\equiv A/2\pi T$ and denoting $|\bold{q}|^2\equiv s$ this yields:
\begin{multline}\nonumber
L^{(T)}_{\rm eff} =
\frac{\mu}{2} D^2 - \theta D - \frac{\kappa}{\beta} \sum\limits_{n\in\mathbb{Z}}\ln\left( 4\pi^2(n + a)^2 T^2 + s  + D \right)\\+\frac{\kappa}{\beta} \sum\limits_{n\in\mathbb{Z}}\ln\left( 4\pi^2(n + 1/2 + a)^2 T^2 + s \right)~,
\end{multline}
where we have now included a $1/\beta$ in the coefficient of the sums to cancel out the contribution from the integral over Euclidean time in $S_E=\int_0^{\beta}dt_E\,L^{(T)}_{\rm eff}$.
%\begin{eqnarray}\nonumber
%L^{(T)}_{bos,eff} = \frac{\mu}{2} D^2 - \theta D - \kappa \ln \det \left( (n + a)^2 T^2 +s  + D \right) + \kappa \ln \det \left( (n + 1/2 + a)^2 T^2 +s \right)~=\\\nonumber
%\frac{\mu}{2} D^2 - \theta D - \kappa \ln \prod\limits_{n\in\mathbb{Z}}\left( (n + a)^2 T^2 + s  + D \right)+ \kappa \ln \prod\limits_{n\in\mathbb{Z}}\left( (n + 1/2 + a)^2 T^2 +s \right)=\\\nonumber
%\frac{\mu}{2} D^2 - \theta D - \kappa \sum\limits_{n\in\mathbb{Z}}\ln\left( (n + a)^2 T^2 +s  + D \right)+ \kappa \sum\limits_{n\in\mathbb{Z}}\ln\left( (n + 1/2 + a)^2 T^2 + s \right)~.
%\end{eqnarray}
Taking a derivative with respect to $s$:
\begin{multline}\label{thermalL}
\frac{d L_{\rm eff}^{(T)}}{ds}=\frac{\kappa}{\beta}\sum\limits_{n\in\mathbb{Z}}f(n)=-\frac{\kappa}{\beta}\sum\limits_{n \in\mathbb{Z}}\frac{1}{4\pi^2(n+a)^2T^2+s+D}\\+\frac{\kappa}{\beta}\sum\limits_{n\in\mathbb{Z}}\frac{1}{4\pi^2(n+{1}/{2}+a)^2T^2+s}~.
\end{multline}
We will treat the two sums separately using contour integration methods. We write (\ref{thermalL}) as $f(n)=-f_1(n)+f_2(n)$, denoting the two summands respectively. Consider: 
$$\oint_{\mathcal{C}} \pi f_1(z) \cot(\pi z)=\oint_{\mathcal{C}}F(z)dz=2\pi i\left(\text{Res} F(z)|_{z=z^+}+\text{Res} F(z)|_{z=z^-}+\sum\limits_{n\in \mathbb{Z}}\text{Res} F(z)|_{z=n}\right)~,$$ 
with $z^{\pm}=-a\pm\frac{i}{2\pi T}\sqrt{s+D}$ and $\text{Res}F(z)|_{z=z^\pm}=\pm\frac{i\cot(\mp\frac{i}{2T}\sqrt{s+D}+\pi a)}{4T\sqrt{s+D}}$ and $\mathcal{C}$ a circle of radius $R$ around the origin with $R\rightarrow\infty$. Given that $F(z)$ decays rapidly enough, the integral evaluates to zero along $\mathcal{C}$ and we obtain:
\begin{equation}
-\sum\limits_{n\in\mathbb{Z}} f_1(n) =-\frac{i}{4T\sqrt{s+D}}\left(\cot\left(\frac{i}{2T}\sqrt{s+D}+\pi a\right)-\cot\left(-\frac{i}{2T}\sqrt{s+D}+ \pi a \right)\right)~.%\frac{1}{(n+a)^2T^2+s+D)}
\end{equation}
For the fermionic sum, the procedure is completely analgous, and (using $\cot(x+\pi/2)=-\tan(x)$) we find:
\begin{equation}
\sum\limits_{n\in\mathbb{Z}} f_2(n) =-\frac{i}{4T\sqrt{s}}\left(\tan\left(\frac{i\sqrt{s}}{2T}+\pi a\right)-\tan\left(-\frac{i\sqrt{s}}{2T}+\pi a\right)\right)~. %\frac{1}{(n+\frac{1}{2}+a)^2T^2+s)}
\end{equation}
We need to integrate these two sums with respect to $s$ to obtain $L_{\rm eff}^{(T)}$. Using the following identities: $\int \frac{\cot(a\sqrt{x}+b)}{\sqrt{x}}dx=2\log \sin (a\sqrt{x}+b)/a$ and $\int \frac{\tan(a\sqrt{x}+b)}{\sqrt{x}}dx=-2\log \cos(a\sqrt{x}+b)/b$, we find:
\begin{equation}
L_{\rm eff}^{(T)}=\frac{\mu}{2}D^2-\theta D-\kappa \, T\log \left(\frac{\cosh\left(\frac{\sqrt{|\bold{q}|^2+D}}{T}\right)-\cos(2\pi a)}{\cosh\left(\frac{|\bold{q}|}{T}\right)+\cos(2\pi a)} \right)~,
\end{equation}
as claimed.

\end{document}